\documentclass[remotesensing,article,accept,pdftex,moreauthors]{Definitions/mdpi} 
\usepackage{tikz}
\firstpage{1} 
\pubvolume{1}
\issuenum{1}
\articlenumber{0}
\pubyear{2025}
\copyrightyear{2025}
\externaleditor{Prasad S. Thenkabail}
\datereceived{3 April 2025} 
\daterevised{17 May 2025} 
\dateaccepted{19 May 2025} 
\datepublished{} 
\hreflink{https://\linebreak doi.org/} 

\usepackage{units}          
\usepackage{subcaption}     
\usepackage{comment}
\usepackage{xspace}
\usepackage{placeins}

\newcommand{\angstrom}{\r{A}ngstr\"{o}m\xspace} 

\definecolor{lightred}{RGB}{230, 180, 179}
\newcommand{\mz}[1]{}
\newcommand{\md}[1]{}
\newcommand{\mdc}[1]{}

\newcommand{\mg}[1]{}
\newcommand{\rg}[1]{}
\newcommand{\pb}[1]{}
\newcommand{\obb}[1]{}

\newcommand\ddfrac[2]{\ensuremath{\frac{\displaystyle #1}{\displaystyle #2}}}
\newcommand{\ud}{\mathrm{d}}
\newcommand{\avg}[1]{\left< #1 \right>}

\usepackage[style=super,nopostdot]{glossaries} 
\makeglossaries

\makeglossaries

\newglossaryentry{brl}
{   name=BRL,
    description={Barcelona Raman LIDAR}
}

\newglossaryentry{pbrl}
{   name=pBRL,
    description={pathfinder Barcelona Raman LIDAR}
}

\newglossaryentry{clue}{
    name=CLUE,
    description={Cherenkov Light Ultraviolet Experiment}
}

\newglossaryentry{orm}{
    name=ORM,
    description={Roque de los Muchachos Observatory}
}

\newglossaryentry{ctao}{
    name=CTAO,
    description={Cherenkov Telescope Array Observatory} 
    }

\newglossaryentry{ctao-n}{
    name=CTAO-N,
    description={Northern site of the Cherenkov Telescope Array Observatory}
}

\newglossaryentry{ctao-s}{
    name=CTAO-S,
    description={Southern site of the Cherenkov Telescope Array Observatory} 
    }

\newglossaryentry{eso}{
    name=ESO,
    description={European Southern Observatory}
}

\newglossaryentry{asl}{
    name=a.s.l.,
    description={Above sea level}
}

\newglossaryentry{aod}{
    name=AOD,
    description={Aerosol Optical Depth}
}

\newglossaryentry{vod}{
    name=VOD,
    description={Vertical Optical Depth}
}

\newglossaryentry{rl}{
    name=RL,
    description={Raman LIDAR}
}

\newglossaryentry{lpp}{
    name=LPP,
    description={LIDAR PreProcessing software}
}

\newglossaryentry{rmsd}{
    name=RMSD,
    description={root-mean-square deviation}
}

\newglossaryentry{magic}{
    name=MAGIC,
    description={Major Atmospheric Gamma-ray Imaging Cherenkov Telescope} 
    }

\newglossaryentry{pbl}{
    name=PBL,
    description={Planetary Boundary Layer}
}

\newglossaryentry{ifae}{
    name=IFAE,
    description={Institut de Física d’Altes Energies}
    }

\newglossaryentry{uab}{
    name=UAB,
    description={Universitat Autònoma de Barcelona}
    }

\newglossaryentry{infn}{
    name=INFN,
    description={Istituto Nazionale di Fisica Nucleare} 
    }

\newglossaryentry{iac}{
    name=IAC,
    description={Instituto de Astrofísica de Canarias} 
    }

\newglossaryentry{ung}{
    name=UNG,
    description={University of Nova Gorica} 
    }

\newglossaryentry{ndf}{
    name=NDF,
    description={Number of Degrees of Freedom} 
    }

\newglossaryentry{rhi}{
    name=RHI,
    description={Range-Height Indication} 
    }

\newglossaryentry{us}{
    name=US,
    description={United States} 
    }

\newglossaryentry{thi}{
    name=THI,
    description={Time-Height Indication} 
    }

\newglossaryentry{rgb}{
    name=RGB,
    description={Red Green Blue} 
    }
\newglossaryentry{msg}{
    name=MSG,
    description={Meteosat Second Generation} 
    }

\newglossaryentry{wgs84}{
    name=WGS84,
    description={World Geodetic System 1984} 
    }

\newglossaryentry{ltcs}{
    name=LTCS,
    description={Laser Traffic Control System} 
    }

\newglossaryentry{rcs}{
    name=RCS,
    description={Range-Corrected Signal} 
    }

\newglossaryentry{llg}{
    name=LLG,
    description={Liquid Light Guide} 
    }

  \newglossaryentry{psd}{
    name=PSD,
    description={Power Spectral Density} 
    }
  
\newglossaryentry{gui}{
    name=GUI,
    description={Graphical User Interface} 
    }
    
\newglossaryentry{fits}{
    name=FITS,
    description={Flexible Image Transport System} 
    }

\newglossaryentry{mdp}{
    name=MDP,
    description={Molecular Density Profile} 
    }

\newglossaryentry{irl}{
    name=IRL,
    description={INFN Raman LIDAR} 
    }

\newglossaryentry{lupm}{
    name=LUPM,
    description={Laboratoire Univers et Particules de Montpellier} 
    }

\newglossaryentry{iact}{
    name=IACT,
    description={Imaging Atmospheric Cherenkov Telescope} 
    }

\newglossaryentry{fov}{
    name=FoV,
    description={Field of View} 
    }

\newglossaryentry{snr}{
    name=SNR,
    description={Signal-to-Noise Ratio} 
    }

\newglossaryentry{lr}{
    name=LR,
    description={LIDAR Ratio} 
    }

\newglossaryentry{enf}{
    name=ENF,
    description={Excess-Noise Factor} 
    }

\newglossaryentry{an}{
    name=AN,
    description={Analog channel} 
    }

\newglossaryentry{pc}{
    name=PC,
    description={Photon-Counting Channel} 
    }

\newglossaryentry{prf}{
    name=PRF,
    description={Pulse Repetition Frequency} 
    }

\newglossaryentry{nsb}{
    name=NSB,
    description={Night-Sky Background} 
    }

\newglossaryentry{psf}{
    name=PSF,
    description={Point Spread Function} 
    }

\newglossaryentry{pao}{
    name=PAO,
    description={Pierre Auger Observatory} 
    }

\newglossaryentry{vrr}{
    name=VRR,
    description={Vibrational-Rotational Raman line}
}

\newglossaryentry{pmt}{
    name=PMT,
    description={Photomultiplier Tube}
}

\newglossaryentry{lst}{
    name=LST,
    description={Large Size Telescope}
}
\newglossaryentry{dac}{
    name=DAC,
    description={Digital-to-Analog Converter}
}

\newglossaryentry{hv}{
    name=HV,
    description={High Voltage} 
    }

\newglossaryentry{lotr}{
    name=LOTR,
    description={Licel Optical Transient Recorder} 
}

\newglossaryentry{vaod}{
    name=VAOD,
    description={Vertical Aerosol Optical Depth} 
    }

\newglossaryentry{pde}{
    name=PDE,
    description={Photon Detection Efficiency}
}
    
\newglossaryentry{uv}{
    name=UV,
    description={Ultra-Violet} 
    }

\Title{A 1.8~m Class Pathfinder Raman LIDAR 
for the Northern Site of the Cherenkov 
Telescope Array Observatory---Performance}

\TitleCitation{A 1.8~m Class Pathfinder Raman LIDAR for the Northern Site of the Cherenkov Telescope Array Observatory---Performance}

\Author{
Pedro José 
 Bauzá-Ruiz~$^{1}$\orcidP, Oscar Blanch~$^{2}$\orcidB, Paolo G. Calisse~$^{3}$, Anna Campoy-Ordaz~$^{1}$\orcidC, 
Sidika Merve Çolak~$^{2}$\orcidI, Michele Doro~$^{4,5,\dagger}$\orcidM, Lluis Font~$^{1}$\orcidF, Markus Gaug~$^{1,}$*\orcidG, Roger Grau~$^{2,\dagger}$\orcidR, Darko Kolar~$^{6}$\orcidK, Camilla Maggio 
~$^{1,\ddagger}$\orcidJ, Manel Martinez~$^{2}$\orcidT, Samo Stanič~$^{6}$\orcidS, Santiago Ubach~$^{1}$\orcidU, Marko Zavrtanik~$^{6}$\orcidZ~and~Miha Živec~$^{6,\dagger}$\orcidX
}

\AuthorNames{Pedro José Bauzá-Ruiz, Oscar Blanch, Paolo G. Calisse, Anna Campoy-Ordaz, Sidika Merve Çolak, Michele Doro, Lluis Font, Markus Gaug, Roger Grau, Darko Kolar, Camilla Maggio, Manel Martinez, Samo Stanič, Santiago Ubach, Marko Zavrtanik and Miha Živec}

\AuthorCitation{Bauzá-Ruiz, P.J. 
; Blanch, O.; Calisse, P.G.; Campoy-Ordaz, A.; Çolak, S.M.; Doro, M.; Font, L.; Gaug, M.; Grau, R.; Kolar, D.; et al.}

\address{%
$^{1}$ \quad Departament de F\'{i}sica, Universitat Aut\`{o}noma de Barcelona and CERES-IEEC, 08193 Bellaterra, Spain; bauza@ieec.cat (P.B.-R.); campoy@ieec.cat (A.C.-O.); lluis.font@uab.cat (L.F.); camilla.maggio.2@gmail.com (C.M.); santiago.ubach@autonoma.cat (S.U.)
 \\
$^{2}$ \quad Institut de Fisica d'Altes Energies (IFAE), 08193 
 Bellaterra, Spain; blanch@ifae.es (O.B.); sidikamerve.colak@autonoma.cat (S.M.Ç.); rgrau@ifae.es (R.G.); martinez@ifae.es (M.M.)\\
$^{3}$ \quad Cherenkov Telescope Array Observatory gGmbH (CTAO gGmbH), Saupfercheckweg 1, \linebreak  69117 Heidelberg,  Germany; paolo.calisse@cta-observatory.org \\
$^{4}$ \quad Department of Physics and Astronomy, University of Padova, I-35131 Padova, Italy; michele.doro@unipd.it\\
$^{5}$ \quad Istituto Nazionale di Fisica Nucleare (INFN), sez. Padova, I-35131 Padova, Italy; michele.doro@unipd.it 
$^{6}$ \quad Center for Astrophysics and Cosmology, University of Nova Gorica, Vipavska 13, 5000 Nova Gorica, Slovenia; darko.kolar@ung.si (D.K.); samo.stanic@ung.si (S.S.); marko.zavrtanik@ijs.si (M.Z.); miha.zivec@ung.si (M.Ž.)\\
}

\corres{Correspondence: markus.gaug@uab.cat}

\firstnote{These 
 authors contributed equally to this work.} 

\secondnote{Current address: CAEN Tools for Discovery  S.p.A., 55049 Viareggio, Italy.}

\abstract{The Barcelona Raman LIDAR (\gls{brl}) will provide continuous monitoring of the aerosol extinction profile along the line of sight of the Cherenkov Telescope Array Observatory (\gls{ctao}). It will be located at its Northern site (\gls{ctao-n}) on the Observatorio del Roque de Los Muchachos. 
This article presents the performance of the pathfinder Barcelona Raman LIDAR (\gls{pbrl}), a prototype instrument for the final \gls{brl}. Power budget simulations were carried out for the \gls{pbrl} operating under various conditions, including clear nights, moon conditions, and dust intrusions. The LIDAR PreProcessing (\gls{lpp}) software suite is presented, which includes several new statistical methods for background subtraction, signal gluing, ground layer and cloud detection and inversion, based on two elastic and one Raman lines. 
Preliminary test campaigns were conducted, first close to Barcelona and later at \gls{ctao-n}, albeit during moonlit nights only. The \gls{pbrl}, under these non-optimal conditions, achieves maximum ranges up to about 35~km, range resolution of about 50~m for strongly absorbing dust layers, and 500~m for optically thin clouds with the Raman channel only, leading to similar resolutions for the LIDAR ratios and \angstrom exponents. Given the reasonable agreement between the extinction coefficients obtained from the Raman and elastic lines independently, an accuracy of aerosol optical depth retrieval in the order of 0.05 can be assumed with the current setup.
The results show that the \gls{pbrl} can provide valuable scientific results on aerosol characteristics and structure, although not all performance requirements could be validated under the conditions found at the two test sites. Several moderate hardware improvements are planned for its final upgraded version, such as gated \glspl{pmt} for the elastic channels and a reduced-power laser with a higher repetition rate, to ensure that the data acquisition system is not saturated and therefore not affected by residual ringing. 
}


\begin{document}

\section{Introduction
}
\label{sec:introduction}

The Cherenkov Telescope Array Observatory (\gls{ctao})~\citep{ctaconcept,ScienceCTA:2019} is
the next-generation observatory of the ground-based Imaging Atmospheric Cherenkov Telescopes (\glspl{iact}) class.
\gls{ctao} will observe high-energy cosmic radiation (``gamma rays'') for research in high-energy astrophysics.
The observatory is composed of more than 60 telescopes at two locations: in the northern hemisphere, \gls{ctao-n} 
is currently under construction at the Observatorio del Roque de Los Muchachos (\gls{orm}, La Palma, Canary Islands, Spain, 28$^\circ$N 17$^\circ$W), and~in the southern hemisphere, \gls{ctao-s} will be constructed at a site belonging to the European Southern Observatory (\gls{eso}), which is close to Cerro Paranal, Chile (24$^\circ$S  70$^\circ$W). 
The telescope arrays are spread over an area of approximately half a square kilometre in the north and three square kilometres in the south and are both located at altitudes of around 2200~m above sea~level.

\glspl{iact} observe the few nanosecond-long flashes of \gls{uv}-blue Cherenkov light emitted by particle cascades generated in the atmosphere by cosmic gamma rays entering the Earth's atmosphere. They use, for~that purpose, large reflectors of up to 28~m diameter and fields-of-view (\gls{fov}) of $4^\circ$--$10^\circ$ for current operating systems. 
By analyzing the distribution of this Cherenkov light received, \glspl{iact} reconstruct the timing, energy, and~direction of the primary gamma ray. 

Because the Cherenkov light is emitted around 5--20~km above the ground, a~dominant contribution to the systematic uncertainty in the gamma-ray energy and flux reconstruction of \gls{iact} results from an inaccurate determination of atmospheric transmittance~\citep{Magic:performance,Hess:performance}. 
For this reason, \gls{ctao} has been programmed~\citep{Gaug:2017Atmo} to continuously monitor the atmosphere along the line of sight of the observing telescopes 
and to, in particular, assess the aerosol extinction profile with Raman LIDARs (\glspl{rl})~\citep{Ballester:2019}, together with the assessment of the Aerosol Optical Depth (\gls{aod}) across the observed \gls{fov}, with a stellar photometer specifically designed for that purpose~\citep{Ebr:2021}. 

Atmospheric conditions at world-class sites for astronomy, such as \gls{ctao-n} and \gls{ctao-s}, are characterized by low aerosol content~\citep{Laken:2016,Fruck:2022igg}, few episodes of dust intrusions~\citep{Lombardi:2010}, 
frequent absence of clouds or presence of cloud cover with negligible optical depth~\citep{Otarola:2017}, and~generally a highly dry and transparent atmosphere~\citep{Hellemeier:2019}, which leads to excellent observing~conditions.  

\gls{iact} science data cannot be reasonably analyzed with \glspl{aod} larger than about 0.7, even if the optical properties of the atmosphere are well characterized~\citep{Schmuckermaier:2023huo,Pecimotika:2023}, because~the influence of the residual systematics on the data becomes too large. Hence, in~this regime, LIDAR-based characterization is not required.
On the other hand, LIDAR-based extinction profiles for \gls{aod} below 0.7 have been successfully used to recalibrate \gls{iact} data~\citep{Schmuckermaier:2023huo} and are foreseen to become an integral part of the \gls{ctao} calibration chain~\citep{Prester:2024}. 
In such conditions, \gls{ctao} requires aerosol profiling with a 
range resolution better than a few hundred meters~\citep {Prester:2024,technicalpaper}. 
LIDARs at \gls{iact} installations shall characterize the entire troposphere and the lower stratosphere.
Since under normal conditions, the~height of the nocturnal planetary boundary layer (\gls{pbl}) at the \gls{ctao-n} site reaches $\lesssim$800~m above ground~\citep{Fruck:2022igg,Sicard:2010} and the Cherenkov light from typical gamma-ray induced particle showers is emitted entirely above it~\citep{HILLAS:JPGPP1990a}, the~fine structure of the \gls{pbl} does not need to be resolved~\citep{Garrido:2013}. Only in the event of Saharan dust 
intrusions (the so-called ``calima'' on the Canary Islands~\mbox{\citep{Lombardi:2008,Fruck:2022igg}}) does the~\gls{pbl} reach higher altitudes, up~to $\sim$6~km above ground~\citep{Sicard:2010,Fruck:2022igg,Barreto:2022}.

High-level  \gls{ctao} science requirements translate into requirements for the determination of atmospheric parameters in terms of clouds and \gls{pbl} properties in addition to operation requirements: 

\begin{enumerate}[label=,leftmargin=0em,labelsep=4mm]
\item \textbf{Cloud properties: 
} The \gls{rl} shall detect clouds in an altitude range from 2~km to 20~km above ground within a cloud Vertical Optical Depth (\gls{vod}) range of 0.01 $<$ \gls{vod} $<$ 0.7. It shall measure \gls{vod} for the detected clouds with an accuracy equal to or better than 0.03 root mean square deviation (\gls{rmsd}) for each laser wavelength. It shall also measure base and top heights for the detected clouds with an accuracy equal to or better than 300~m \gls{rmsd}. 
\item \textbf{PBL properties:} The \gls{rl} shall detect the \gls{pbl} with heights ranging from 0.5~km to 9~km above ground and Vertical Aerosol Optical Depths (\glspl{vaod}) ranging from 0.03 to 0.7. It shall provide \glspl{vaod} for the detected \gls{pbl} with an accuracy equal to or better than 0.03~\gls{rmsd} for each laser wavelength. It shall also measure the heights of detected \glspl{pbl} with an accuracy equal to or better than 300~m \gls{rmsd}. {It shall measure the extinction \angstrom exponent with an accuracy better than 0.3~\gls{rmsd}}.
\item \textbf{Pointing capability:} The \gls{rl} shall have a range of pointing directions starting from 25$^\circ$ or lower elevation angles up to zenith and be applicable to all azimuth angles.
\item \textbf{Measurement time:} To limit interference with \gls{ctao} science observations, the~RL shall measure the aerosol extinction profile to the required accuracy along any line-of-sight within the pointing limits within one minute or less.
\end{enumerate}

These requirements may already be achieved with elastic lines only if reasonable efforts are made to continuously maintain an absolute LIDAR calibration~\citep{Fruck:2022igg,Gaug:2022}, and~external information is available about the size distribution of aerosols~\citep{Lombardi:2008,Rodriguez2009,Berjon:2019}. 
However, the~availability of Raman lines allows for improved accuracy obtained directly from the data rather than on the basis of regular calibration procedures. This led us to develop a custom-designed four-wavelength \gls{rl}~\citep{technicalpaper}, which is described hereafter.

\subsection{Pathfinder Barcelona Raman~LIDAR}

For the \gls{ctao-n} purposes, 
the Institut de Fisica d'Altes Energies (\gls{ifae}, Barcelona, Spain), in~collaboration with the Autonomous University of Barcelona (\gls{uab}, Barcelona, Spain), the~Italian National Institute for Nuclear Physics (\gls{infn}, Padova, Italy) and the University of Nova Gorica (\gls{ung}, Nova Gorica, Slovenia) designed a \gls{rl}~\citep{doro:2013,Ballester:2019,technicalpaper}, which we hereafter call the Barcelona Raman LIDAR (\gls{brl}).

A pathfinder instrument for the \gls{brl} was assembled at the UAB Campus during the years 2014--2019. We label this instrument 
``pathfinder Barcelona Raman LIDAR (\gls{pbrl})''. 
An exhaustive description of the technical details and design requirements of the \gls{pbrl} was presented in a sister manuscript~\citep{technicalpaper}. This work focuses on the performance of the instrument.  The~general idea is to apply the lessons learnt from the \gls{pbrl} in order to upgrade it to its final version, the~\gls{brl}, which will be delivered to the \gls{ctao} as the observatory RL for \gls{ctao-n}.

\begin{figure}[H]
\includegraphics[width=0.75\linewidth]{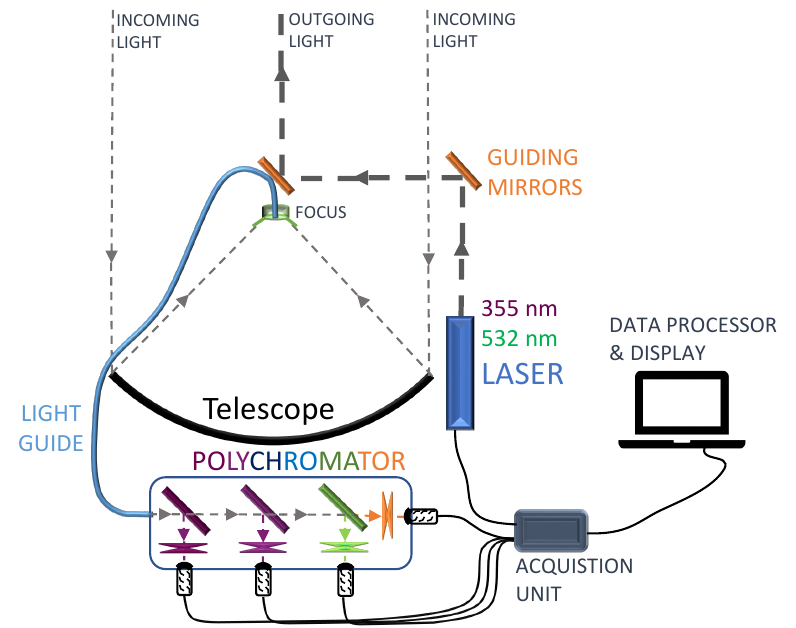}
\caption{A 
 schematic drawing of the \gls{pbrl} and its main components: the receiver, comprising the telescope, the~polychromator, and~the data acquisition unit, and~the transmitter, comprising the laser and guiding mirrors. {The figure has been adopted from~Ballester et~al.~\cite{technicalpaper}}.
\label{fig:lidar_scheme}}
\end{figure}

\begin{figure}[H]
    \raisebox{6.1cm}{%
    \includegraphics[angle=-90,width=.33\linewidth]{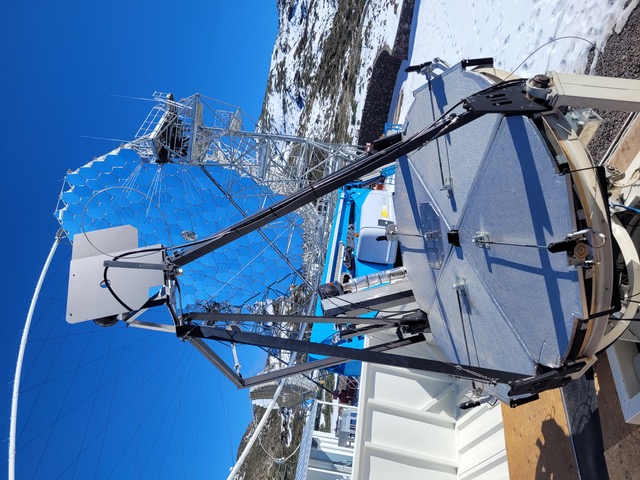}} \hfill
    \includegraphics[width=0.65\linewidth,trim={26.5cm 12.5cm 9cm 2cm},clip]{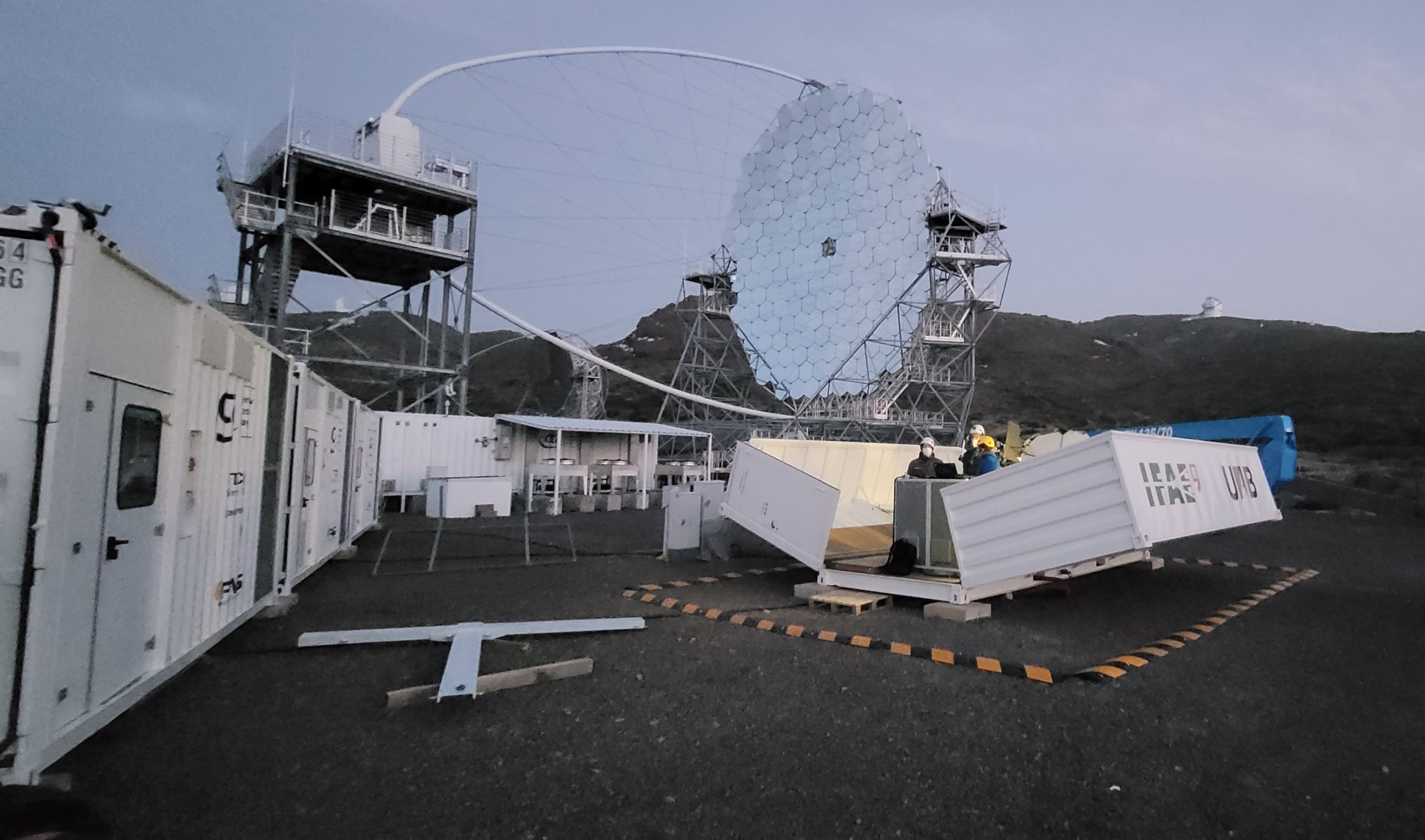}     
    \caption{Pictures of the \gls{pbrl} during its test campaign at the \gls{orm} next to \gls{ctao}'s first Large Size Telescope (\gls{lst}-1). {On the left side, a~close-up of the LIDAR is shown with closed mirror protection petals. The~\gls{lst}-1 telescope is seen in the background. On~the right side, the~protective enclosure container is visible 
, as well as the full area within the \gls{lst}-1, which had been temporarily lent to the \gls{pbrl}}.\label{fig:pBRL_LST}}
\end{figure}

The \gls{pbrl} is shown schematically in Figure~\ref{fig:lidar_scheme}, and~a picture of the instrument is shown in Figure~\ref{fig:pBRL_LST}. The~\gls{pbrl} was built by refurbishing a foldable commercial transport container containing a steerable telescope equipped with a 1.8~m diameter parabolic mirror of 1.8~m focal length, originally belonging to the Cherenkov Light Ultraviolet Experiment (\gls{clue})~\citep{Alexandreas:1995}. 
The mirror is a float-glass solid substrate with aluminum coating, which was recoated with a quartz protection in 2022~\citep{technicalpaper}. The~\gls{pbrl} hosts a pulsed 10~Hz  Brilliant  Nd:YAG laser from the company Quantel~\citep{webquantel} emitting at 1064~nm with frequency doubling and tripling at wavelengths of 532~nm and 355~nm. The~laser light is guided via a set of two mirrors to be made coaxial. 
This allows us to measure aerosol scattering and extinction through the two elastic and two vibro-rotational (\gls{vrr}) anti-Stokes Raman backscattering on N$_2$~\citep{Zenteno:2021} at 387~nm and 607~nm. The~latter was not used for the \gls{pbrl} but~is considered for the final \gls{brl} design. 
The focused light on the focal plane is transported  via an 8~mm diameter, 3.2~m long Liquid Light Guide (\gls{llg}) of the Lumatec Series 300~\citep{web.lumatec}, with a fused silica window and fluoropolymer tubing, 
to a 
custom-designed optical system for
wavelength separation and light detection  (the so-called ``polychromator''). The~polychromator is equipped with lens-doublets to focus the light beam and dichroic mirrors to progressively separate the four wavelengths of interest, and~10~nm wide interference filters to filter light at four photomultiplier tubes (\gls{pmt}) units. The~\gls{pmt} signals are digitized in four commercial Licel Optical Transceiver Recorder (\gls{lotr}) units.

\vspace{-9pt}

\subsection{Datataking~Campaigns}

The first commissioning tests were carried out in 2018--2020 at the \gls{uab} Campus. 
After turning off internal street lighting, the~area was sufficiently dark on those nights when the \gls{pbrl} was tested. 
During that time, the~system was gradually commissioned, complemented with \gls{lotr} and control hardware and software. 
In 2020, the~primary mirror, degraded by being inoperative since the time when \gls{clue} was operating (until the year 2000), was realuminized and quartz-coated~\citep{technicalpaper}. Only one data set, taken on 7 December 2020, was obtained at the \gls{uab} Campus after this intervention. 
Shortly after, the~\gls{pbrl} 
was shipped to the \gls{orm}, the~\gls{ctao-n} site. 
It was temporarily installed next to the Large Size Telescope~(\gls{lst})-1, the~first installation of \gls{ctao-n} (see Figure~\ref{fig:pBRL_LST}) near~its final future location.

The \gls{pbrl} was deployed at \gls{orm} from February 2021 to May 2022. During~this period, the~\gls{pbrl} operated for a total of 33~twilight time slots during moon nights and accumulated approximately 20 h of data. 
Its operation was, due to restrictions imposed by the Instituto de Astrofísica de Canarias (\gls{iac}), which managed the \gls{orm} site, and~the operation scheme of the \gls{lst}-1 telescope, mainly confined to the astronomical twilight periods of moonlit nights, when the rest of the observatory, and~particularly the \gls{lst}-1, did not operate. This constraint, along with the need to protect the equipment, resulted in reduced LIDAR performance, as the system was designed to operate during dark nights or moderate moonlight when the \gls{ctao} would be collecting data. 
The eruption of the Cumbre Vieja (Tajogaite) volcano in September 2021 and the subsequent deposition of volcanic ash also led to a suspension of observatory activities, including \gls{pbrl} testing. Operation resumed in January 2022 and lasted until May before~the LIDAR was finally dismantled and transported back to the \gls{uab} Campus.

The paper is structured as follows. In~Section~\ref{sec:simulation}, we characterize the \gls{pbrl} via the power budget simulation method, computing the expected return power and the signal-to-noise ratio and verifying the expectation to meet the \gls{ctao} requirements. In~\mbox{Sections~\ref{sec:evaluation_signal}--\ref{sec:rebinning}}, we describe in detail the signal reconstruction procedures that we have developed, 
with special consideration on statistical soundness. 
Section~\ref{sec:profiles} presents the inversion methods used for the characterization of the ground layer and clouds. 
In Section~\ref{sec:results}, we show results on the performance of the \gls{pbrl} using several case studies of data taken during measurement campaigns close to Barcelona and at the \gls{orm}, highlighting cases of clear sky, the~presence of calima, and atmospheric volcanic ash. 
We discuss and compare our \gls{pbrl} with other current systems in Section~\ref{sec:discussion} and draw our conclusions in Section~\ref{sec:conclusion}.


\section{Methods 
}
\label{sec:methods}
{As part of the \gls{rl}'s in-kind contribution to the \gls{ctao}, we have developed an advanced LIDAR analysis software suite, which incorporates several new and innovative methods using robust statistics, a~significantly improved likelihood-based gluing technique including methods for the correction of baseline ringing, a~new rebinning {algorithm}, and~a novel way of layer detection and treatment. Most of these are built as updates on existing literature methods, while some are to be considered as novel contributions. Those novel techniques will be introduced in this section.}

\subsection{Performance Simulation Using the Return Power Budget~Method}
\label{sec:simulation}

In order to estimate the \gls{brl} capabilities and to check whether its design~\citep{technicalpaper} was able to meet the \gls{ctao} requirements, we first carried out a computational analysis to assess the sensitivity and performance capabilities. We developed a power-budget (or link-budget, see, e.g.,~\cite{Measures1984}) simulation program, originally developed in~\citet{Eizmendi:2011}, to~calculate the expected return power and signal-to-noise ratio (\gls{snr}) for single laser shots at each \gls{brl} laser wavelength and at various altitudes while considering different atmospheric conditions. We also computed the time required to achieve a given minimum \gls{snr} because of the short time allowed to operate the \gls{brl} during standard \gls{ctao} operations (see Section~\ref{sec:introduction}).

\subsubsection{Return~Power}
\label{sec:budget_power}
The computation of the return power 
takes into account molecular and aerosol backscattering for the case of the elastic lines~\citep{tomasi} (355~nm, 532~nm) and the two $\textrm{N}_2$ Raman lines (387~nm and 607~nm). Average atmospheric density profiles~\citep{Fruck:2022igg} for the \gls{orm} were assumed, as~well as average aerosol extinction profiles for the clear night or typical cases of nights affected by calima~\citep{Barreto:2022}. 
\angstrom exponents  
  ($k = -\ln\left(\alpha^\mathrm{aer}_{\lambda_1}/\alpha^\mathrm{aer}_{\lambda_2}\right)/\ln\left(\lambda_1/\lambda_2\right)$.) of $k=1.45$ have been assumed for the clear night and $k=0.32$ for the case of strong calima 
~\citep{Schmuckermaier:2023huo}.  

At full overlap, the~return power budget $P_{\lambda_0}(R)$ for the elastic lines as a function of range $R$ and wavelength $\lambda_0$, is then computed using the standard LIDAR equation 
~\citep{Ansmann:1992}:
\begin{equation}\label{eq:return_power_elastic}
    P_{\lambda_0}(R) = \frac{E(\lambda_0)\: c\: A}{2R^2}\left(\beta_{\lambda_0}^{mol}(R) + \beta_{\lambda_0}^{aer}(R)\right)\cdot\exp\left(-2\int_0^R \alpha_{\lambda_0}^{mol}(r) + \alpha_{\lambda_0}^{aer}(r) dr\right)~,
\end{equation}
with  $E(\lambda_0)$ the pulse energy emitted by the laser at wavelength $\lambda_0$, $c$ the speed of light, and~$A=2.5~\mathrm{m}^2$ the effective geometric telescope area.  The~molecular and aerosol volume backscatter coefficients are $\beta_{\lambda_0}^{mol}(R)$ and $\beta_{\lambda_0}^{aer}(R)$, whereas $\alpha_{\lambda_0}^{mol}(r)=8\pi/3 \:\cdot\beta_{\lambda_0}^{mol}(r)$ is the molecular volume extinction coefficient and $\alpha_{\lambda_0}^{aer}(r) = \textit{LR}(r) \cdot \beta_{\lambda_0}^{aer}(r)$ the aerosol volume extinction coefficient, related through the LIDAR ratio \textit{\gls{lr}}$(r)$. We have not applied any overlap correction in Equation~(\ref{eq:return_power_elastic}), because~our LIDAR has been optimized for a low region of full overlap, plus near-range optics in the final design~\citep{technicalpaper}. The~analysis of the return power signals will start only after full overlap is~reached.  

The return power $P_{\lambda_R}(R)$ for the anti-Stokes Raman wavelength $\lambda_R$ is given by the RL equation~\citep{Ansmann:1992}:

\vspace{-9pt}
\begin{adjustwidth}{-\extralength}{0cm}
\centering 
\begin{equation}\label{eq:return_power_raman}
    P_{\lambda_R}(R) =  \frac{E(\lambda_0)\: c\: A}{2R^2} 
   n_\mathrm{N_2}(R) \cdot  \ddfrac{d\sigma_{\lambda_R}(\pi)}{d\Omega}\exp\left(-\int_0^R\left(\alpha_{\lambda_0}^{mol}(r) + \alpha_{\lambda_0}^{aer}(r) + \alpha_{\lambda_R}^{mol}(r) + \alpha_{\lambda_R}^{aer}(r)\right) dr\right)~,
\end{equation}
\end{adjustwidth}
where $n_\mathrm{N_2}(R)$ is
the number density of the nitrogen molecule and $d\sigma_{\lambda_R}(\pi)/d\Omega$ the range-independent differential Raman scattering cross-section in the backward direction, and~the rest of the symbols follow the definition in Equation (\ref{eq:return_power_elastic}).


The background power received $P_{bg}(\lambda)$ around wavelength $\lambda$ is calculated as follows:

\begin{equation}\label{eq:return_power_bkg}
    P_{bg}(\lambda)=L_{bg}\,A\;\Omega \,\Delta\lambda ~,
\end{equation}
where $L_{bg}$ is the night sky irradiance ($2.7\cdot 10^{-13}~\mathrm{W}\,\mathrm{cm}^{-2} \,\mathrm{nm}^{-1}\,\mathrm{sr}^{-1}$ for the clear moon-less night~\citep{BennEllison:technote,Eizmendi:2011}, ~$3.0\cdot 10^{-11} \,\mathrm{W}\,\mathrm{cm}^{-2}\,\mathrm{nm}^{-1}\,\mathrm{sr}^{-1}$ for the case of observations under moderate moonlight~\citep{KrisciunasSchaefer:1991,Noll:2012}), $\Omega=2\pi \cdot (1 - \cos(\theta_t/2))$ is the solid angle from the telescope \gls{fov} $\theta_t=4.4~\mathrm{mrad}$ -- defined as the ratio between the LLG diameter of 8~mm and the telescope focal length of 1.8~--  and $\Delta\lambda=10~\mathrm{nm}$ is the interference filter bandwidth of every channel~\citep{technicalpaper}.

From the return power, the~photon rate is computed as the sum of the contributions produced by the backscattered laser light ($\mathcal{R}_\mathrm{sig}$) and the background light ($\mathcal{R}_\mathrm{bg}$): 
\begin{align}
    & \mathcal{R}_\mathrm{sig}(\lambda,R)  = \xi(\lambda) \cdot P_\lambda(\lambda,R)\cdot\lambda/hc~,\nonumber \\
    & \mathcal{R}_\mathrm{bg}(\lambda)  = \xi(\lambda) \cdot P_{bg}(\lambda) \cdot \lambda/hc~,  \label{eq:RsigRbg}
\end{align}
where $P_\lambda$ and $P_{bg}$ are obtained from Equation (\ref{eq:return_power_elastic}) or Equation (\ref{eq:return_power_raman}) and~Equation~(\ref{eq:return_power_bkg}), respectively, $\xi(\lambda)$ are the efficiency factors that take into account the mirror reflectivity, the~transmission of the \gls{llg}, the~polychromator and the \gls{pmt} photon-detection efficiency (see Table~\ref{tab:calibpower}), and~$h$ is the Planck~constant.

\begin{figure}[H]
\includegraphics[width=0.49\textwidth]{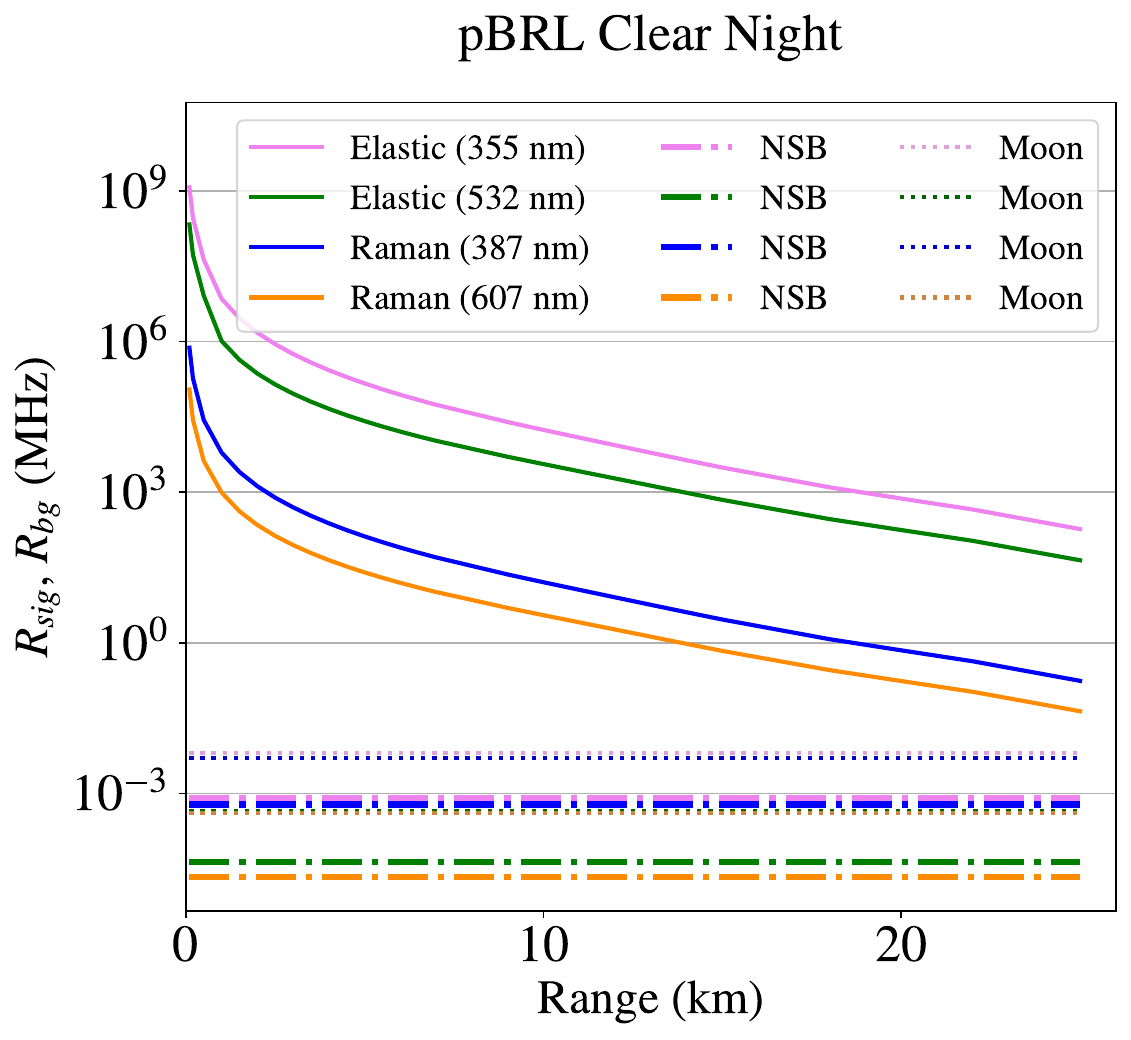}
\includegraphics[width=0.49\textwidth]{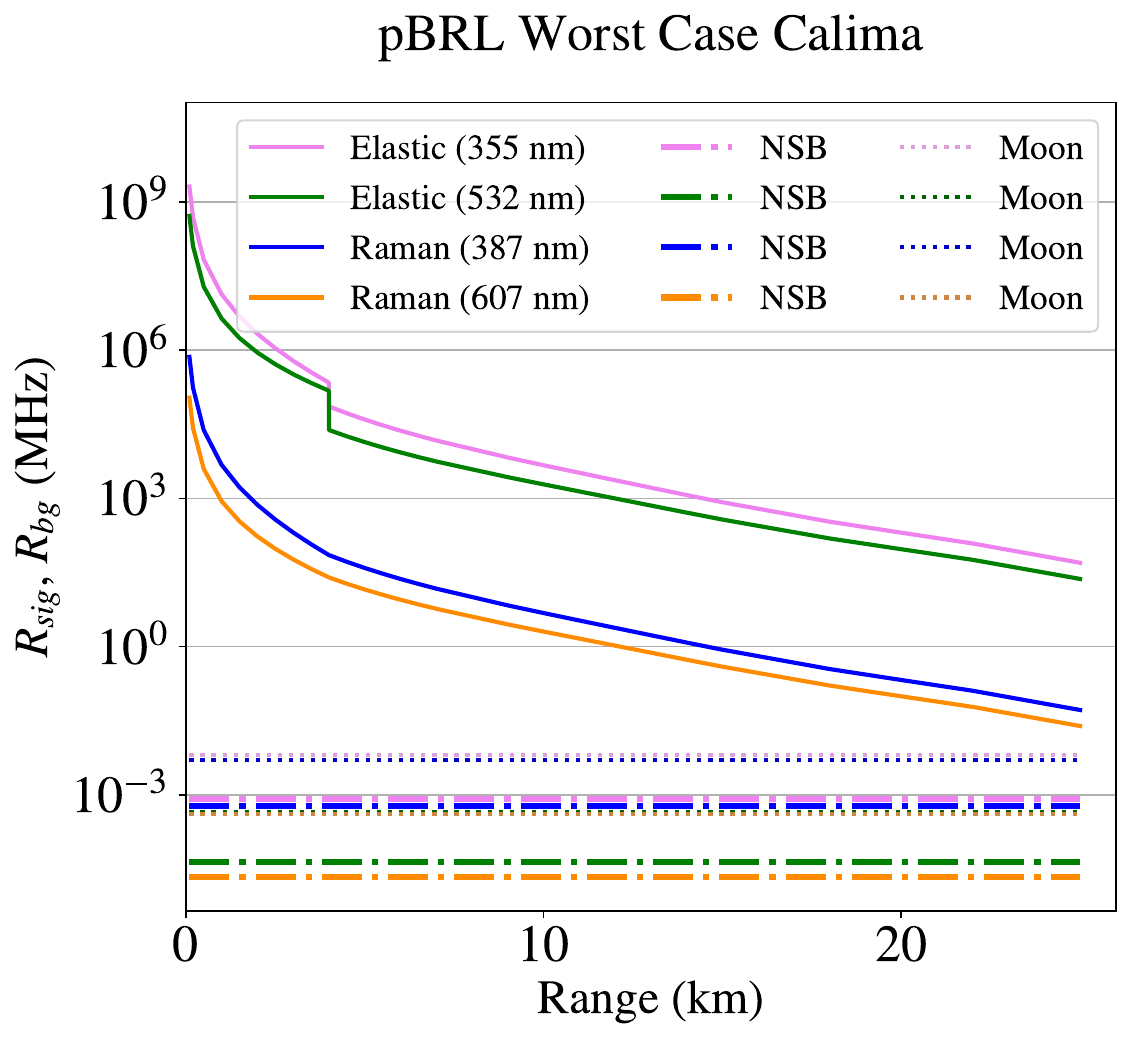}
\caption{Expected signal and background rates of the \gls{pbrl}, in~two extreme cases: a clear night at the \gls{orm}, and~a dust intrusion scenario with a \gls{vaod} of 0.7, the~maximum allowed for science data taking. {The step at 4~km in the elastic lines of the latter stems from the assumed sharp transition from dust to free troposphere at the boundary layer~height. }
\textit{NSB} 
stands for the normal Night-Sky-Background found at the \gls{orm}~\citep{BennEllison:technote}, and~\textit{Moon} 
for a night sky illuminated by a first or third quarter moon.\label{fig:return_powers}}
\end{figure}

The photon rates for the four wavelengths of interest of the \gls{brl} are reported in Figure~\ref{fig:return_powers} for two extreme cases: the best-case scenario of a clear night with only background aerosol presence~\citep{Fruck:2022igg} and the worst-case scenario with the presence of calima. One can see that in either case we expect large rates above the background up to ranges above 30~km. 

\vspace{-12pt}

\begin{table}[H]
\caption{Parameter values assumed for the power budget calculations of the \gls{pbrl}.  \label{tab:calibpower} }
\begin{tabular}{ccp{9cm}}
\toprule
\textbf{Parameter} & \textbf{Value}  & \textbf{Comments} \\
\midrule
$A$   & 2.3~m$^2$  &  1.8~m primary mirror minus shadows
\\
$\eta_\mathrm{dig.}$ & $0.90~\pm~0.05$  & photon counting efficiency of the readout \\
\midrule
\multicolumn{3}{l}{355 nm channel}\\
$E$        & 80~mJ  & energy per pulse \\
$N_0$      & 1.4$\cdot$10$^{17}$ & photons per pulse \\
$l$   &  7.5~m  & digitization length for 12-bit, 20~MS/s sampling rate \\
$\rho$        & 0.95  &  mirror reflectivity, after~re-aluminization, otherwise $<$0.3 
\\
$\zeta$ &  $0.34~\pm~0.04$ & combined \gls{llg} and polychromator transmissions \citep{technicalpaper} 
\\
\textit{\gls{pde}} & $0.42~\pm~0.03$  & \gls{pmt} photon-detection efficiency \citep{toyama} \\
$\xi$ &  0.13~$\pm$~0.02   & combined channel efficiency $\xi=\rho\cdot\zeta\cdot\textit{\gls{pde}}$ \\
\midrule
\multicolumn{3}{l}{387 nm channel}\\
$E$        & 80~mJ  & energy per pulse (at 355~nm) \\
$N_0$      & 1.4$\cdot$10$^{17}$ & photons per pulse \\
$l$   &  7.5~m & digitization length for 12-bit, 20~MS/s sampling rate \\
$\rho$        & 0.96  &  mirror reflectivity, after~re-aluminization,  otherwise $<$0.3  
\\
$\zeta$ &  $0.31~\pm~0.04$   & combined \gls{llg} and polychromator transmissions \citep{technicalpaper} 
\\
\textit{\gls{pde}} & $0.43~\pm~0.03$  & photon-detection efficiency \citep{toyama} \\
$\xi$ &  0.12~$\pm$~0.02   & combined channel efficiency  $\xi=\rho\cdot\zeta\cdot\textit{\gls{pde}}$  \\
\midrule
\multicolumn{3}{l}{532 nm channel}\\
$E$        & 128~mJ  & energy per pulse  \\
$N_0$      & 3.4$\cdot$10$^{17}$ & photons per pulse \\
$l$   &  3.75~m & digitization length for 16-bit, 40~MS/s sampling rate \\
$\rho$        & 0.97  &  mirror reflectivity, after~re-aluminization,  otherwise $<$0.3  
\\
$\zeta$ &  $0.31~\pm~0.03$   & combined \gls{llg} and polychromator transmissions \citep{technicalpaper} 
\\
\textit{\gls{pde}} & $0.13~\pm~0.03$  & photon-detection efficiency \citep{toyama} \\
$\xi$ &  0.035~$\pm$~0.009  &  combined channel efficiency  $\xi=\rho\cdot\zeta\cdot\textit{\gls{pde}}$ \\

\bottomrule
\end{tabular}
\end{table}

\begin{table}[H]\ContinuedFloat
\caption{{\em Cont.}}
\begin{tabular}{ccp{9cm}}
\toprule
\textbf{Parameter} & \textbf{Value}  & \textbf{Comments} \\
\midrule

\multicolumn{3}{l}{607 nm channel}\\
$E$        & 128~mJ  & energy per pulse (at 532~nm) \\
$N_0$      & 3.4$\cdot$10$^{17}$ & photons per pulse \\
$l$   &  3.75~m &  digitization length for 16-bit, 40~MS/s sampling rate \\
$\rho$        & 0.97  &  mirror reflectivity, after~re-aluminization,  otherwise $<$0.3 
\\
$\zeta$ &  $0.14~\pm~0.02$ & combined \gls{llg} and polychromator transmissions  \citep{technicalpaper} 
\\
\textit{\gls{pde}} & $0.04~\pm~0.01$  & photon-detection efficiency \citep{toyama} \\
$\xi$ &  0.05~$\pm$~0.01  &  combined channel efficiency $\xi=\rho\cdot\zeta\cdot\textit{\gls{pde}}$   \\
\bottomrule
\end{tabular}
\end{table}
\unskip

\subsubsection{Signal to Noise~Ratio}
We compute the \gls{snr} for one single laser shot for the case of a photon-counting \gls{lotr} adapted from Rocadenbosch et al.~\cite{Rocadenbosch:1998}:
\begin{equation}\label{eq:snr}
    \textit{SNR}_1(R) = \left(\mathcal{R}_\mathrm{sig}(R) - \mathcal{R}_\mathrm{bg}(R)\right)\cdot\ddfrac{1}{\textit{ENF}_\textit{PMT}} \cdot \sqrt{\ddfrac{2\Delta r}{c}\frac{1}{(\mathcal{R}_\mathrm{sig}(R)+\mathcal{R}_\mathrm{bg}(R))}} \quad,
\end{equation}
where $\textit{\gls{enf}}_\textit{\gls{pmt}} \approx 1.08$ is the excess noise factor of our \glspl{pmt}~\citep{Mirzoyan:2017,Eschbach:PhD}, which accounts for the fluctuations in the signal produced by the PMT for each photoelectron, and~$\Delta r$ is the range resolution required (which might include rebinning of the data). 

In the far range, where the signal becomes background-dominated, the~time required to reach a fixed minimum $\textit{\gls{snr}}_\mathrm{goal}$ can be written as follows:
\begin{equation}
    T_{\textit{SNR}_\mathrm{goal}}(R)=\left(\frac{\textit{SNR}_\mathrm{goal}}{\textit{SNR}_1(R)}\right)^2\frac{1}{\textit{PRF}
    }  
    \quad,
\end{equation}
where \textit{\gls{snr}} is computed with Equation~(\ref{eq:snr}) and \textit{\gls{prf}} is the laser pulse repetition frequency.

\begin{figure}[H]
\includegraphics[width=0.49\textwidth]{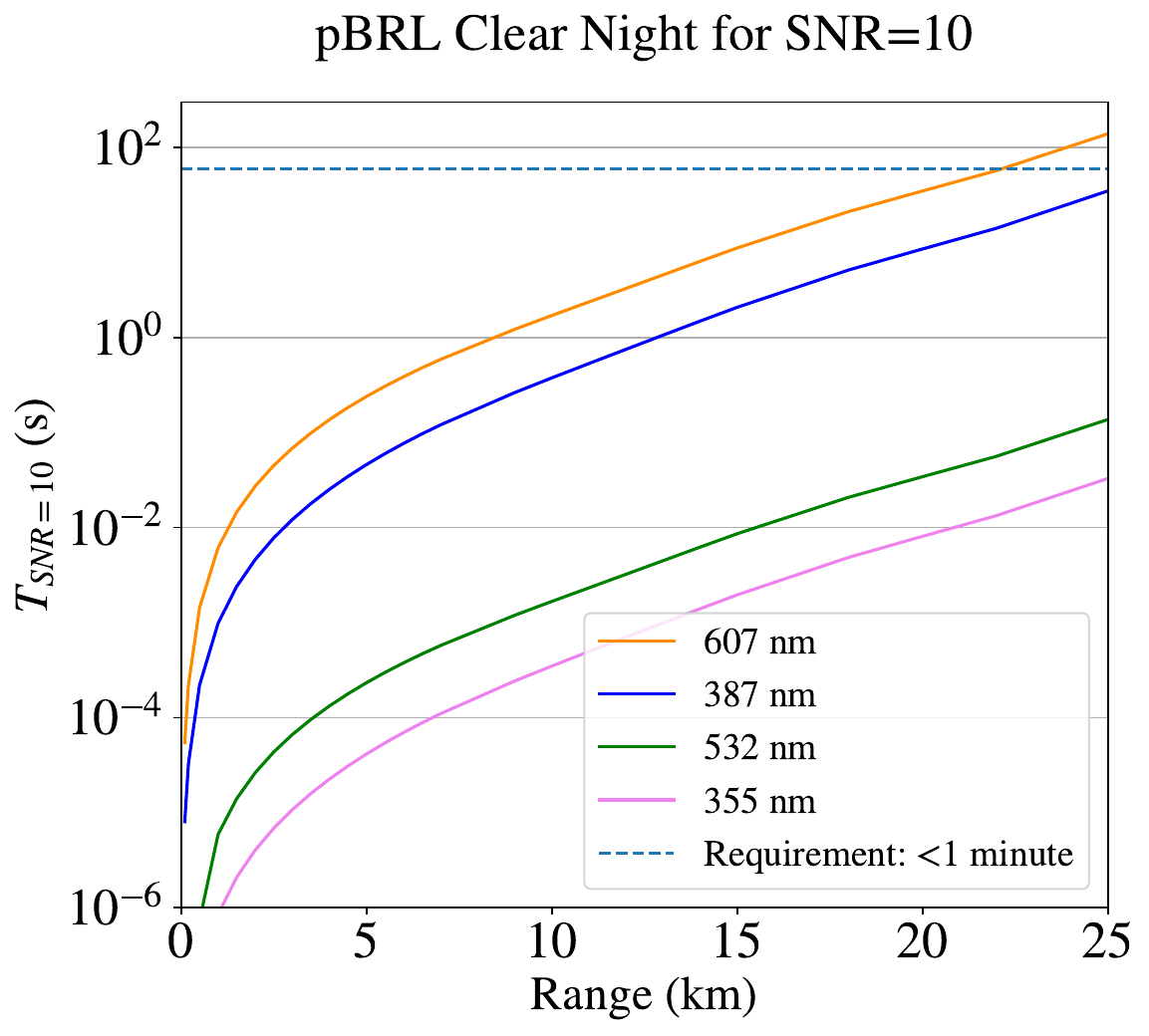}
\includegraphics[width=0.49\textwidth]{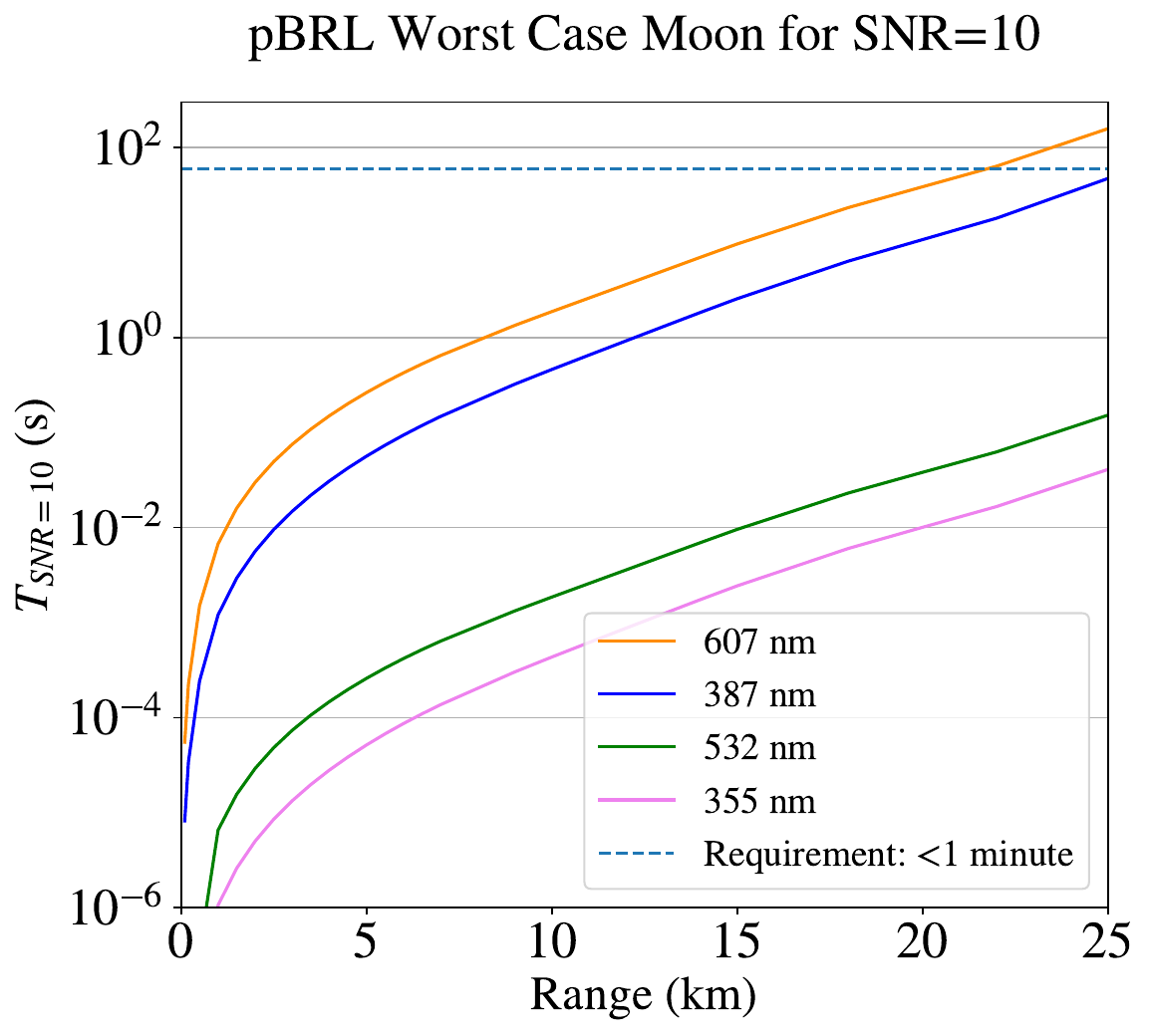}
\caption{Required observation times to reach a $\textit{\gls{snr}}_\mathrm{goal}\geq 10$ with a range resolution of 300~m for the \gls{pbrl} with 10~Hz \gls{prf} under two example conditions: a clear night at the \gls{orm} (Left), and~a moon-lit night with the maximum moon phase allowed by \gls{ctao} for science data taking (Right). The~requirement for observation times less than one minute is marked as a horizontal dashed~line.  \label{fig:obstimes}}
\end{figure}

In Figure~\ref{fig:obstimes}, we report the
$T_{\textit{\gls{snr}}_\mathrm{goal}}(R)$
 needed to reach $\textit{\gls{snr}}_\mathrm{goal}\geq 10$ for the four wavelengths of interest and for two standard observation cases: that of a clear night and that for a night sky illuminated by a first or third quarter moon, a~typical limit for \gls{iact} science data taking. 
 One can see that in either case, the~desired $\textit{\gls{snr}}_\mathrm{goal}$ is reached within a minute up to a 25~km range. In~both cases, a~\textit{\gls{prf}} of 10~Hz was assumed, according to the specifications of the \gls{pbrl} laser~\citep{technicalpaper}.
 
\begin{figure}[H]
    \centering
        \includegraphics[width=0.47\linewidth]{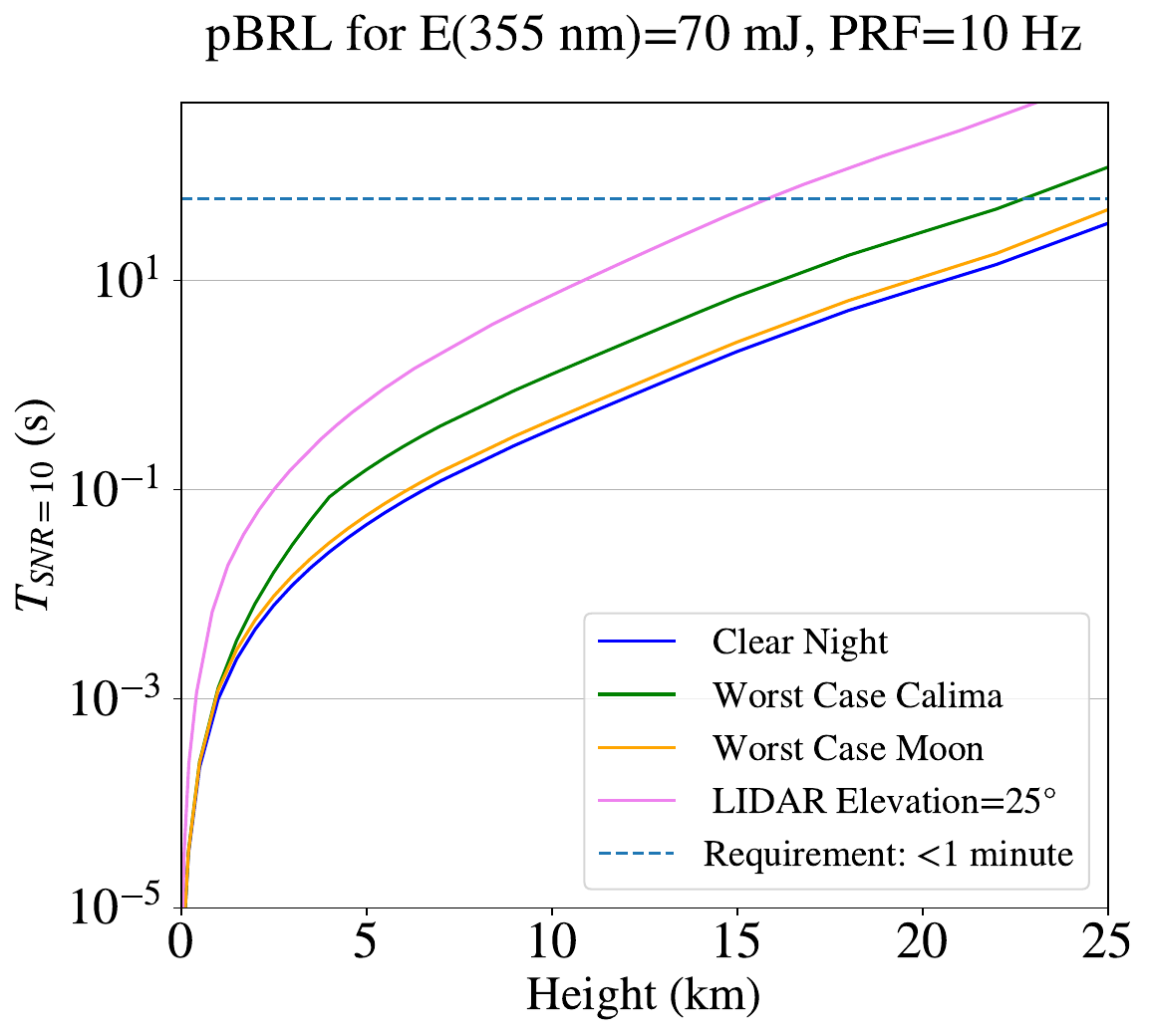}
        \includegraphics[width=0.52\linewidth]{Figures/chap2/ObsTime_Clear_70.0-mJ_120.0-mJ_10.0-Hz.pdf}
    \caption{Required time to reach a $\textit{\gls{snr}}_\mathrm{goal}\geq 10$ for the 387~nm Raman line under different conditions for the \gls{pbrl} with a laser power of 70~mJ at 355~nm (left) 
    and an alternative solution based on a laser with lower power of only 10~mJ, but~a  higher repetition rate of 200~Hz (right). The~requirement for observation times less than one minute is marked as a horizontal dashed line.\label{fig:linkbudget-comparison} }
\end{figure}

In Figure~\ref{fig:linkbudget-comparison} we focus on $T_{\textit{\gls{snr}}_\mathrm{goal}}(R)$ as a function of height for the weak $387$~nm Raman line under different atmospheric conditions, laser power and pointing elevation: a vertically pointing \gls{rl} during the clear night, the~presence of \textit{calima}, and~moderate moon is shown, and~one that points at the lowest required elevation of $25^\circ$. We evaluated a case with a slow \gls{prf}~=~10~Hz and a large laser pulse power of 70~mJ with that of a faster \gls{prf}~=~200~Hz and a lower laser pulse power of 10~mJ. One can see again that within a minute of integration time, both solutions reached the required $\textit{\gls{snr}}_\mathrm{goal}\geq 10$, with~the latter proposal performing slightly better.  For~very low-elevation observations, this time interval allows for a precise determination of the return power only up to about (15--20)~km. 

\subsection{Pre-Process~Analysis}
\label{sec:evaluation_signal}

As introduced above, the~backscattered signal is transported from the focal plane via a liquid light guide (\gls{llg}) to an optical bench and collected by \glspl{pmt} (see Figure~\ref{fig:lidar_scheme}). The~\gls{pmt} signal is read with commercial Licel modules. The~\gls{lotr} type TR20-12bit delivers averaged signals from an analog (\gls{an}) and a photon-counting (\gls{pc}) detection chain. Two newer versions TR40-16bit that were purchased 
also provide an average of the squared analog (\gls{an}2) signal. 
Due to the presence of the two readout modes and a very wide range of return power (see Figure~\ref{fig:return_powers}), special care must be taken to maintain precision and accuracy throughout the entire range. This requires careful signal preparation,  frequent sanity checks, and~the use of robust statistics. 
We describe the procedures applied to guarantee this performance. These include several novel statistical approaches, which have not been discussed in the literature to our knowledge so far. 
In the following, we use the term \textit{line} 
for one of the four wavelengths of interest and \textit{channel} 
for an \gls{lotr} readout channel. We have therefore two (\gls{an},\gls{pc}) or three (plus \gls{an}2) channels per line for a given \gls{lotr} module.

The open-access \gls{brl} Pre-Processing (\gls{lpp}) software~\citep{web.pbrlsw} 
converts raw data files produced by the \gls{lotr} into processed data products, primarily aerosol extinction profiles for different photon wavelengths. The~\gls{lpp} is divided into two components: the front-end component is in charge of providing a Graphical User Interface (\gls{gui}), allowing the expert user to perform debugging of the software and graphical analyses;  
the back-end component is in charge of not only implementing the functionalities directly required by the front-end but also implementing communication and interaction with the different external components involved in the complete task execution process, such as a database.
The functionalities of the \gls{lpp} include the following: 

\begin{itemize}
    \item \textbf{LPP analysis:  
} The actual core of the program.
 \item \textbf{Graphical User Interface (GUI):
 } Implementation of a user-friendly interface with a finalized design using front-end technologies. 
      \item \textbf{Logging and configuration:
      } A comprehensive logging system with database entries and the generation of log files. Configuration options through YAML files, including auto-update capabilities and \gls{gui} configurability.
    \item \textbf{Testing:
    } Execution of Continuous Integration (CI) tests covering \gls{gui}, HTTP interactions, database connections, and~specific functionalities. 
    In addition, ample possibilities are provided for manual testing on diverse datasets for all \gls{lpp} analysis steps.
    \item \textbf{Molecular Density Profile ({MDP}):
    } A downloader component for molecular density profiles from the European Centre for Medium-Range Weather Forecasts (ECWMF) or the Global Forecast System (GFS) at a given location on Earth and time. 
   \item \textbf{Licel to {FITS} converter:
   } An automatic file format converter from the Licel raw data format~\cite{web.licel.dataformat} to the \gls{fits} file format~\citep{web.fits}, used as a standard for \gls{ctao}. 
\end{itemize}

\subsubsection{Raw Data Sanity~Checks}
\label{sec:sanitychecks}

Before the actual data analysis begins, the~\gls{lpp} performs automated sanity checks to ensure the integrity of the measurements. These checks help to detect hardware failures, misconfigurations, and~transmission errors that could compromise the data. By~implementing these safeguards, the~system can identify issues early, 
ensuring early warnings to users about possible issues and that only high-quality data are~processed.


An initial routine verifies that the LIDAR system has detected a sufficient number of \gls{lotr} readout channels. 
The absence of active channels may indicate hardware disconnection or~miscommunication. 
If no channels are detected, the~system immediately stops the analysis and logs an error, prompting the user to inspect the hardware and connections. 
In addition, the~\gls{lpp} inspects each channel for zero-valued data arrays, indicating issues such as a disconnected signal cable or an inactive or too low \gls{pmt} high-voltage setting. If~all channels produce zero data, the~analysis is stopped and the user is asked to review the system’s connections and data acquisition processes. If~only certain channels are affected, they are flagged and excluded from further analysis to prevent biased~results. 

The efficiency of the \gls{pc} channels is evaluated by analyzing the percentage of data with non-zero counts. Channels where fewer than a configurable percentage (default: 20\%) of data register photon counts are flagged as~unreliable.

Oscillations in detected signals can introduce noise and distort results. To~investigate this effect, the~\gls{lpp}  performs spectral analysis of the amplitude channels to detect oscillations that exceed predefined thresholds. The~\gls{lpp} generates Power Spectral Density (\gls{psd}) plots to highlight irregular oscillations. If~oscillations are found, the~system logs an error and suggests checking the signal stability or~hardware. 

\begin{figure}[H]
\includegraphics[width=1.0\textwidth]{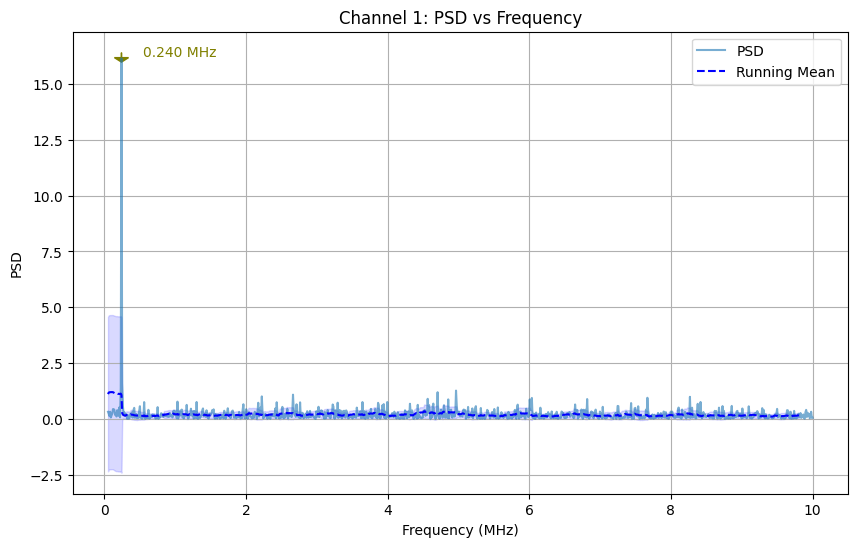}
\caption{Power Spectral Density (\gls{psd}) vs. Frequency plot for Channel~1 (355 nm), showing a detected oscillation at approximately 0.24~MHz, probably from a switching power supply. The~light blue full lines show the measured \gls{psd} spectrum, while the dashed dark blue line shows the running mean over a window of 20 bins. The~light blue area displays the RMSD of the running mean. Both are used to reveal strong local deviations from the underlying trend.  The~olive arrow marks the frequency where the oscillation was~detected. \label{f:leading_freq}}
\end{figure}

Figure~\ref{f:leading_freq} shows the \gls{psd} as a function of frequency for a readout channel affected by an oscillation of the ground line with a frequency near 0.24~MHz. The~supply line of that channel was fixed later with appropriate low-pass filters.

\subsubsection{Time Offset~Adjustments}
\begin{figure}[H]
\centering
\includegraphics[width=0.58\textwidth]{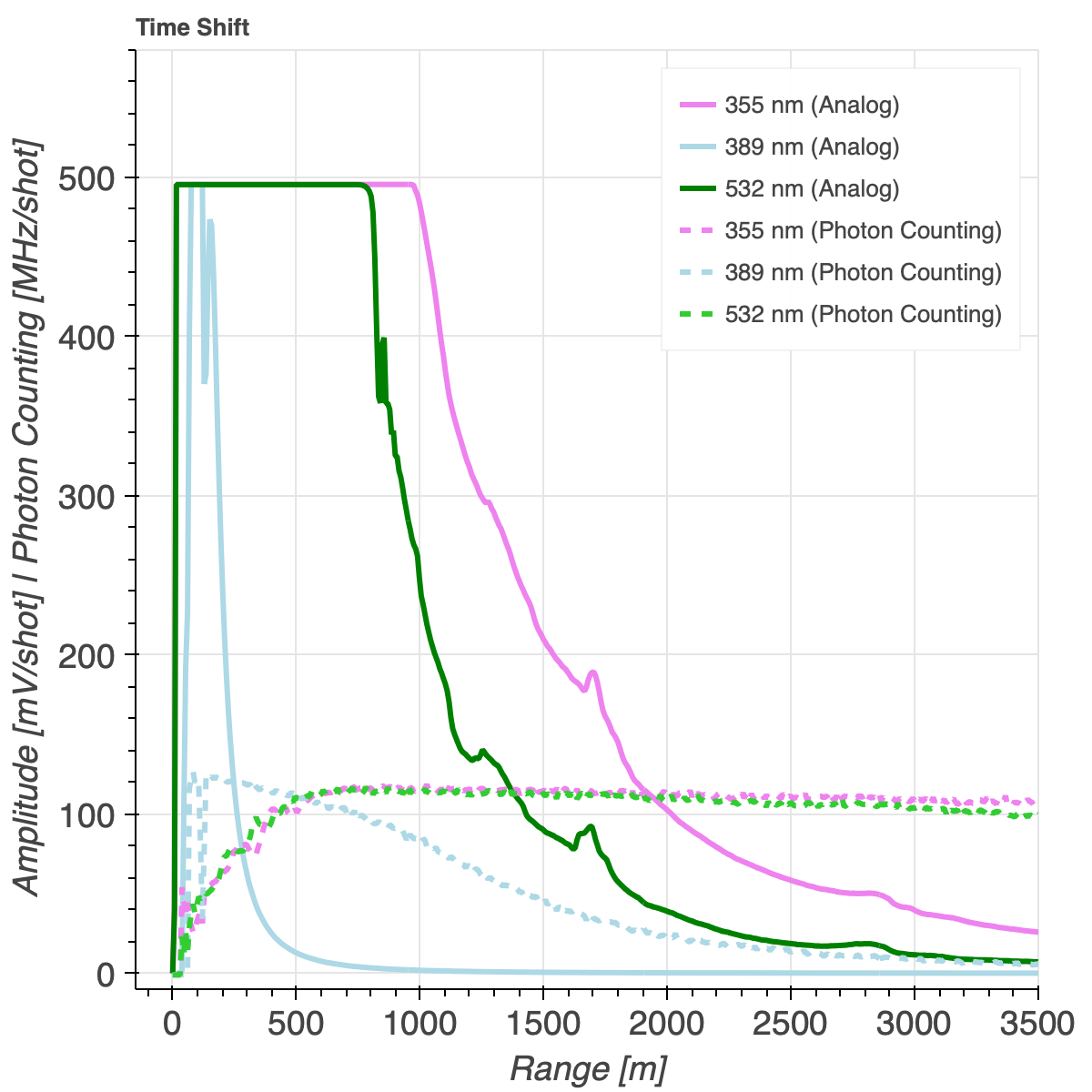}
\begin{minipage}[t]{.4\textwidth}
\vspace{-6.5cm}
\includegraphics[width=1\textwidth,trim={0 10.5cm 0 0.2cm},clip]{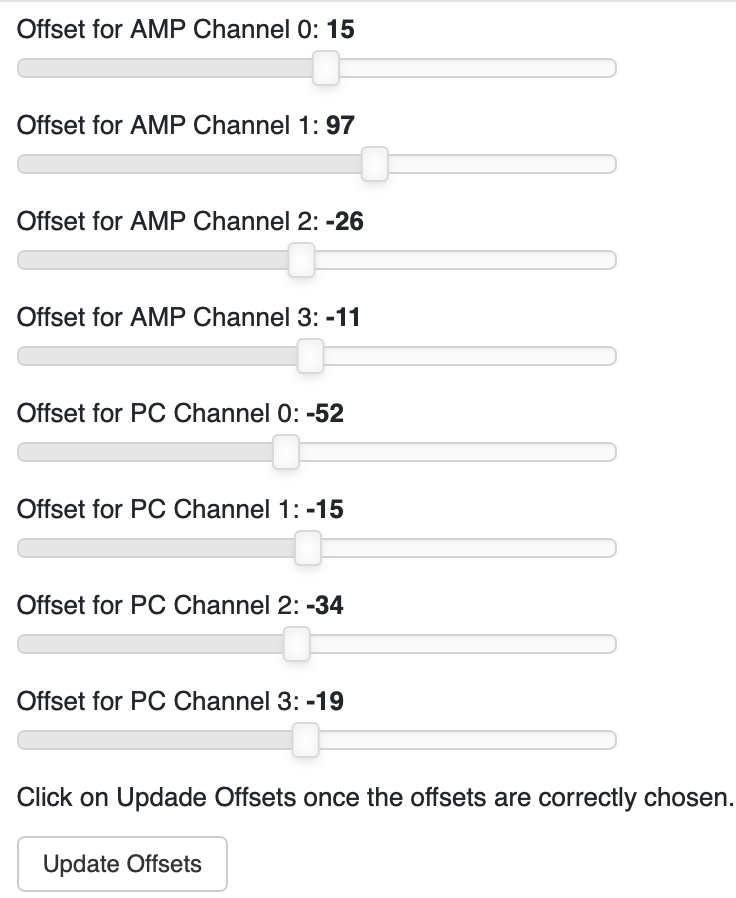}
\includegraphics[width=1\textwidth,trim={0 4.0cm 0 6.7cm},clip]
{Figures/chap3/TimeOffsets_Selectable.png}
\end{minipage}
\caption{An 
 example of the interactive time-delay adjustment of the \gls{lpp}. In~this range selection, the~two elastic \gls{an} channels can be time-adjusted through the atmospheric feature visible at $\sim$1700~m. The~offsets (in meters) on the right sliders can be continuously adjusted. The~feature is too weak, however, to~produce a visible signal drop in the Raman channel and also allow for the adjustment of the time delay of that channel. Note the strong \gls{an} signal saturation of the elastic lines below $\sim$1000~m. 
 }
\label{f:ts_structures}
\end{figure}

\glspl{lotr} have shown to exhibit different internal time delays between analog (\gls{an}) and photon-counting (\gls{pc}) channels of the same module, which change for different \gls{lotr} versions.  We found that each line needed separate adjustments, which can be achieved interactively with the online display of the \gls{lpp} (see Figure~\ref{f:ts_structures}), and, once found, fixed through correct settings in a configuration file.  
This procedure allows for a precise relative 
adjustment by means of adequate atmospheric features visible in more than one channel.  
We found time offsets between the \gls{an} and \gls{pc} channels of the same \gls{lotr} of the order of $\sim$0.5~$\upmu$s for the TR-20 and $\sim$50~ns for the newer TR-40, with~the \gls{an} signal arriving earlier. Between~two \glspl{lotr}, time differences of the same order have been observed.  
Finally, an~absolute time calibration needs to be performed for all channels synchronously, which is achieved through the diffuse backscatter signal of the laser light from the first dichroic guide mirror (see Figure~\ref{fig:pBRL_LST}), visible in some channels. This procedure will be modified for the final \gls{brl}, in~which the stray light from diffuse reflection of laser light from the guiding mirrors will be isolated from the rest of the~system.

\subsubsection{Photon Background and Offset~Determination}
The calculation of amplitude offsets and the photon background of Equation~(\ref{eq:return_power_bkg}) requires some dedicated attention for two reasons: 

\begin{enumerate}
\item Ion feedback from the photocathode or atmospheric muons traversing the photomultiplier can create spurious high signals, particularly in the amplitude channel, even in the absence of backscattered laser light.
\item The exact ranges of signal-free 
data are unknown a priori. 
Contamination of those regions used for background estimation with signal (e.g., from~spurious reflection of laser light on the guide mirrors or late atmospheric backscatter) must be avoided. 
\end{enumerate}

To ensure a robust computation of the backgrounds even in the presence of these two nuisances, we elaborated a scheme that uses robust statistics and continuous tests for Poissonian behavior of the photon~background. 

\begin{figure}[H]
\includegraphics[width=1.0\textwidth,trim={0 0 0 1.5cm},clip]{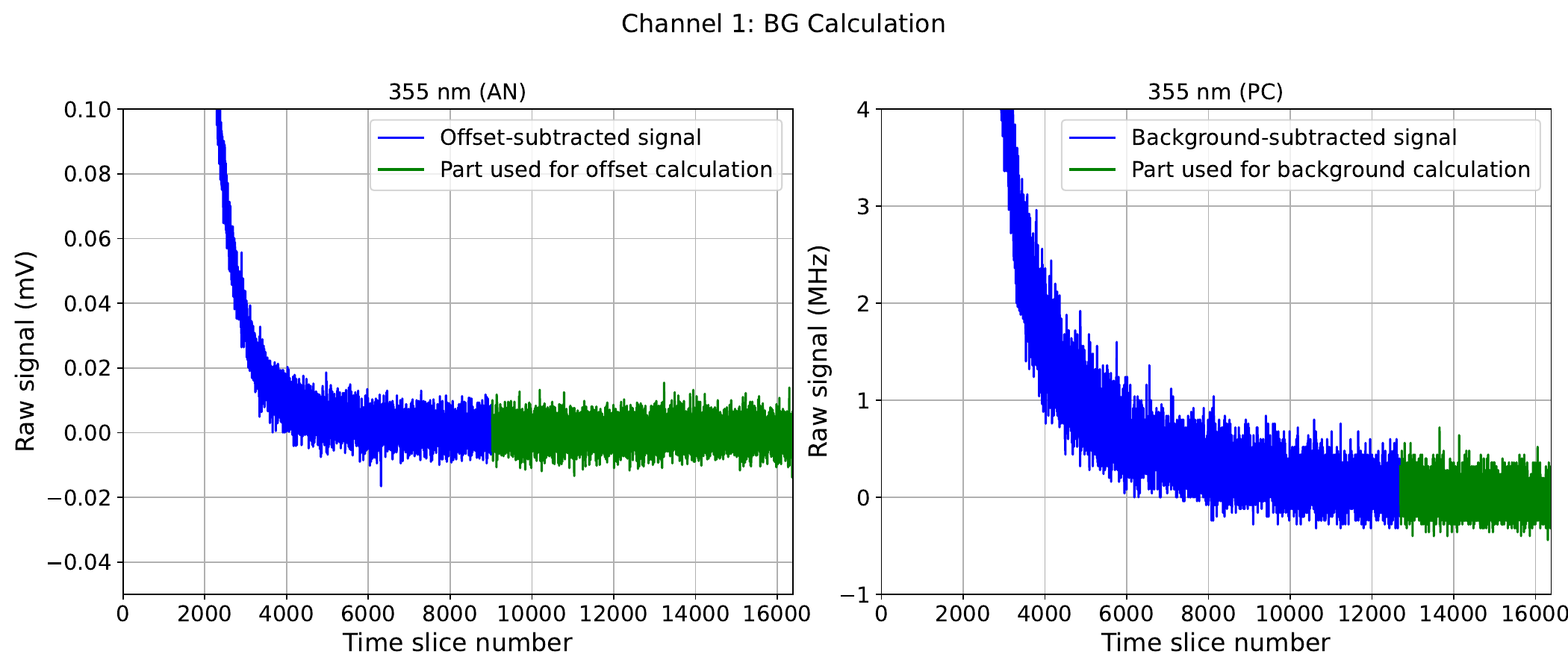}
\caption{Background 
 calculation for the Elastic Channel (355~nm). The~background-subtracted signal is shown with the region used for the determination of the background (green) and regions not considered due to signal leakage (blue). On~the left side, for~the \gls{an} channel,
on the right side, for~the \gls{pc} channel, the~signal is similarly segmented, and no sample region before the signal is available. }
\label{f:bg_calc1}
\end{figure}

\begin{figure}[H]
\centering
\includegraphics[width=1.0\textwidth,trim={0 0 0 1cm},clip]{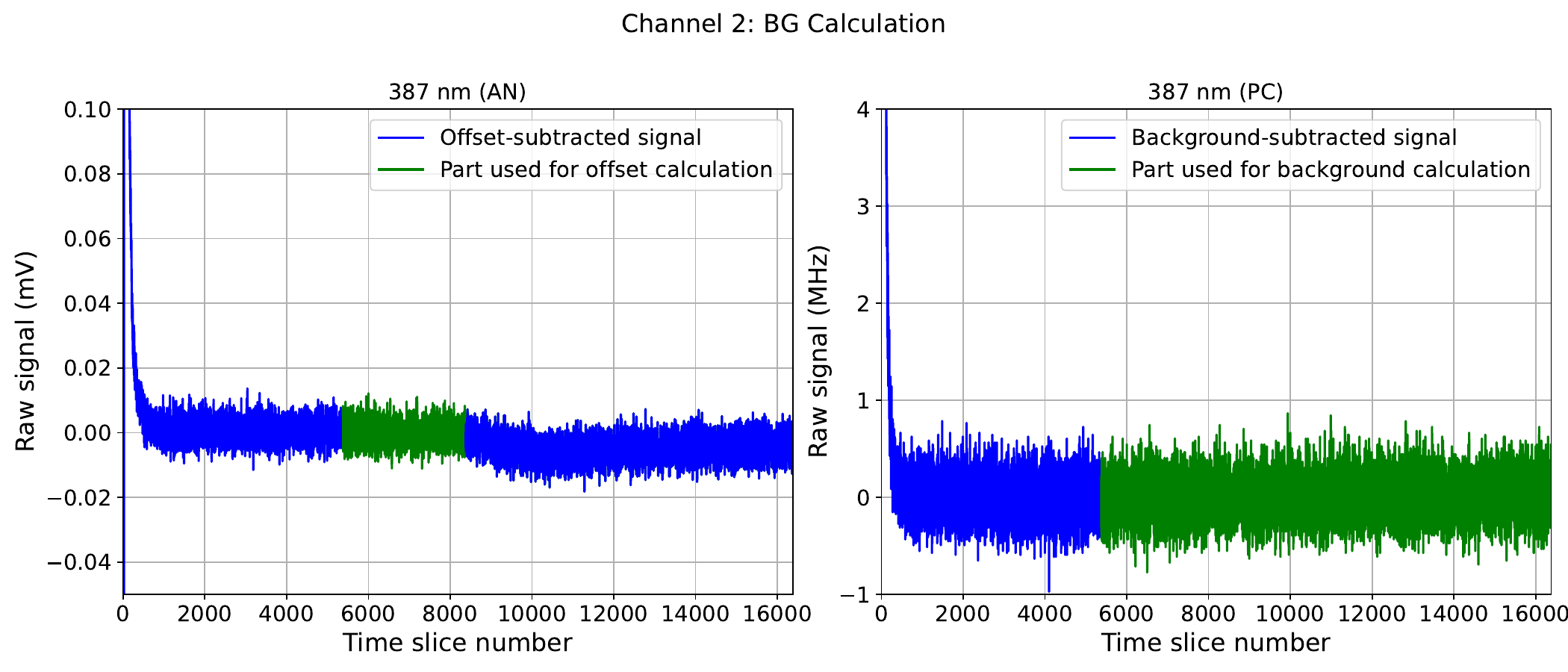}
\caption{Background 
 calculation for the Raman Channel (387~nm). Same scheme as in Figure~\ref{f:bg_calc1}. On~the left side, the~\gls{lotr} \gls{an} signal exhibits a small drop in its electronic baseline right after half the readout range, at~a range of about 65,000~m. Only this single \gls{lotr} is affected. 
On the right side, the~\gls{pc} region is shown.
}
\label{f:bg_calc2}
\end{figure}

Both amplitude offset and photon background are searched before and after 
the backscatter signal. In~both cases, a~generous initial search range is set (up to 400~ns at the beginning of the trace and the last 500~$\upmu$s recorded. A~search procedure, outlined below, is repeated until an optimal background search region is found (see Figures~\ref{f:bg_calc1} and~\ref{f:bg_calc2}).

\unskip

For each set of $n$ data points in a tentative background sample, the~distribution of counts is trimmed~\citep{Wilcox:2003} with $\alpha=0.025$ of the data clipped at both extremes because~of a possible presence of outliers (see point 1). From~that, the~trimmed mean $\mu_t$ is calculated together with the unbiased distribution variance
\begin{equation}
s^2_n \approx \ddfrac{s^2_w}{\left(1-2\,\alpha\right)^2}~,
\end{equation}
where $s^2_w$ is the sample variance of the \textit{winsorized} distribution~\citep{Winsor:1947,Wilcox:2003}.  
In the case of PC channels, the~total number of photo-electrons found in the background sample,
\begin{equation}
  N_\mathrm{ph,tot} =   \mu_t \cdot N_{sh} / f_{s}
\end{equation}
is calculated, with~$N_{sh}$ being the number of laser shots and $f_{s}$ the \gls{lotr} sampling frequency. That number is then compared with the corresponding scaled Poisson variance
\begin{equation}
 s^2_\mathrm{ph,tot} =  s^2_n \cdot N_{sh}^2 / f_{s}^2~.   
\end{equation}

We require that $s^2_\mathrm{ph,tot}\lesssim N_\mathrm{ph,tot}$, i.e.,~its Poissonian expectation. Any excess variance must come from signal leakage and shall be discarded from the background calculation.
If $s^2_\mathrm{ph,tot}>1.03\cdot N_\mathrm{ph,tot}$, that is, the~variance exceeds by more than 3\%, signal contamination is assumed and the window size for the background calculation is reduced (use, e.g.,~Figure~\ref{f:bg_calc1} for guidance): in the search region before
the backscatter signal, the~size of the background search range is reduced by a single time slice; for the search region after 
the signal, the~search region is reduced by 20\%.  This is performed iteratively until the condition is met, 
or the sample range falls below a minimum number of background time slices of $n= 2/(0.03)^2 \approx 2000$. 
In the latter case, the~channel is declared unreliable. 
The amplitude offset is then obtained from the same window. Once the mean background count and its variance are defined, it is straightforward to compute the mean background signal rate and its standard deviation:
\begin{align}
\mathcal{R}_b &:= \mu_t \\
s_{\mathcal{R}_b} &:= \sqrt{\ddfrac{s_n^2}{n-1}} 
\end{align}

The resulting $<$1.5\% systematic error allowed by the procedure corresponds roughly to the accuracy, with~which an absolute calibration of the LIDAR is possible~\citep{Fruck:2022igg,Gaug:2022}.

This procedure allows also to possibly reveal technical issues with \gls{lotr} channels, like the amplitude channel of one TR20-16bit,  which exhibits an unexpected electronic amplitude offset that decreases after half the time slices are reached (see Figure~\ref{f:bg_calc2}). In~this case,  we resort to computing the background only using the time slices before the count jump, at~exactly half the time slice range. To~avoid any significant leakage of the signal into these samples, we used the corresponding \gls{lotr} for a Raman~channel.

\subsection{Analog to Photon-Counting Signal~Gluing}
\label{sec:gluing}

In the context of \gls{pbrl} data analysis, signal gluing is performed to match the \gls{lotr} \gls{an} and \gls{pc} signals, taking into account their signal and background variances, dead-time corrections, and signal~saturation. 

We implemented two methods: one based on a $\chi^2$-minimization between the two signal channels, with~sliding windows and different window sizes, 
extending the methods described by~\citet{Whiteman:1992,Lange:2012,Li:2023}, and~a second {algorithm} based on maximization of a likelihood function
that combines probability density functions from both channels, building on the method of~\citet{Veberic:2012}.

\subsubsection{$\chi^2$-Based~Gluing}
\label{sec:chi2gluing}
In this procedure, the~photon-counting channel is first corrected for pulse pile-up~\mbox{\cite{Whiteman:1992,Donovan:1993,Gao:2013}}, assuming a non-paralysable~\citep{Newsom:2009} system, following:
\begin{equation}
\mathcal{R}_{i,\mathrm{corr}} = \ddfrac{\mathcal{R}_{i,\mathrm{obs}}}{1-\tau \mathcal{R}_{i,\mathrm{obs}}},  \label{eq:nonparalysable}
\end{equation}
\noindent
where $\mathcal{R}_{i,\mathrm{obs}}$ is the recorded photon rate (normally expressed in units of MHz) in range bin~$i$, $\mathcal{R}_{i,\mathrm{corr}}$ is the pileup-corrected one, and~$\tau$ is the resolving time of the discriminator-counter combination (the ``dead time'') of the system. According to the recommendations of Licel~\citep{web.licel.pmt}, we used $\tau=3.70\times 10^{-3}$~$\upmu$s for TR-20 \gls{lotr} modules operated at a sampling rate of $f_s=20$~MSamples/s and $\tau=3.06\times 10^{-3}$~$\upmu$s for the more recent TR-40 models, with~$f_{s}=40$~MSamples/s~\citep{technicalpaper}. However, with~the more elaborate likelihood-gluing method (Section~\ref{sec:Lgluing}), we found that the deadtime was rather of the order of $\tau \approx 8$~ns, which we use in the~following.

The constant photon background $\mathcal{R}_b$ is subtracted \textit{after} 
the pileup correction, and~a general \gls{pc} efficiency $\epsilon$ can be introduced, leading to a background- and efficiency-corrected photon count rate and variance of the following:
\begin{align}
\mathcal{R}_{i,\mathrm{chi2}} &= \ddfrac{1}{\epsilon}\cdot \left(\ddfrac{\mathcal{R}_{i,\mathrm{obs}}}{1-\tau \mathcal{R}_{i,\mathrm{obs}}} - \ddfrac{\mathcal{R}_b}{1-\tau \mathcal{R}_b} \right) \label{eq:Ribgcorr}\\
s^2_{U,L}(\mathcal{R}_{i,\mathrm{chi2}}) &= \ddfrac{1}{\epsilon^2}\cdot \left(\ddfrac{f_{s}^2}{N_\textit{sh}^2} \cdot  \ddfrac{\vert X_i-\mu_{U,L}(X_i)\vert^2}{\left(1-\tau \mathcal{R}_{i,\mathrm{obs}}\right)^4} + \ddfrac{s^2_{\mathcal{R}_b}}{\left(1-\tau \mathcal{R}_b\right)^4}\right) \quad,  \label{eq:Ribgcorrvar}
\end{align}
\noindent
where Poissonian statistics have been assumed for the total accumulated number of photoelectrons $X_i=N_{sh} \mathcal{R}_{i,\mathrm{obs}}/f_{s}$ for the (asymmetric) variances $s^2_{U,L}$. 
Some 
  care must be taken when calculating the Poissonian variances of the
  measured photo-electron counts $X_i$ in a given sampling bin $i$.
  The 68.3\% Confidence Interval (CI) does not correspond to $[X_i -
  \sqrt{X_i},X_i+\sqrt{X_i}]$, due to the asymmetry of the Poissonian
  probability mass function, particularly for low rates. We use the
  prescription of \citet{Garwood:1936} (implemented as method
  \texttt{exact-c} in \texttt{statsmodel}'s function \texttt{confint\_poisson})
  for an improved asymmetric CI $[\mu_L,\mu_U]$, so that the Poissonian
  probability $P_L =P(X \geq X_i|\mu=\mu_L)= 0.1587$ and $P_U =P(X \leq
  X_i|\mu=\mu_U)= 0.1587$ and $\mu_L$ and $\mu_U$ are obtained from the
  $[0.1587,1-0.1587]$ quantiles of the gamma-function with $X$ as shape
  parameter (although that method still suffers from slight
  over-coverage~\citep{Blaker:2000,Swift:2009}).

The raw amplitudes $A_{i,\mathrm{obs}}$ have a contribution from the same photon background, scaled to AN voltages: $g\, \mathcal{R}_b$, where $g$ is the signal photo-electron gain, and~the electronic baseline voltage $B$. The~sum of both has been measured as background in the AN channel as $A_b = B + g\,\mathcal{R}_b$, 
hence: $B = A_b - g\,\mathcal{R}_b$. 
For the amplitude part, we assume linear scaling between registered amplitudes $A_{i,\mathrm{obs}}$ above measured offset $A_b \pm s_{A_b}$ and the background-corrected photon count rates, leading to the ``virtual amplitude count rates'':
\begin{align}
\mathcal{R}^\prime_{i,\mathrm{chi2}} &= f_s\cdot\ddfrac{ \left(A_{i,\mathrm{obs}} - A_b - 
O\right) }{g}  \label{eq:Riprimebgcorr}\\
s^2(\mathcal{R}^\prime_{i,\mathrm{chi2}}) &= f^2_{s}\cdot \ddfrac{\left(A_{i,\mathrm{obs}}-A_b+ g\,\mathcal{R}_b\right)}{g\,N_\textit{sh}}\cdot \textit{ENF}^2 + f^2_{s}\cdot \ddfrac{s^2_{A_b}}{g^2}   \quad,
\label{eq:Riprimebgcorrvar}
\end{align}   
\noindent
with $\textit{\gls{enf}}\approx 1.08$  denoting the excess noise factor of the \glspl{pmt} used~\citep{Mirzoyan:2017,Eschbach:PhD}.
The gain $g$ and an additional possible offset $O$ are then free fit parameters that minimize:
\begin{equation}
\chi^2_{j,n} = \sum_{i=j}^{j+n} \ddfrac{\left(\mathcal{R}_{i,\mathrm{chi2}}-\mathcal{R}^\prime_{i,\mathrm{chi2}}\right)^2}{s^2_{U,L}(\mathcal{R}_{i,\mathrm{chi2}})+s^2(\mathcal{R}^\prime_{i,\mathrm{chi2}})- f^2_s/N_\textit{sh}\cdot \left(A_{i,\mathrm{obs}}-A_b+ g \mathcal{R}_b\right)/g}\quad, \label{eq:chi2}
\end{equation}
\noindent
where $j$ denotes the starting range index of a fit window and $n$ is the window length. Note that the correlated part of both the AN and PC variances has been subtracted from the denominator of Equation~(\ref{eq:chi2}) and $s^2_{U}(\mathcal{R}_{i,\mathrm{chi2}})$ is chosen when $\mathcal{R}_{i,\mathrm{chi2}}>\mathcal{R}'_{i,\mathrm{chi2}}$, otherwise $s^2_{L}(\mathcal{R}_{i,\mathrm{chi2}})$ is~used.

Since Equation~(\ref{eq:nonparalysable}) is only valid if $\mathcal{R}_{i,\mathrm{obs}} < 1/\tau$, a~minimum value for $j$ is obtained immediately, apart from possibly excluded ranges affected by clouds. We chose as a starting condition $\mathcal{R}_{i,\mathrm{obs}} < 1/(3\tau)$, to~be on the safe side. At~the same time, any saturated signal $A_{i,\mathrm{obs}}^\mathrm{sat}$ must be removed from the gluing procedure. 
Note that by construction, the~fit offset~$O$ should come out within a few standard deviations $s_{A_b}$ of zero. We have restricted the minimizer to vary within $\pm 10\,s_{A_b}$. Similarly, the~\gls{pc} efficiency $\epsilon$ was previously fixed to 0.9 (see also the results from the next Section~\ref{sec:Lgluing}).

The core idea is to find a range and location in the indices (a window) that provide a stable solution for the fit parameters $g$ and~$O$, and~a $\chi^2/(n-2)\sim 1$. 

The {algorithm} then slides through the data samples in windows of width $n$, until~$\mathcal{R}_{j+n,\mathrm{obs}} < s_{\mathcal{R}_b}$. 
For each window, Equation~(\ref{eq:chi2}) is minimized with respect to the parameters $g,O$.

We selected five test windows $n$, logarithmically spaced between ranges $\Delta r(n)=3000$~m and $\Delta r(n)=30,000$~m. These values can be dynamically selected from a configuration file, though, for~each channel separately. 
For each $\Delta r$, the~{algorithm} starts keeping track of how many valid points fall within each window. 
The {algorithm} then identifies the closest starting index $j$
and iterates through the data while checking whether both the photon-counting and amplitude signals lie above their predefined minimum thresholds: $\mathcal{R}_{i,\mathrm{obs}} > \Delta \mathcal{R}_b$ and $A_{i,\mathrm{obs}} > 4 s_{A_b}$. 
If both conditions are met, 
the window is extended until the $\Delta r$ has been reached. 
After finding all indices within a given window $\Delta r$ that meet these criteria, they are stored as a series of valid intervals for each test window~size.

The {algorithm} parallelizes the processing of each $\Delta r$ gluing window size, taking advantage of  \texttt{Python3} 
’s multiprocessing module, which allows it to spread the work over multiple CPU cores. The~parallelization  is achieved using the following~steps:
    \begin{enumerate}
        \item Identify the maximum range and prepare the data: the last index within the valid gluing range is stored; this sets the upper bound for processing. Then, the gluing window sizes are divided into batches of window sizes. Each batch contains one $\Delta r$ value for sequential processing, with~each batch running in parallel.
        \item Create a pool of processes: the multiprocessing pool is initialized with the number of CPU cores available, allowing the {algorithm} to utilize all processing resources effectively. The~batches are created as tuples of $\Delta r$ values and the corresponding channel ID, then these are the input for each parallel process.
        \item Parallel execution: the pool distributes the batches across the CPU cores, calling the batch process function for each batch. 
        This function iterates over each $\Delta r$ in the batch and executes the minimization of the $\chi^2$, Equation~(\ref{eq:chi2}). This setup allows each core to handle a different $\Delta r$ in parallel, greatly speeding up the computation.
        \item Collect results: Once all batches have been processed, the~results from each batch are stored and flattened into individual results, which contain the minimization results.
    \end{enumerate}

\begin{figure}[H]
\begin{adjustwidth}{-\extralength}{0cm}
\centering
\includegraphics[width=0.855\linewidth]{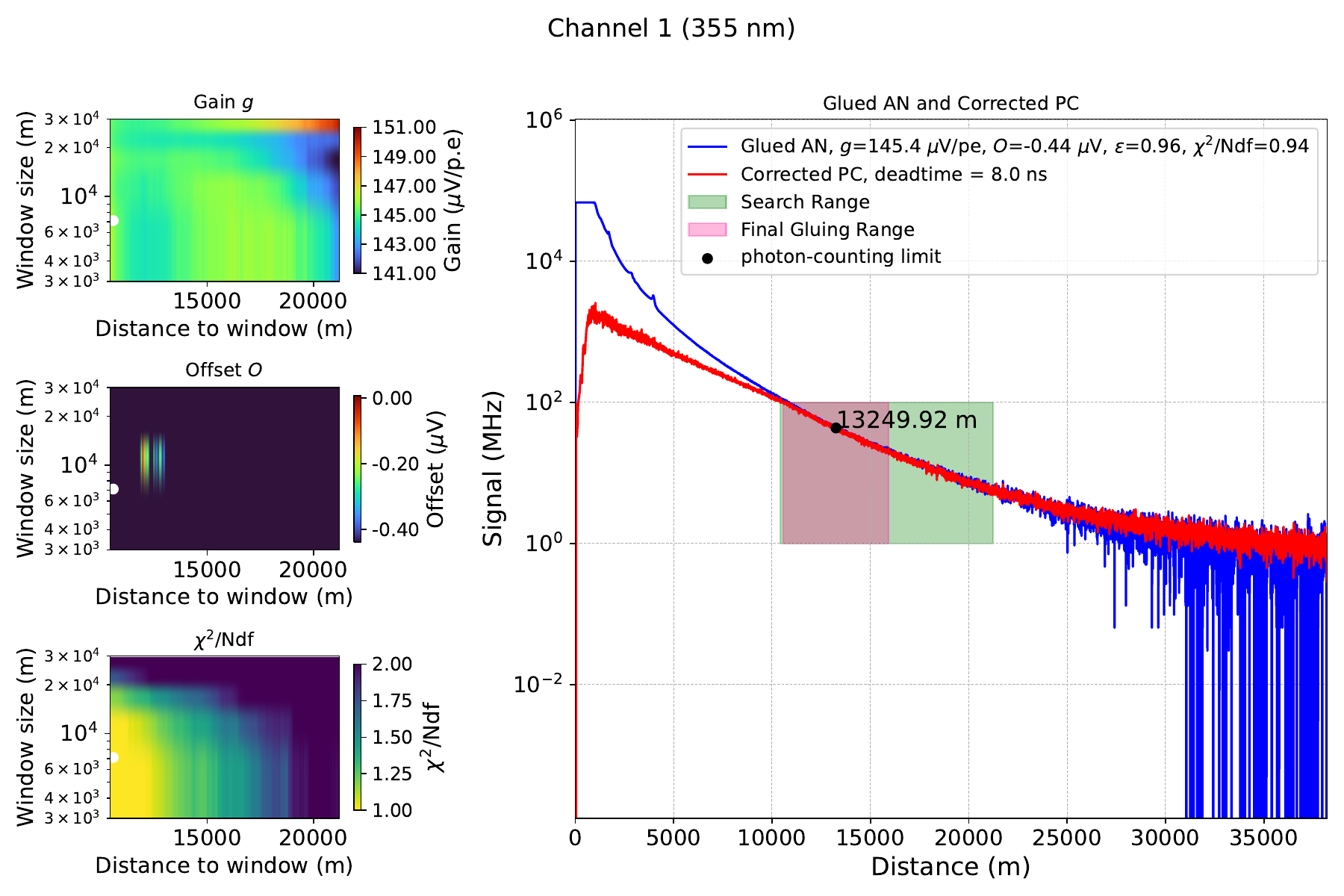}
\includegraphics[width=0.855\linewidth]{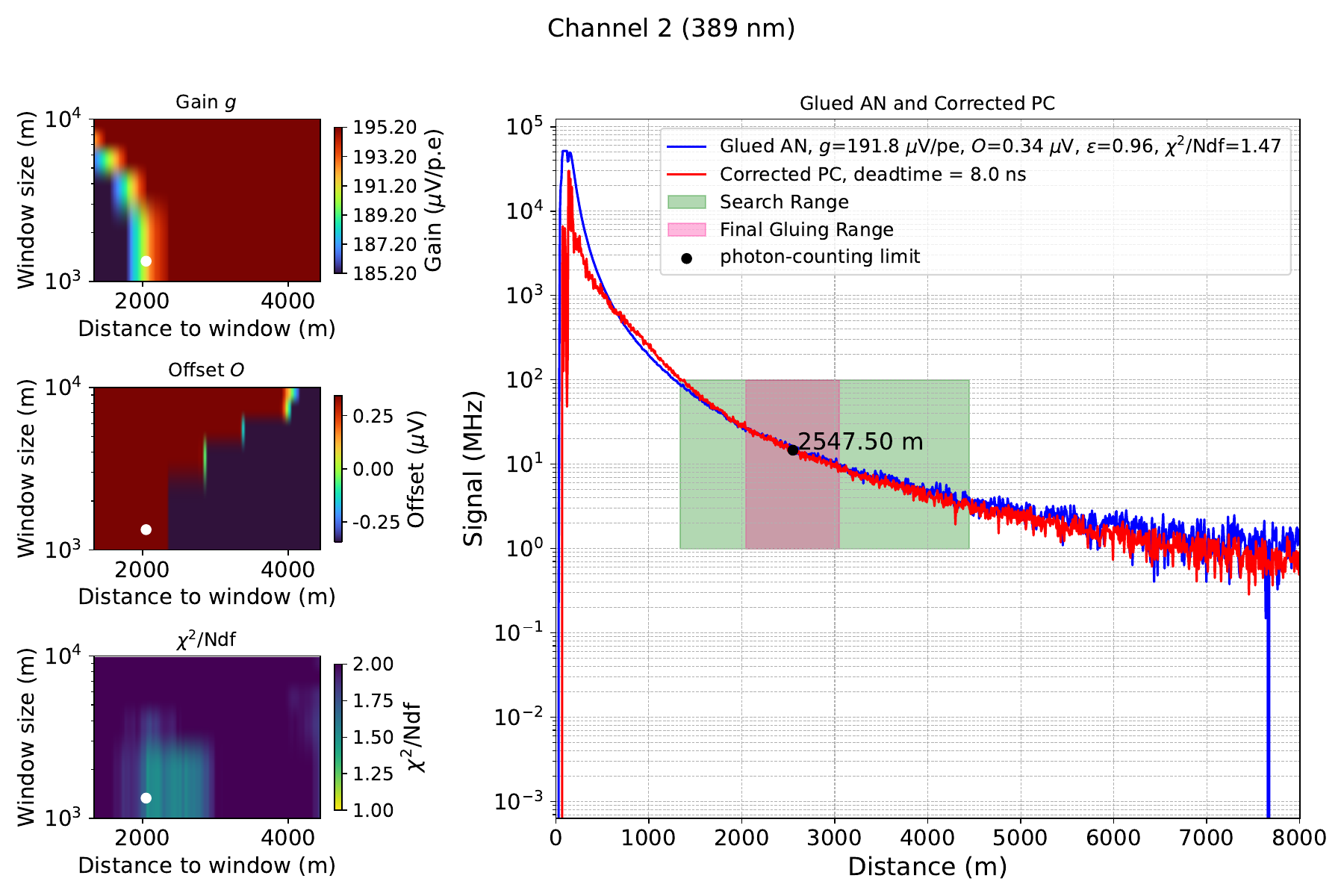}
\end{adjustwidth}
\caption{
Visualization 
 of the $\chi^2$-based gluing process applied to an  355~nm elastic channel with amplitude saturation registered by an \gls{lotr} with 20~MSamples/s (\textbf{top}) and a 387~nm Raman channel registered by the same \gls{lotr} (\textbf{bottom}). The~graphs on the right side show scaled amplitude (blue) and deadtime-corrected \gls{pc} (red). The~green regions display the search range, through which windows of different sizes have been slid, and~the final selected windows (pink). The~black points show the transition between scaled amplitude and deadtime-corrected \gls{pc} for the glued signal. 
On the left side, the~fitted \gls{an} gain $g$, the~additional \gls{an} offset $O$ (see Equation~(\protect\ref{eq:Riprimebgcorr})), and~the reduced $\chi^2$ of the fit (Equation~(\protect\ref{eq:chi2}) divided by $n-i-2$). The~white dots highlight the final chosen solution. 
\label{f:gluing-ch1}
}
\end{figure}

The absolute minimum reduced $\chi^2/\textit{\gls{ndf}}$ obtained is then searched and its absolute offset value $|O|$ tested to lie below a predefined threshold, until~a reference minimum $\chi^2/\textit{\gls{ndf}}$ is found. From~that result on, the~{algorithm} tries to enlarge the window, as~long as the $\chi^2/\textit{\gls{ndf}}$ does not exceed the reference by 10\% or falls anyhow below 1.1 and $|O|$ remains within one standard deviation of the PC background. The~result of that procedure is then a range that is as large as possible within the given constraints and has yielded an acceptable $\chi^2$ from the minimization. The~centre of that range is used to switch from the \gls{an} to the \gls{pc} signal (see the black points labeled ``photon-counting limit'' in Figure~\ref{f:gluing-ch1}).

The corresponding gluing constants $g$ and~$O$ 
are further used to construct the amplitude part of the signals (Equation~(\ref{eq:Riprimebgcorr})) and their uncorrelated variances (Equation~(\ref{eq:Riprimebgcorrvar})). Additional correlated uncertainties $>3\%$ or even $\gg 3\%$ (depending on the final gluing range chosen) must be assumed due to the uncertainties of the gluing parameter $g$. 
For the photon-counting part, Equations~(\ref{eq:Ribgcorr}) and~(\ref{eq:Ribgcorrvar}) are used. Here, a~correlated uncertainty from the selected PC efficiency $\epsilon$ of $>3\%$ applies. These correlated uncertainties can be greatly reduced by averaging gluing results over many data sets and studying and correcting their temperature dependencies and general ageing over time~\citep{Fruck:2022igg}. However, such a study is beyond the scope of this~work.

\subsubsection{Likelihood-Based~Gluing}
\label{sec:Lgluing}

Following and expanding the approach of \citet{Veberic:2012}, we construct a likelihood of the detection process in both channels, which can be expressed as
\begin{align}
    \mathcal{L} &= \prod_{i=j}^{j+n} \textit{PDF}   \left(p_i,\delta,\epsilon|\mathcal{R}_{i,\mathrm{obs}}\right) \cdot \textit{PDF}   \left(p_i,g,r_b,a_b,\gamma^2|A_{i,\mathrm{obs}},\textit{ENF}\right)   \nonumber\\
    & \quad {} \quad \times \textit{PDF}   \left(r_b|\avg{\mathcal{R}_b},s_{\mathcal{R}_b}\right)
    \times \textit{PDF}   \left(a_b|\avg{A_b},s_{A_b}\right)\label{eq:L}
\end{align}
\noindent
where $p_i$ is the (unknown) total number of incident photoelectrons (summed over all $N_\textit{sh}$) in channel $i$, and $\delta=f_s\tau/N_\textit{sh}$ and $\gamma$  
are the electronic noise of the signal transmission and digitization.  
The product runs from a suitable start index, which excludes amplitude saturation and ensures that $\mathcal{R}_{i,\mathrm{obs}} < \tau$. Note that $\tau$ becomes a fit parameter in this method through $\delta$ 
as~well as the photon detection efficiency $\epsilon$ of the PC channel. 
{We denote the following parameters, which refer to an average per laser shot with capital letters, and~those that refer to total accumulated over all shots with lower cases letters; that is, $a_i=N_\textit{sh} A_{i,\mathrm{obs}}$ and the background of the (summed) amplitude channels is $a_b$, similarly for the total photon background over all laser shots, $r_b$. 
}

Moreover, we assume that the AN channels measure voltages with an expectation of $E[{a_i}] = g p_i + N_\textit{sh} B$, where $B$
corresponds to the electronic pedestal. The~variance of ${a_i}$ can be described by~\citep{Veberic:2012}:
\begin{align}
    s_{a,i}^2 &= N_\textit{sh} \gamma^2 + \textit{F}^2  g^2 p_i ,
    \label{eq:sigmai21}
\end{align}
\noindent
where we have, in~addition to the approach of \citet{Veberic:2012}, added the excess noise contribution to the signal otherwise correlated with the \gls{pc} channel: $\textit{F}^2 = \textit{\gls{enf}}^2 -1$.  

The constant photon background $r_b$ now becomes an unknown quantity, measured on $n_\mathrm{off}$ background data fields, just as the background in the \gls{an} channel $a_b = {N_\textit{sh} B} + g \beta r_b$, which has been measured to $N_\textit{sh}(\avg{A_b} \pm s_{A_b})$. Note that $s_{A_b}$ here denotes the uncertainty of the mean $\avg{A_b}$, while $\gamma^2$ is the variance of the {electronic noise} per time slice. 
In addition, an~amplitude background efficiency parameter $\beta$ has been introduced that takes into account the efficiency of the amplitude offset measurement to measure the photon background, in~addition to the electronic baseline.

Defining $m_i = N_\textit{sh} \mathcal{R}_{i,\mathrm{obs}} /f_s$ 
and using Poissonian statistics for the \gls{pc} part and Gaussian statistics for the amplitudes, we define the following likelihood: 

\begingroup
\allowdisplaybreaks
\begin{align}
    \ln\mathcal{L} &= \sum_{i=j_\mathrm{on}}^{j_\mathrm{on}+n_\mathrm{on}} \Big[ - \frac{\epsilon p_i}{1+\delta \epsilon p_i} + m_i \ln\left(\ddfrac{\epsilon p_i}{1+\delta \epsilon p_i}\right) 
    - \nonumber\\ 
    & {} \quad- \ddfrac{\ln \left(N_\textit{sh} \gamma^2 + F^2 g^2
    p_i \right)}{2} - \ddfrac{\left(a_i-g( p_i- \beta r_b) -a_b\right)^2}{2\left(N_\textit{sh}\gamma^2 +  F^2 g^2 p_i\right)} \Big] \nonumber\\
  & {}  \qquad 
     + \sum_{i=j_\mathrm{off}}^{j_\mathrm{off}+n_\mathrm{off}} \Big[ -\ddfrac{\epsilon r_b}{1+\delta\epsilon r_b} + m_{i}\ln\left(\ddfrac{\epsilon r_b}{1+\delta\epsilon r_b}\right) 
    \Big] - \frac{\left(a_b - N_\textit{sh}\avg{A_b}\right)^2}{2 N_\textit{sh} s^2_{A_b}}
  \label{eq:logL}
\end{align}
\endgroup

This likelihood has $n_\mathrm{on}$ independent parameters $p_i$, apart from the outer parameters $g, \delta,\epsilon,r_b$ and $a_b$. 
Like in \citet{Veberic:2012}, the~parameter $\gamma$ can be previously determined from the residuals of a linear fit to $m_i/N_\textit{sh}$ as a function of $A_{i,\mathrm{obs}}$ over a range, where $m_i/N_\textit{sh}$ contains their lowest 2\%. We checked that $\gamma^2$ obtained this way is almost identical to $s^2_{A_b}$. Then it was fixed for the next~steps.  

To numerically maximize the likelihood, we follow the same approach as \citet{Veberic:2012}, by~minimizing $-2\ln{\mathcal{L}}$ with respect to the set of $p_i$ (with the minimum located at $\widehat{p}_i$), for~a given set of external parameters and minimize then the ``outer'' negative likelihood to find the solutions ($\widehat{g}, \widehat{\delta}, \widehat{\epsilon}, \widehat{\beta}, \widehat{a}_b, \widehat{r}_b|\widehat{p}_i$), which provide a global maximum of Equation~(\ref{eq:logL}).  Initialized with the prescriptions of \citet{Veberic:2012}, $\epsilon_\mathrm{init} = 0.95$, $\beta_\mathrm{init} = 1$, ${a_b}_\mathrm{init} = N_\textit{sh}\avg{A_b}$, and~${r_b}_\mathrm{init} = N_\textit{sh}\avg{\mathcal{R}_b}$, the~likelihood always converges in less than forty evaluations of the outer~function.  


An issue with Equation~(\ref{eq:logL}) needs extra attention: By construction, the~combination of parameters $\epsilon, p_i,g$ and $\beta$ is degenerate or may lead to $p_i \rightarrow \infty$, when simultaneously $g\rightarrow 0$ and $\epsilon\rightarrow 0$. This has to do with the fact that the PMT excess noise variance multiplies with $g^2$, but~only with $p_i$, while all other ingredients of the likelihood, Equation~(\ref{eq:logL}), show a linear combination of $\epsilon$ or $g$ and $p_i$. We suspect that this may be the reason why \mbox{\citet{Veberic:2012}} did not treat PMT excess noise at all in their likelihood. 
To remedy this behavior, we maximize $\ln \mathcal{L}$ twice: first, in~a setup similar to \citet{Veberic:2012}'s approach, in~which the excess noise contribution $F$ is set to zero. With~the corresponding results, we pick $\widehat{\epsilon}$ and fix it during a second maximization of $\ln \mathcal{L}$, this time with $F$ set to its correct value. The~latter maximization does not diverge anymore to infinite values of $p_i$.

\begin{figure}[H]
\begin{minipage}[b]{.484\linewidth}
\includegraphics[width=0.99\linewidth]{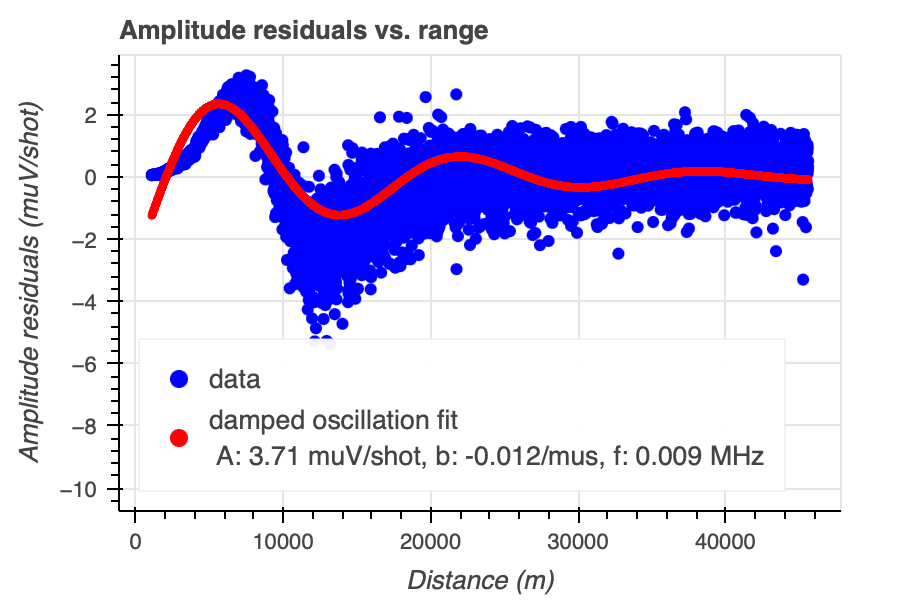}
\includegraphics[width=0.99\linewidth]{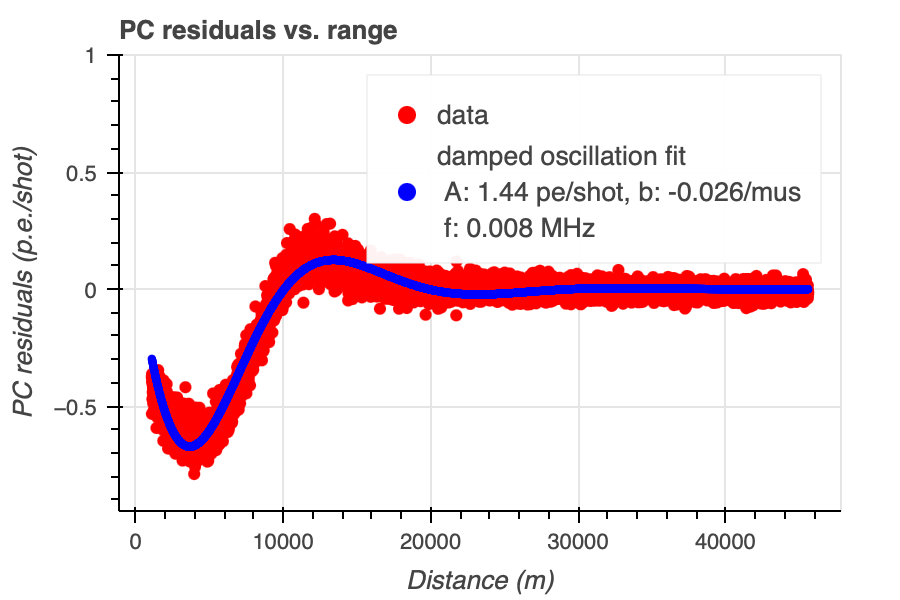}
\end{minipage}
\begin{minipage}[b]{.484\linewidth}
\includegraphics[width=0.99\linewidth]{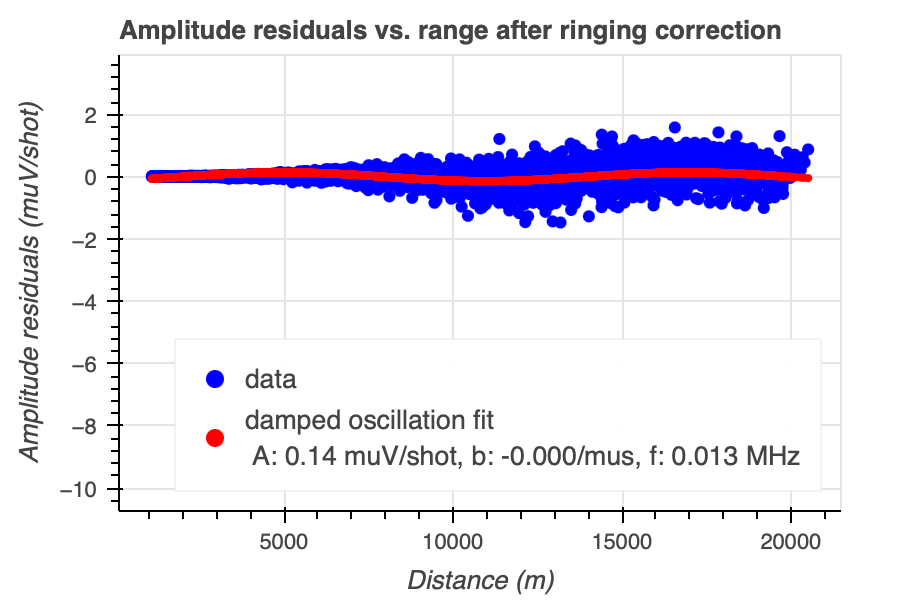}
\includegraphics[width=0.99\linewidth]{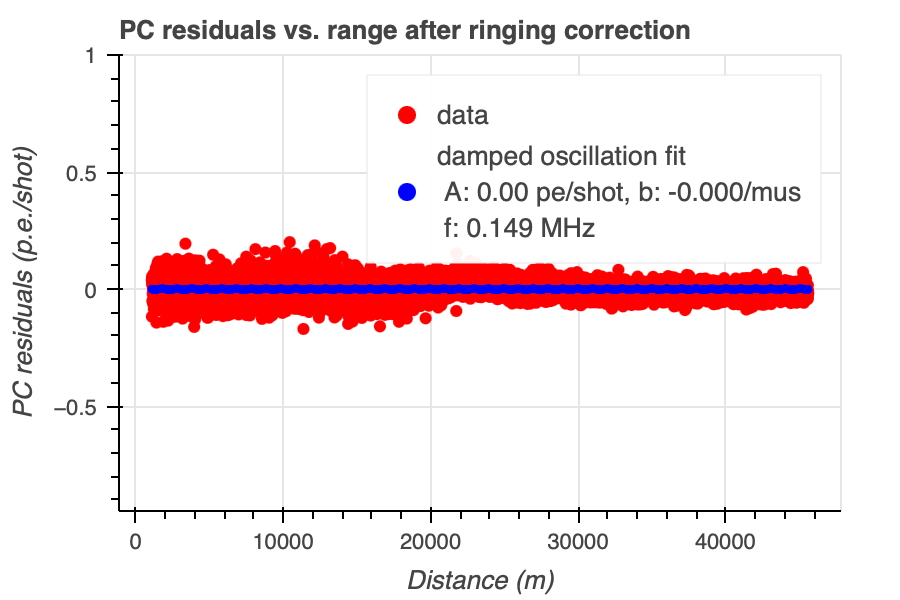}
\end{minipage}
\caption{Visualization 
 of the \gls{pc} oscillations found during the likelihood-based gluing process on the 355~nm elastic channel with 
a strongly over-driven \gls{lotr}. Only the \gls{pc} counting residuals on the lower left side had been corrected to obtain 
both figures on the right side. 
\label{f:gluing-L_ringing}}
\end{figure}

Finally, our data themselves have shown ringing of the PC baseline, after~injecting signal pulses into the \gls{lotr}, which exceed the maximum amplitude of 500~mV, recommended by the provider. To~highlight this behavior, 
Figure~\ref{f:gluing-L_ringing} shows residuals of $a_i{/N_\textit{sh}}$ and $m_i{/N_\textit{sh}}$ at the location of the likelihood maximum:
\begin{align}
r(a_i) &= \ddfrac{\left(a_i - \widehat{a}_b\right) - \widehat{g} \left(\widehat{p}_i- \widehat{\beta} \widehat{r}_b\right)}{N_\textit{sh}} \\
r(m_i) &= \ddfrac{m_i - \widehat{\epsilon}\widehat{p}_i/\left(1+\widehat{\delta}\widehat{\epsilon}\widehat{p}_i\right)}{N_\textit{sh}}\label{eq:resm}
\end{align}
as a function of range $R$. The~residuals have been fitted to a damped oscillation function
\begin{equation}
r = A \cdot \exp\left(-\ddfrac{2b}{c}\cdot R\right) \cdot \cos\left(\ddfrac{4\pi f_r }{c}\cdot R + \phi\right) \quad, \label{eq:dampedoscillation}
\end{equation}
where $A$ the initial ringing amplitude, $b$ the damping factor, $c$ the speed of light, 
$f_r$ the ringing frequency and $\phi$ the phase. We tested a subtraction of either of the fitted residuals from the amplitude or photon-counting signal, but~only the latter could effectively remove the oscillation from \textit{both} residuals (Figure~\ref{f:gluing-L_ringing}, right). For~this reason, we believe that the amplitude channel of the \gls{lotr} is still sufficiently linear, even in the case of strong over-driving of the \gls{lotr}, but~the photon-counting baseline gets affected by ringing. Nonetheless, ringing affects the PC signal over a range in which the amplitude signal is normally used; hence, only the gluing procedure is slightly affected, but~to a much smaller extent, the~glued signal~itself. 

Whenever a fit (Equation~(\ref{eq:dampedoscillation})) to the photon-counting residuals after gluing (Equation~(\ref{eq:resm})) shows an amplitude $A$ larger than 0.2~photo-electrons per shot, the~detected ringing is therefore subtracted from the photon-counting signal, and~the gluing procedure is repeated. During~the oscillation fits, the~damping factor $b$ is limited from below to $b>0.01~\upmu$s$^{-1}$, ensuring that the posterior corrections alter the signal at high ranges by always less than $0.2 \cdot \exp(-0.07 \cdot R(\mathrm{km}))$ photoelectrons per shot. However, in~the cases studied with our data, the~final amplitude and damping factors are orders of magnitude smaller (see Figure~\ref{f:gluing-L_ringing}). 

The complete fitting procedure leads to single photoelectron gains compatible with those obtained from the $\chi^2$-based gluing procedure, but~the dead times appear to be significantly larger (8--9~ns) than those provided by Licel (3--4~ns). The~discrepancy is independent of the ringing correction and probably due to the limited bandwidth of a 10~m coaxial cable signal line connecting each PMT with its \gls{lotr} and the resulting increase in the pulse width of a single~photoelectron.

\begin{figure}[H]
\begin{adjustwidth}{-\extralength}{0cm}
\centering
\begin{minipage}[b]{.44\linewidth}
\includegraphics[width=0.93\linewidth,trim={1.5cm 0cm 2cm 0},clip]{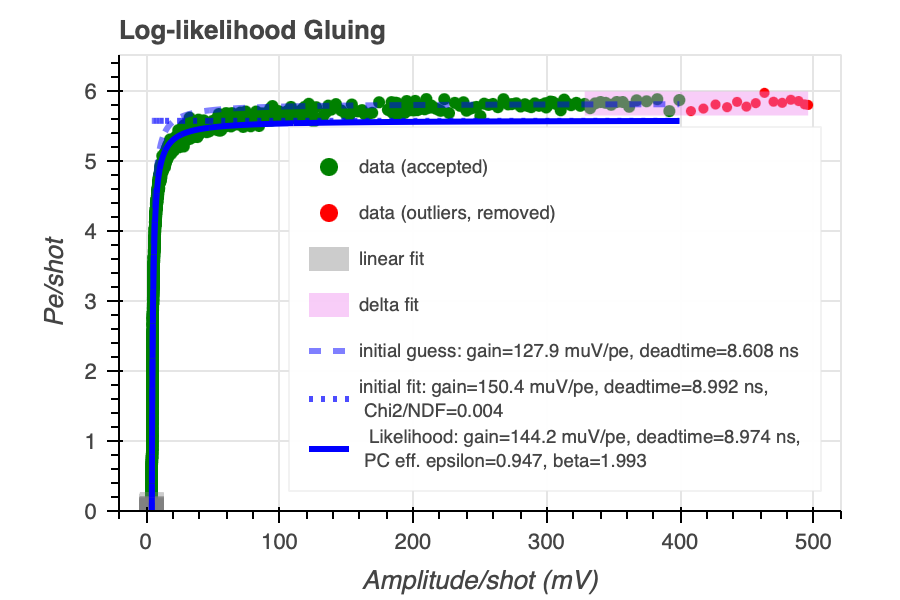}
\includegraphics[width=0.93\linewidth,trim={1.5cm 0cm 2cm 0},clip]{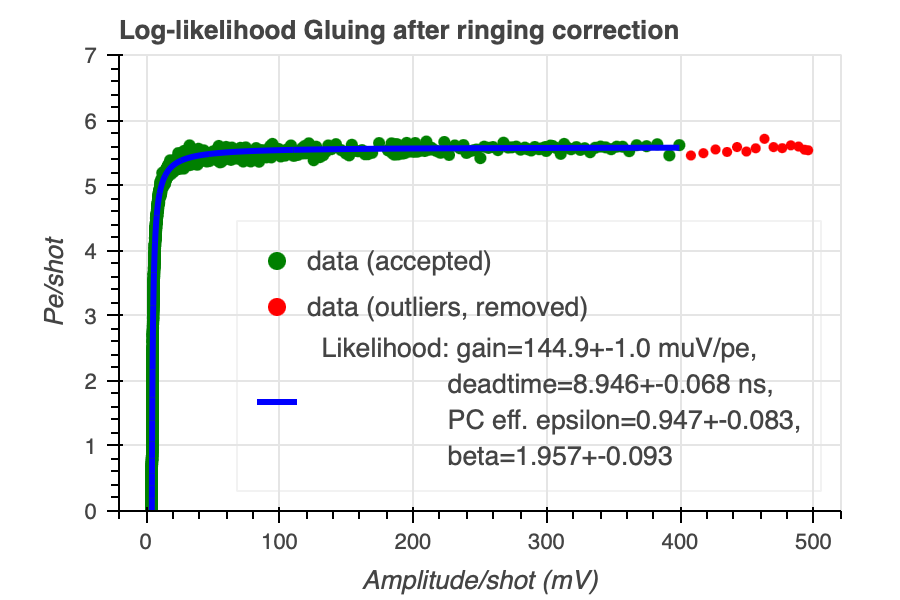}
\end{minipage}
\includegraphics[width=0.55\linewidth]{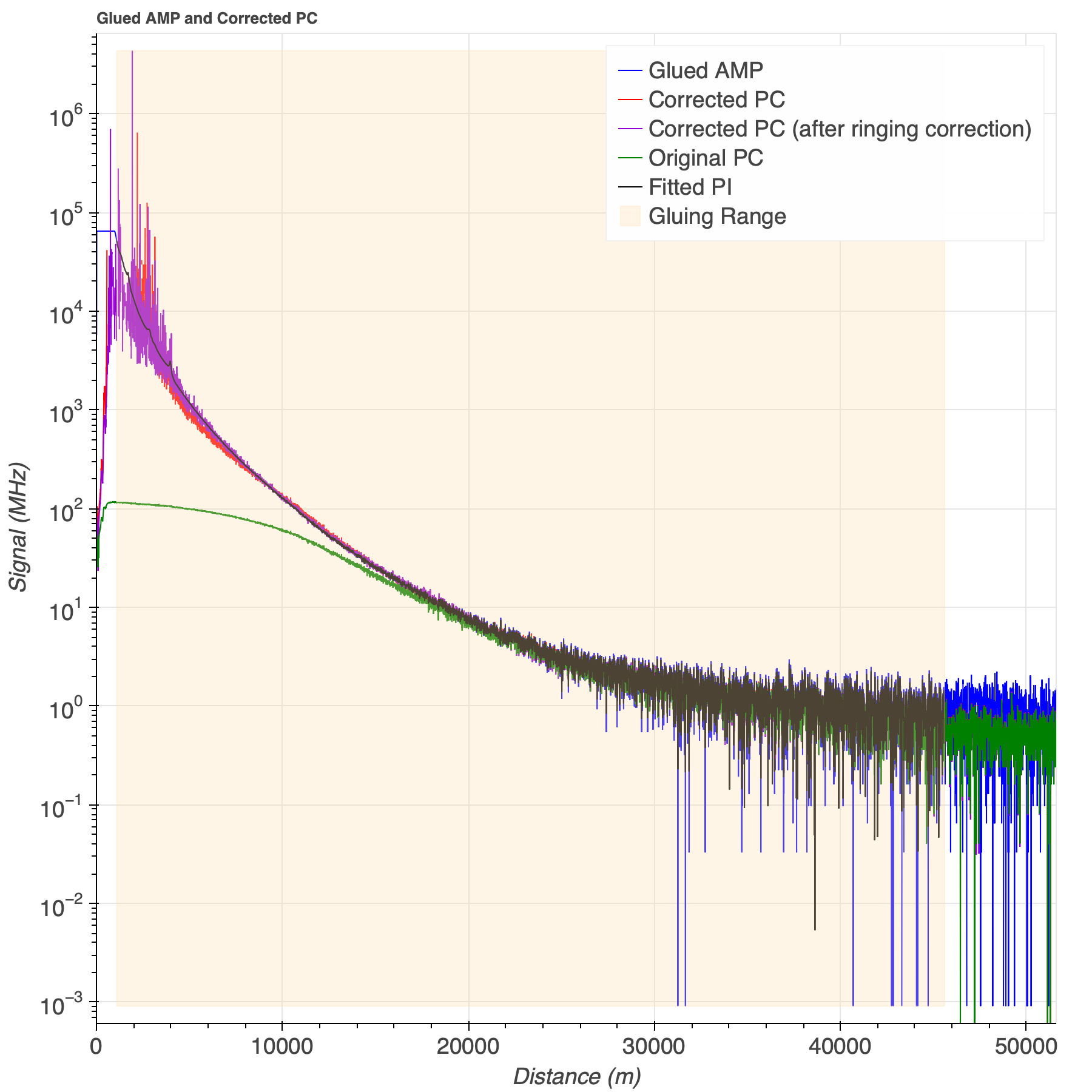}
\end{adjustwidth}
\caption{Visualization 
 of the likelihood-based gluing process applied to a 355~nm elastic channel with amplitude saturation registered by an \gls{lotr} with 20~MSamples/s. The~graphs on the left side show the number of photo-electron counts per shot ($m_i/N_{sh}$) as a function of the registered amplitudes per shot ($a_i/N_{sh}$). Green points show accepted data, and~red points show data that were previously excluded from all fits. 
The top left graph shows data before the ringing correction was applied, and~the different initial guess fits suggested by~\citet{Veberic:2012}, and~a full likelihood-based solution obtained from the minimization of Equation~(\ref{eq:logL}). Below, the ringing correction had been applied, and the likelihood minimization repeated. Now, the~solution describes the data correctly. The~graph on the right side shows scaled amplitude (blue) and dead-time-corrected \gls{pc}, before~(red) and after (violet) ringing correction. In~black, the~solution for the \gls{pc} expectation is shown. The~pale yellow region displays the region used for the likelihood gluing. 
\label{f:gluing-Lch0}}
\end{figure}

\begin{figure}[H]
\begin{adjustwidth}{-\extralength}{0cm}
\centering
\begin{minipage}[b]{.44\linewidth}
\includegraphics[width=0.95\linewidth,trim={1.5cm 0cm 2cm 0},clip]{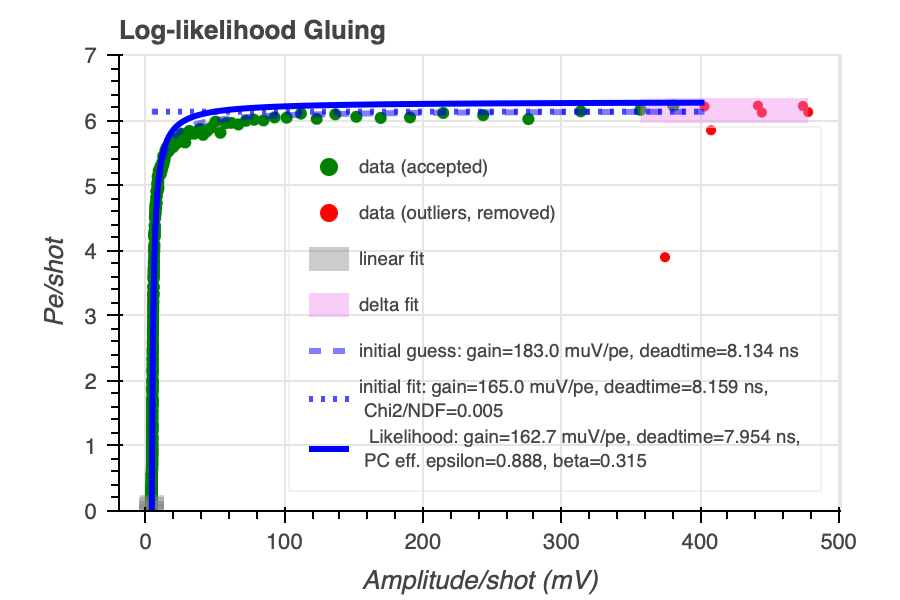}
\includegraphics[width=0.95\linewidth,trim={1.5cm 0cm 2cm 0},clip]{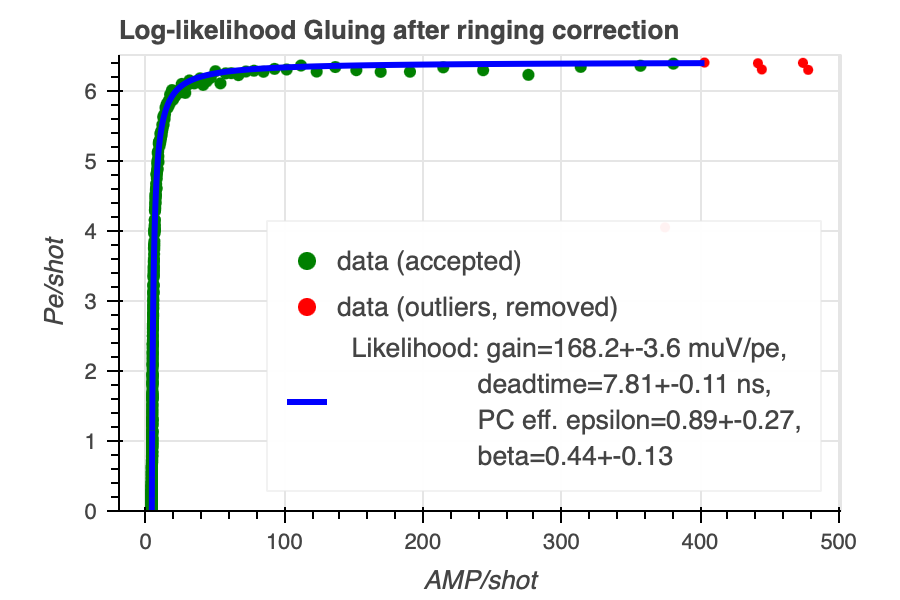}
\end{minipage}
\includegraphics[width=0.55\linewidth]{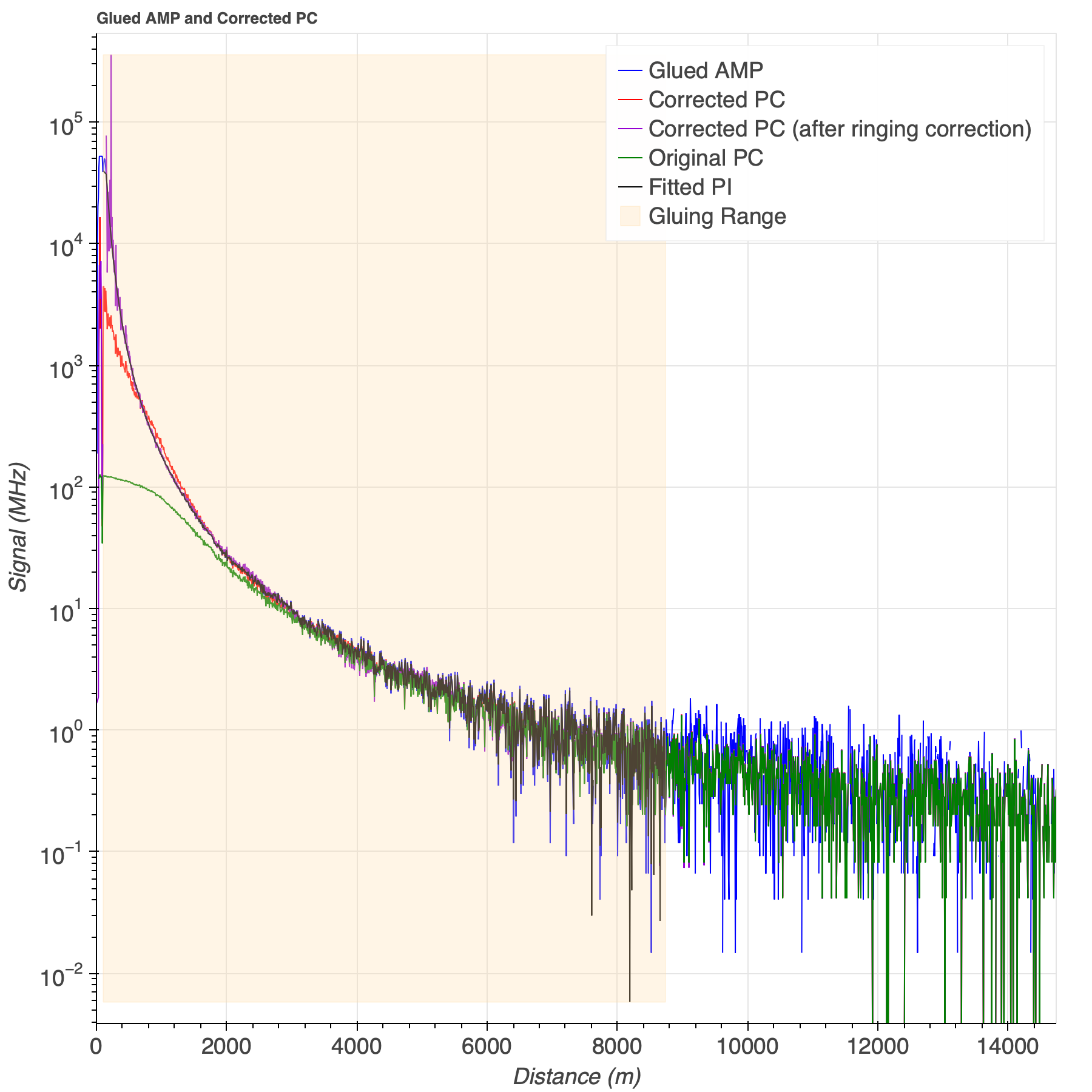}
\end{adjustwidth}
\caption{Visualization 
 of the likelihood-based gluing process applied to an  387~nm elastic channel with amplitude saturation registered by an \gls{lotr} with 20~MSamples/s. See Figure~\protect\ref{f:gluing-Lch0} for details.
\label{f:gluing-Lch1}}
\end{figure}

Figures~\ref{f:gluing-Lch0} and~\ref{f:gluing-Lch1} show scaled amplitudes $a_i{/N_\textit{sh}}$, dead-time-corrected photon counts $m_i{/N_\textit{sh}}$, and~fitted photon count expectations $\widehat{p}_i{/N_\textit{sh}}$, before~and after the ringing correction has been applied and together with the different fits (from initial guesses to the likelihood-based solutions before and after ringing corrections applied). {One can observe that the solution before ringing correction is rather poor and does not describe well the data for amplitudes larger than $\sim$20~mV. After~correction, the~bias for large amplitudes has disappeared. }
On the right-hand side, the~different signals are shown, which appear to match well, particularly after the ringing correction has been~applied.

\unskip

A rather lengthy and CPU-intensive construction of a Hessian matrix $H_{ij} = \partial\ln\mathcal{L}/\partial x_i\partial x_j$ from the likelihood Equation~(\ref{eq:logL}), with~$x_i$ and $x_j$ being combinations of the parameters ($g,\delta,\epsilon,\beta,a_b$ and $r_b$) around the likelihood maximum ($\widehat{g}, \widehat{\delta}, \widehat{\epsilon}, \widehat{\beta},  \widehat{a}_b, \widehat{r}_b$), profiled on all $p_i$, allows to obtain an estimate of the covariance matrix $H^{-
1}_{ij}$ of the outer parameters. In~elastic channels, we find uncertainties $\sqrt{H^{-1}_{ii}}/\widehat{x_i}<0.2\%$ for all parameters, except~$\widehat{\beta}$, which exhibits an uncertainty of $\approx$0.5\%. Correlations between parameters $\sqrt{H^{-1}_{ij}/(\widehat{x}_i\widehat{x}_j)}$ yield values $<$0.3\% in all cases, except~$\beta$ and $r_b$, which are correlated by $<0.5\%$. In~the case of Raman lines, the~uncertainties are of order $\sqrt{H^{-1}_{ii}}/\widehat{x_i}<8\%$ for the single photoelectron gain $g$ and the PC efficiency $\epsilon$ (and correlations between both parameters of the same order), and~$\sqrt{H^{-1}_{ii}}/\widehat{x_i}<4\%$ for $\beta$. Whereas uncertainties $>1\%$, obtained from one single data set, are unacceptable to meet the \gls{brl} performance requirements, these can be reduced by averaging gluing results from various data sets taken during similar conditions, e.g.,~during the same night with stabilized and constant \gls{pmt} gain and \gls{lotr} discriminator~thresholds.

The glued signal with this method is then constructed from the recorded amplitudes: 

\vspace{-6pt}
\begingroup\makeatletter\def\f@size{9}\check@mathfonts
\def\maketag@@@#1{\hbox{\m@th\normalsize\normalfont#1}}
\begingroup
\allowdisplaybreaks
\begin{align}
{x}_i &= \ddfrac{a_i - \widehat{a}_b}{\widehat{g}} - \left(1-\widehat{\beta}\right)\widehat{r}_b \\
\mathcal{R}^\prime_{i,\mathrm{logL}} &= \ddfrac{f_s}{N_\textit{sh}}  \cdot {x}_i \label{eq:RiprimeLgluing}\\
s^2_{U,L}(\mathcal{R}^\prime_{i,\mathrm{logL}}) &= f_s^2\cdot \left(\ddfrac{\vert {x}_i-\mu_{U,L}({x}_i)\vert^2}{N_\textit{sh}^2}\cdot \textit{ENF}^2 + \ddfrac{s^2_{A_b}}{\widehat{g}^2} + \left(1-\widehat{\beta}\right)^2s^2_{\mathcal{R}_b}  + \ddfrac{1}{12}\ddfrac{(500~\mathrm{mV})^2}{2^{2N_\mathrm{bits}}N_\textit{sh}\,\widehat{g}^2}\right)~, 
\label{eq:RiprimeLgluingvar}
\end{align}   
\endgroup
\endgroup
where the last entry corresponds to the resolution of the amplifier ($N_\mathrm{bits}=12$ or $N_\mathrm{bits}=16$, depending on the \gls{lotr} model). The~\gls{pc} channels are converted to
\begin{align}
{x}_i &= \ddfrac{N_\textit{sh}}{f_s} \cdot \mathcal{R}_{i,\mathrm{obs}}\\
\mathcal{R}_{i,\mathrm{logL}} &= \ddfrac{1}{\widehat{\epsilon}}\cdot \left(\ddfrac{\mathcal{R}_{i,\mathrm{obs}}}{1-\widehat{\delta}{x}_i} \right) - \ddfrac{f_s}{N_\textit{sh}}\cdot\widehat{r}_b \label{eq:RiLgluing}\\
s^2_{U,L}(\mathcal{R}_{i,\mathrm{logL}}) &= \ddfrac{1}{\hat{\epsilon}^2}\cdot \left(\ddfrac{f_{s}^2}{N_\textit{sh}^2} \cdot  \ddfrac{\vert {x}_i-\mu_{U,L}(X_i)\vert^2}{\left(1-\widehat{\delta}{x}_i\right)^4} \right) + s^2_{\mathcal{R}_b} \quad.
\label{eq:RiLgluingvar}
\end{align}

Finally, the~transition between \gls{an} and \gls{pc} regime is chosen to be the first entry $i$, at~which $s^2_U(\mathcal{R}_{i,\mathrm{logL}})< s^2_U(\mathcal{R}^\prime_{i,\mathrm{logL}})$.

\subsubsection{Range-Corrected~Signal}

The logarithm of the Range-Corrected-Signal (\gls{rcs}) is constructed from the glued \gls{pc} rates $\mathcal{R}_{i,\mathrm{meth}}$, where $\mathrm{meth}$ can denote the $\chi^2$-gluing method, or~the likelihood-based one with or without ringing correction:
\begin{align}
\mathcal{S}_{i,\mathrm{meth}} &= \ln\left(R_i^2 \cdot \mathcal{R}_{i,\mathrm{meth}}      \right)~, \\
s^2_{U,L}(\mathcal{S}_{i,\mathrm{meth}}) &= \ddfrac{s^2_{U,L}(\mathcal{R}_{i,\mathrm{meth}})}{\mathcal{R}^2_{i,\mathrm{meth}}} + \ddfrac{4}{12} \left(\ddfrac{w}{R_i}\right)^2 ~
\end{align}
and $w=c/(2f_s)$ is the sampling bin~width. 

\begin{figure}[H]
\centering
\hspace{-4cm}
{\small 355 nm}
\hspace{4.5cm}
{\small 387 nm}
\hspace{4.5cm}
{\small 532 nm}
\begin{adjustwidth}{-\extralength}{0cm}
\centering
\includegraphics[width=0.31\linewidth]{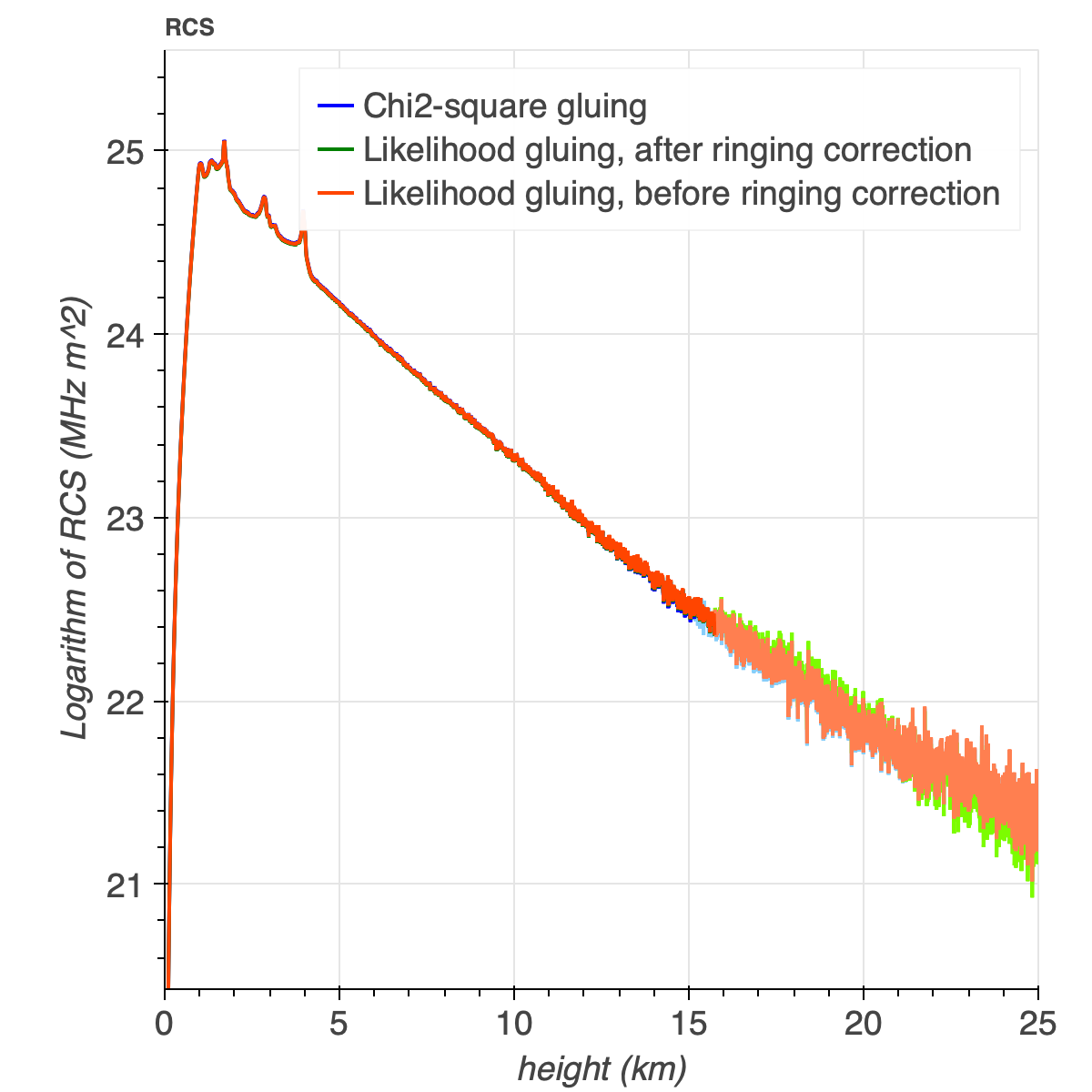}
\includegraphics[width=0.31\linewidth]{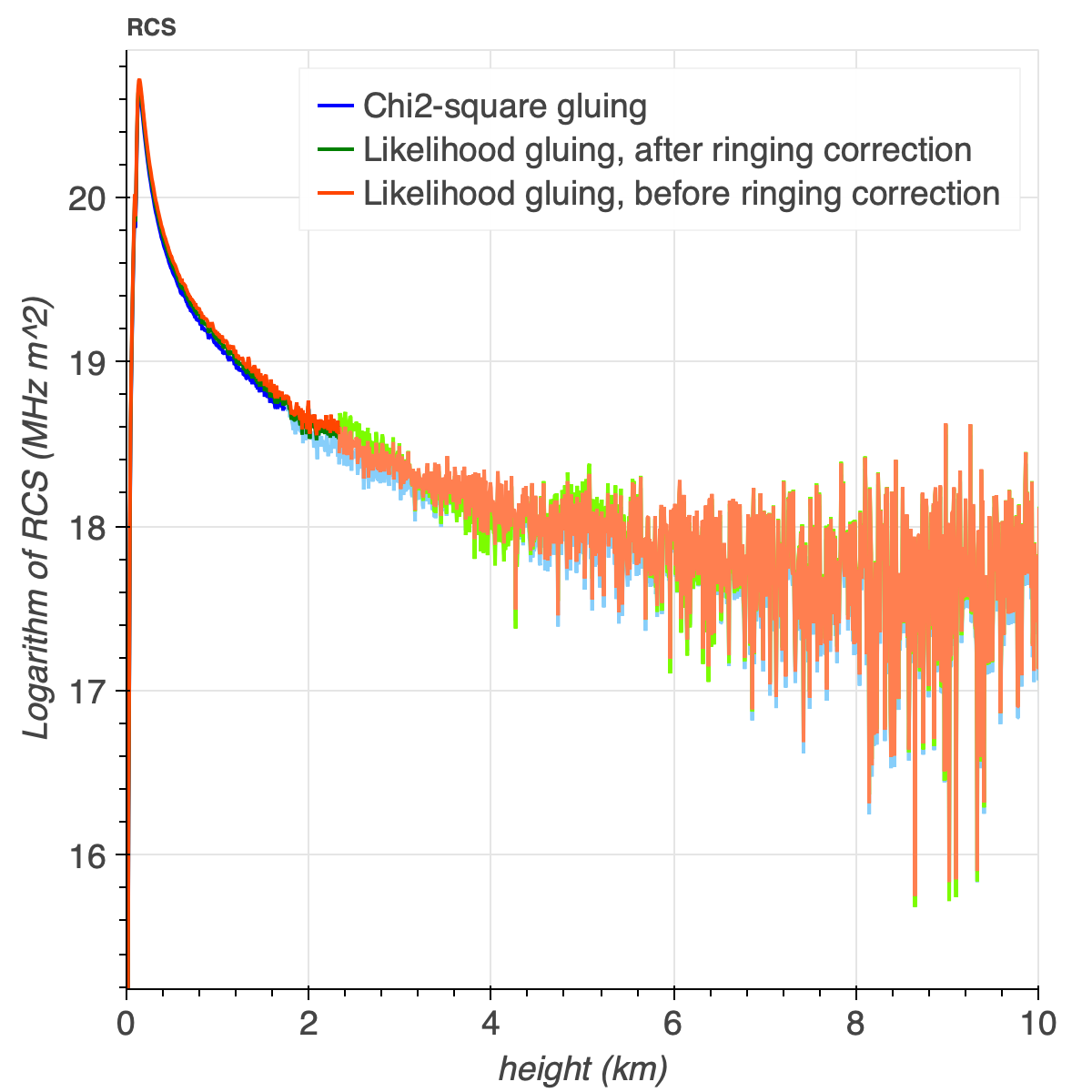}
\includegraphics[width=.31\linewidth]{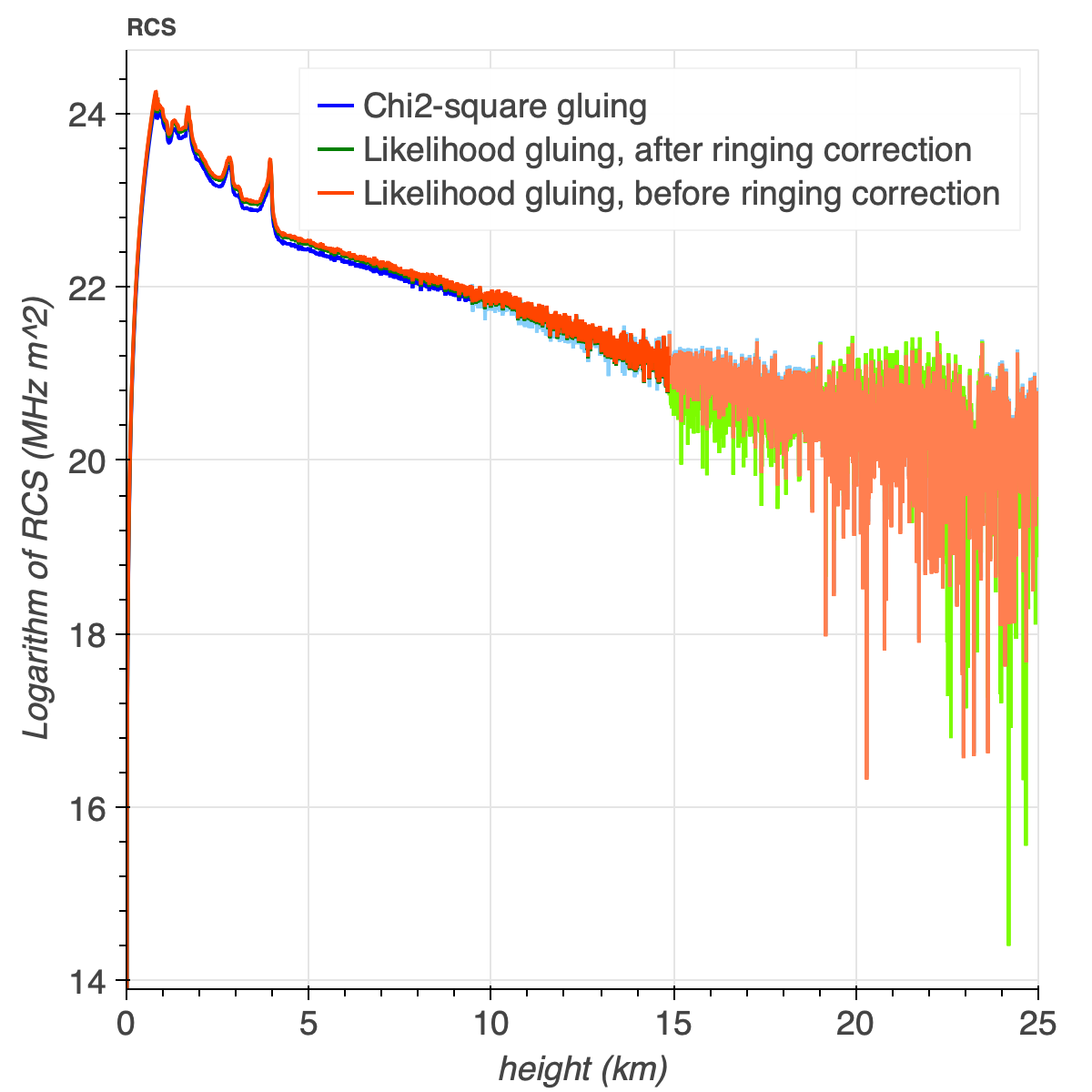}
\includegraphics[width=0.31\linewidth]{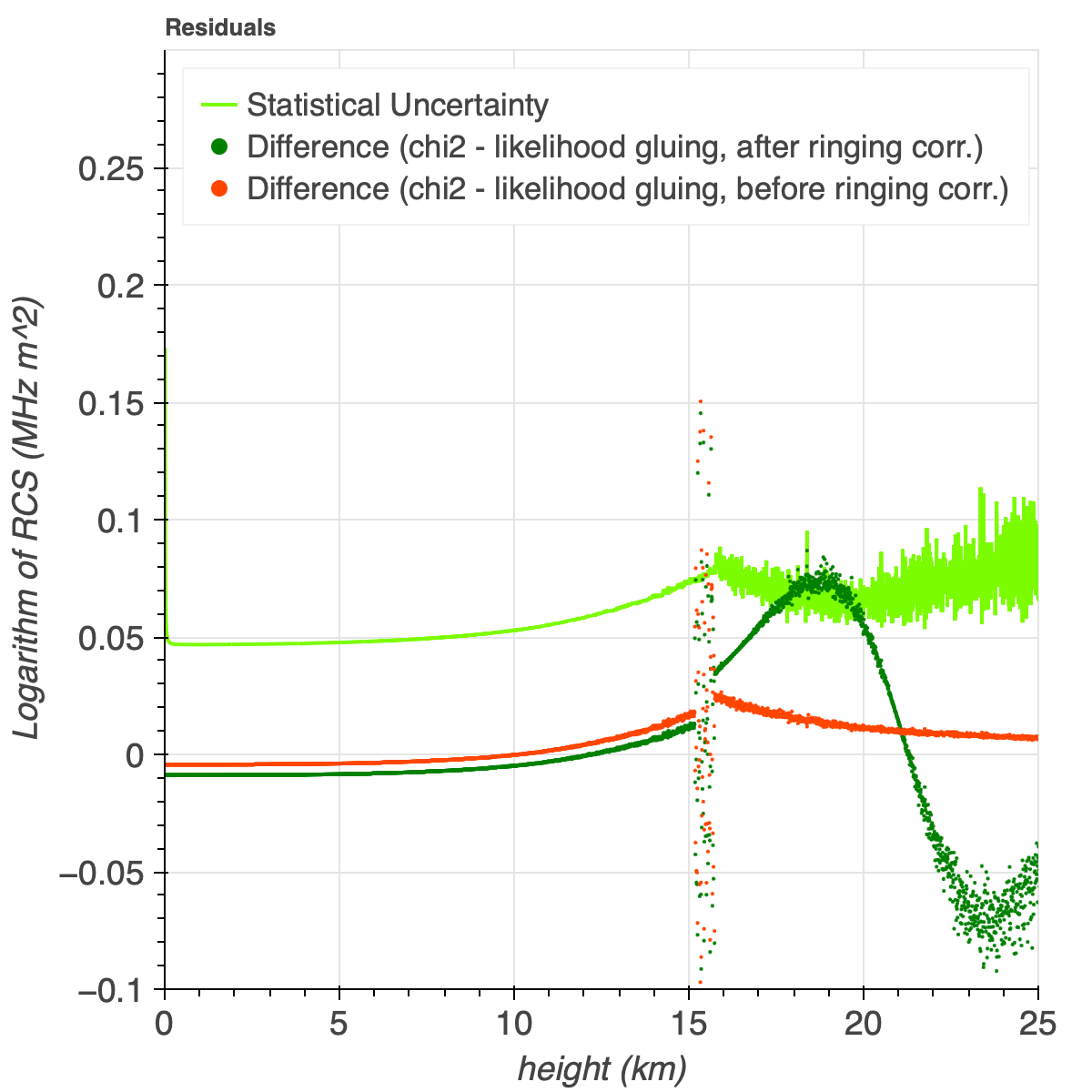}
\includegraphics[width=0.31\linewidth]{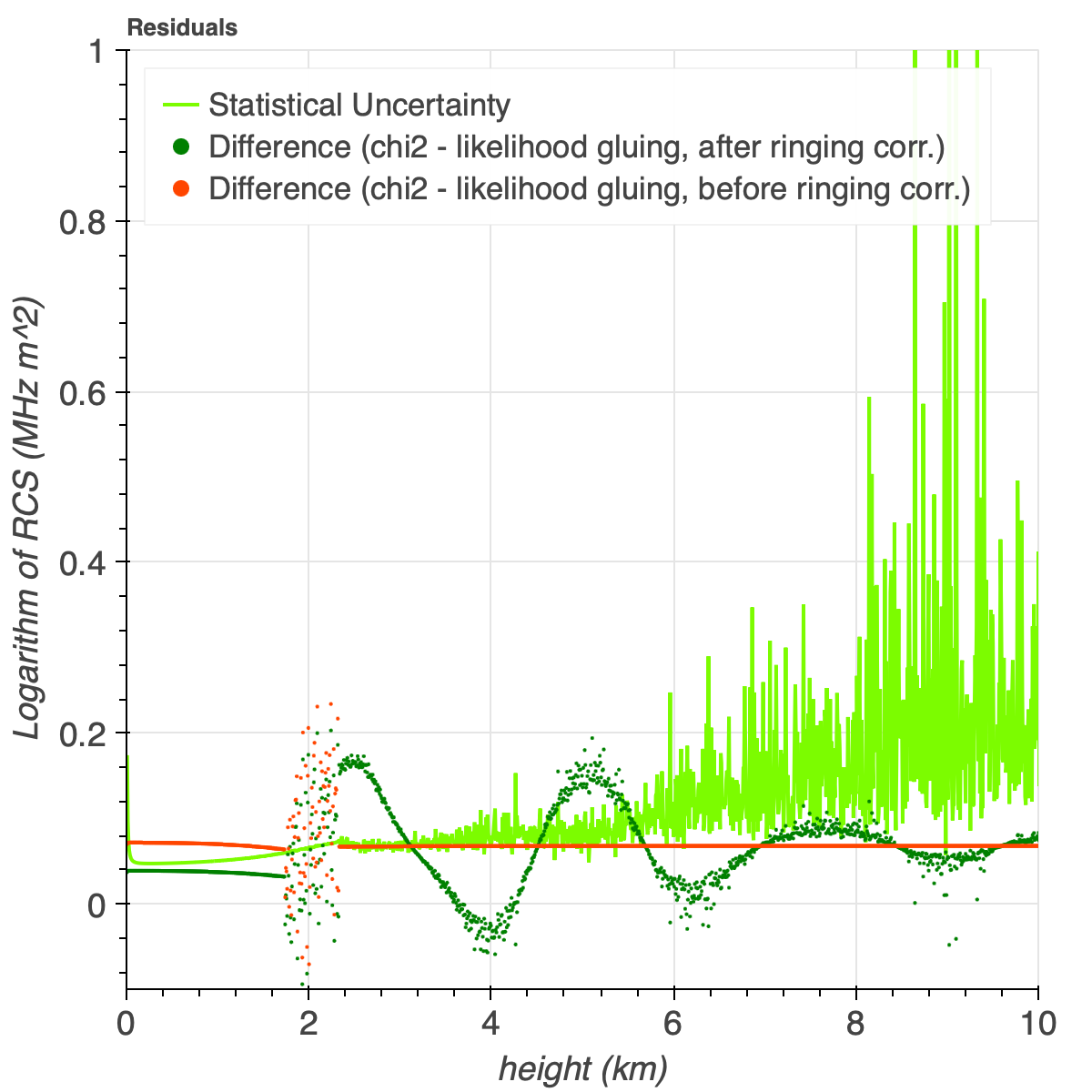}
\includegraphics[width=.31\linewidth]{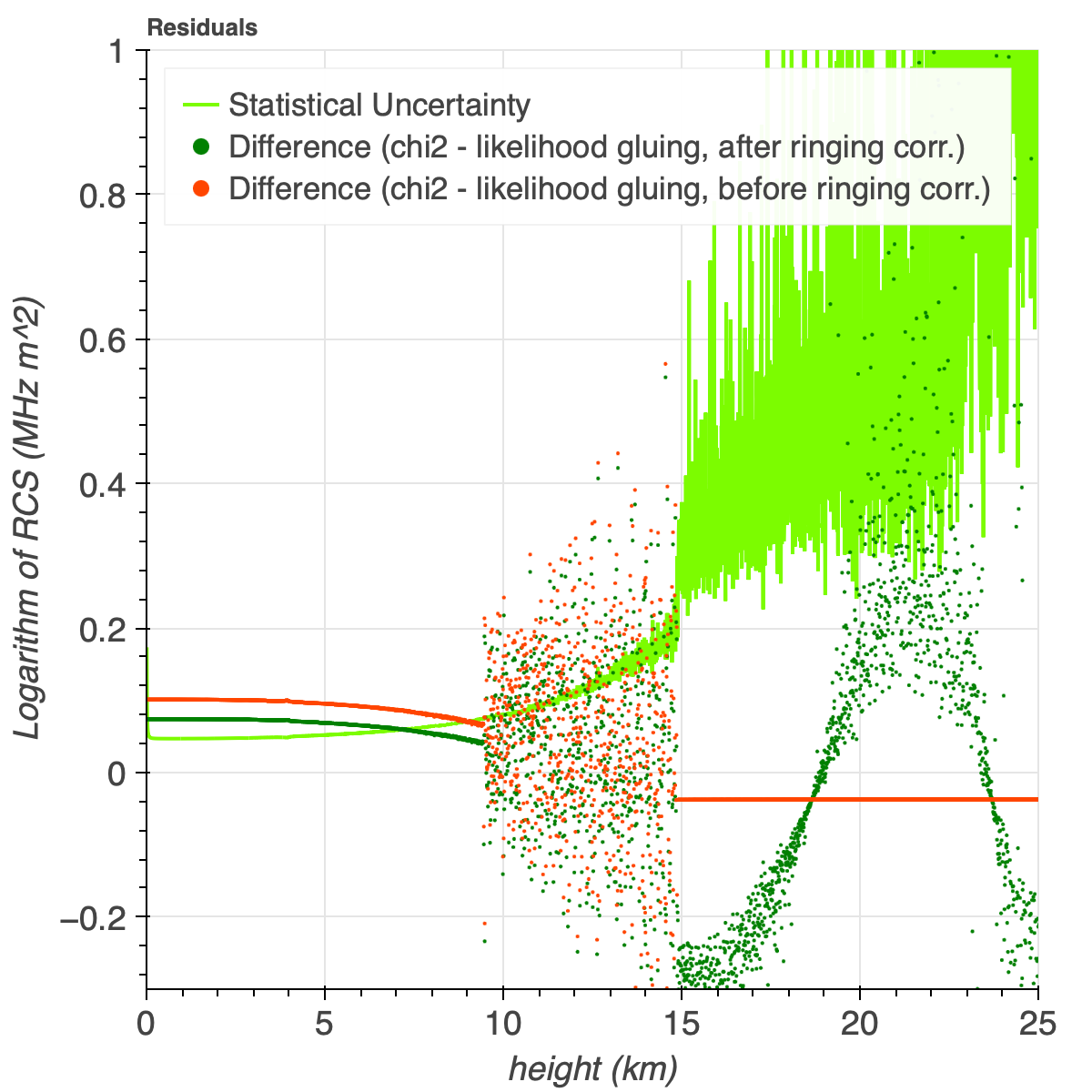}
\end{adjustwidth}
\caption{\textbf{Top}: Logarithms 
 of the range-corrected-signals, $\mathcal{S}_i$, obtained with the $\chi^2$-based gluing method (blue), the~likelihood-based method, before~(red) and after (green) a ringing correction has been applied. No re-binning has yet been applied here; negative signals are not shown. \textbf{Below}: The residuals between the signals from the upper plots obtained with likelihood-based methods (before and after ringing correction) and those  from the $\chi^2$-based gluing are shown, as~well as the statistical uncertainty of the likelihood-based $\mathcal{S}_i$. The~left columns show a 355~nm elastic line with little photon background, the~central column a 389~nm Raman line, and~the right column a 532~nm elastic line with very strong photon background from the street lighting of the \gls{uab} university~campus. \label{fig:RCS}}
\end{figure}

Figure~\ref{fig:RCS} compares $\mathcal{S}_{i}$ from the three gluing algorithms for different scenarios: a rather background-free elastic line (Figure~\ref{fig:RCS} left), a~Raman line (Figure~\ref{fig:RCS} centre) and an elastic line dominated by strong photon background (Figure~\ref{fig:RCS} right). We can see that the three methods produce comparable results, with~deviations $\lesssim 25\%$ for all cases and ranges. {Note that the different transition ranges from amplitude to photon-counting signals between the two gluing methods generate a region of apparently stronger fluctuations in the residuals.} The strongest deviations are found when comparing the ringing-corrected signals with those without such a correction, on~the overdriven \glspl{lotr}. As~expected, the~closer the photon-counting signal range approaches ground, the~stronger such differences become visible. We believe, however, that the ringing-corrected glued signal describes the true photon rates better, and~that it is actually the (traditional) reference $\chi^2$ method that deviates. Nevertheless, such systematic deviations may mask aerosol transmission retrievals of the same magnitude, and~a future upgrade of the system shall solve the issue by hardware rather than software. Apart from the ringing correction, deviations of $\lesssim 10\%$ are found, particularly at low ranges. The~355~nm elastic line currently fulfils the $<3\%$ requirement (see Section~\ref{sec:introduction}), whereas the Raman line seems to be close to reaching that level, if~the ringing correction can be trusted. Only the case of a 532~nm elastic line with unusually strong background light contamination clearly fails this test, with~systematic deviations $\lesssim 10\%$ between the different gluing methods, even if the ringing correction can be~trusted.  

\subsection{Dynamic~Rebinning}
\label{sec:rebinning}

The dynamic rebinning algorithm
 addresses the presence of negative values produced during background subtraction in the glued signal. 
The {algorithm} operates by dynamically adjusting the bin widths  structure to eliminate negative values, while at the same time maintaining fine binning where a strong signal (for instance, due to a cloud) is~visible.

The glued \gls{pc} rates $\mathcal{R}_i$ are segmented into discrete windows where values fall below a predefined threshold $T$. This threshold is chosen to isolate regions that contain signals with too small values $\mathcal{R}_i<T$ (by default $T=3 \cdot r_b$ is chosen). The~{algorithm} returns a series of such windows $W_j$:
\begin{equation}
    W_j \text{ boundaries: } [i_1, i_2], \quad \text{where } \mathcal{R}_i < T, \, \forall i \in [i_1, i_2].
\end{equation}

    If all $\mathcal{R}_i \geq T$,
    the windows remain unchanged and correspond to the sampling window of the \gls{lotr}.
    Wherever windows are found with $\mathcal{R}_i < T$, the~{algorithm} proceeds to the re-binning~phase.

\newpage
In windows containing negative values, adjacent bins are averaged iteratively in pairs until the averaged signal in the new window is positive or the new window contains the entire $W_j$. For~each bin $i$, the~signal rate $\mathcal{R}_i$, bin width $w$, and~variance are updated as follows:

\vspace{-6pt}
\begin{adjustwidth}{-\extralength}{0cm}
\centering 
\begin{equation}
    \mathcal{R}_{\text{re-binned}}(j) = \frac{1}{N} \sum_{i \in W_j} \mathcal{R}_i, \quad
    w_{\text{re-binned}}(j) = \sum_{i \in W_j} w(i), \quad
    s^2_{\text{re-binned}}(\mathcal{R}_j) = \frac{1}{N^2} \sum_{i \in W_j} s^2(\mathcal{R}_i).
\end{equation}
\end{adjustwidth}
Here
, $N$ is the step size, representing the number of original bins combined during the re-binning process. This normalization ensures that the signal density and associated variances remain consistent with the original bin structure and 
that the total signal content is preserved:
\begin{equation}
    \sum_{j} \mathcal{R}_{\text{re-binned}}(j) = \sum_{i} \mathcal{R}_i.
\end{equation}

For windows with an odd number of bins, the~last bin is summed with the previous pair of bins, forming a ``three-bin'' group:
\begin{equation}
    \mathcal{R}_{\text{re-binned}}(j) = \mathcal{R}_{i-2} + \mathcal{R}_{i-1} + \mathcal{R}_i.
\end{equation}
This ensures consistency in bin width distributions while maintaining the stability of the signal~representation.

If a window continues to contain negative values, even after rebinning (e.g., at~the very end of the summation procedure), it gets replaced by a default baseline signal value, corresponding to $P(X>0|\mu=\mu_0)=0.5$ yielding $\mu_0=0.69$ photoelectrons in the entire window:
\begin{align}
    \mathcal{R}_{\text{baseline}} &= \epsilon \cdot \ddfrac{0.69\,f_{s}}{ N_\textit{sh}}  \\
    s^2_{U,L}(\mathcal{R}_{\text{baseline}}) &= 
    \epsilon^2\cdot\frac{\vert 0.69 - \mu_{U,L}(0)\vert^2 f_{s}^2 }{N_\textit{sh}^2} \\
    s^2_L(\mathcal{R}_{\text{baseline}}) &= 
    \epsilon^2\cdot\frac{0.69^2 f_{s}^2 }{N_\textit{sh}^2} \\
    s^2_U
    (\mathcal{R}_{\text{baseline}}) &= 
    \epsilon^2\cdot\frac{\left( 1.84 - 0.69 \right)^2 f_{s}^2 }{N_\textit{sh}^2}
\end{align}

Once all windows are processed, the~rebinned signal values, widths, and~variances are integrated to form a new range-corrected signal. This updated signal is now free from negative values and apt for logarithmic transformation:
\begin{equation}
    \ln\left(\mathcal{R}_{\text{re-binned}} \cdot R^2 \right), \quad \mathcal{R}_{\text{re-binned}} > 0.
\end{equation}

\begin{figure}[H]
\centering
\includegraphics[width=0.49\linewidth]{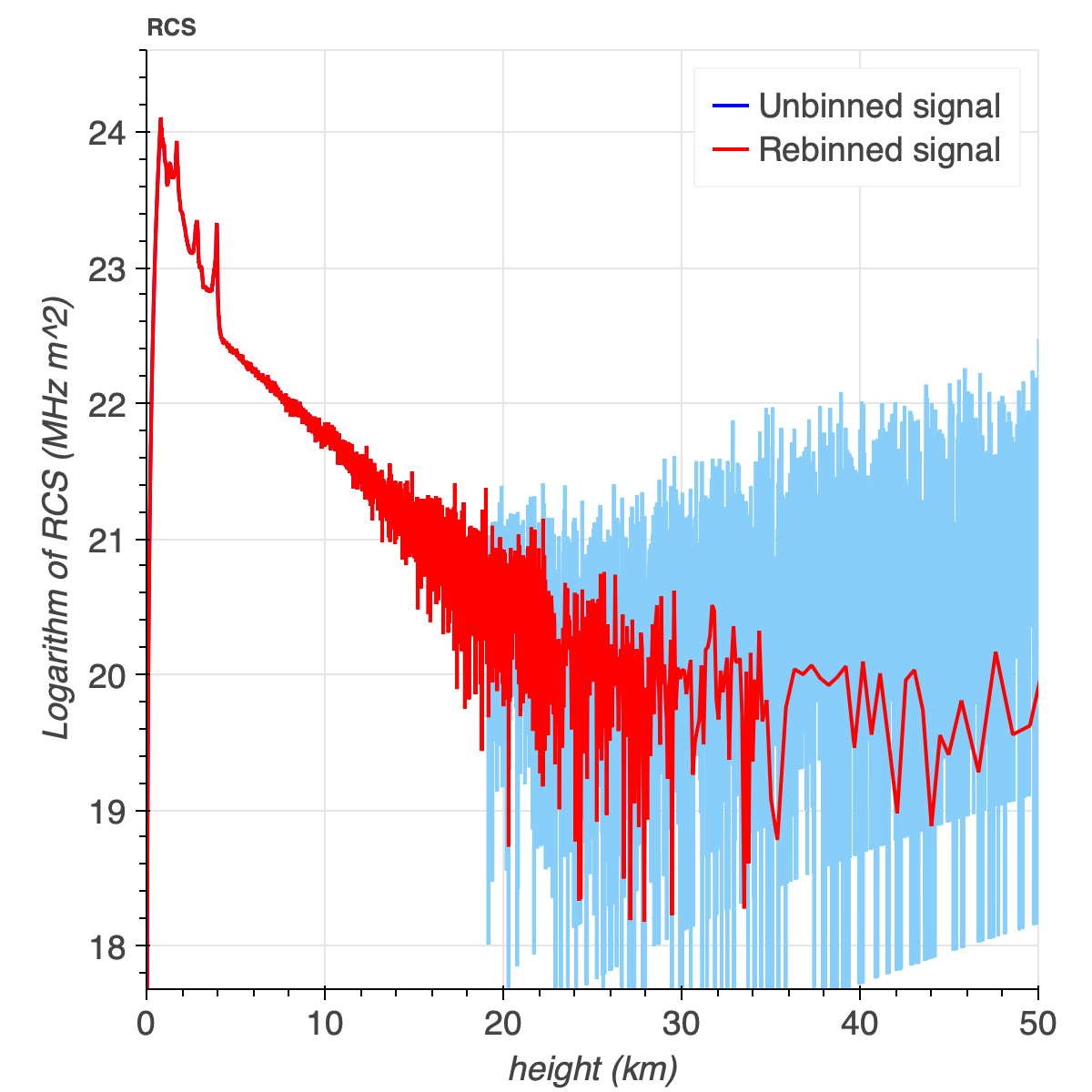}
\includegraphics[width=0.49\linewidth]{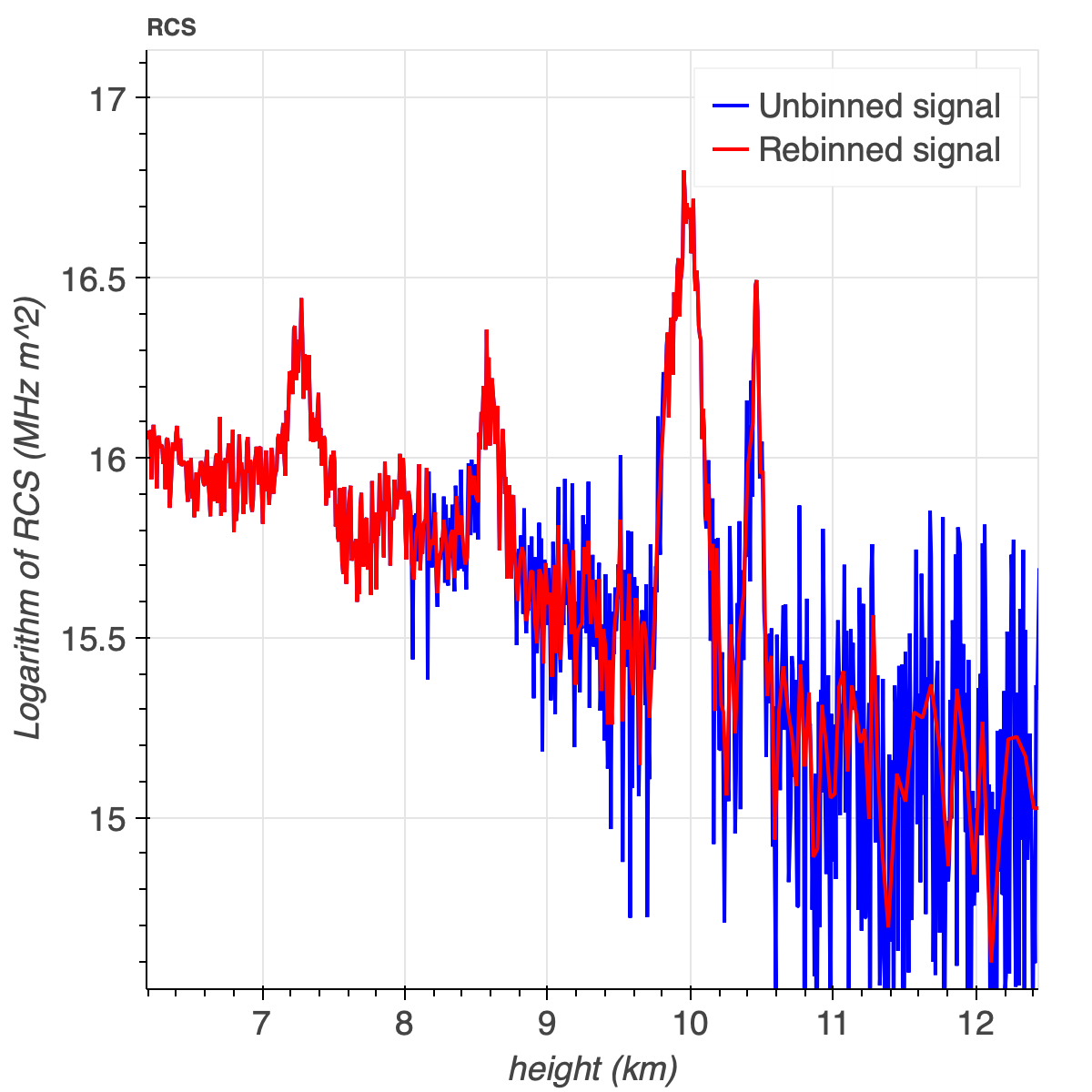}
\caption{
Two examples of a rebinned signal $\mathcal{S}$. On~the left side, the~range from $\sim$20~km to $\sim$35~km could be recovered for analysis, compared to the unbinned signal. On~the right side, the~details of a signal with a complicated structure of clouds are shown. 
\label{fig:rebinning}
}
\end{figure}

Figure~\ref{fig:rebinning} highlights several aspects of the performance of the presented rebinning {algorithm}. One can see that the rebinning {algorithm} correctly adjusts the binning windows so that the structures are kept with fine bins, whereas the regions in between and after the clouds are rebinned with wider~windows.

\subsection{Profile~Retrieval} \label{sec:profiles}

In this section, we describe the methodologies developed and used to retrieve the molecular profile expectation of the signal, the~determination of the end-of-ground-layer height, and~the cloud search {algorithm}. We present the inversion methods used to determine the extinction coefficients of the ground layer and~clouds.

\label{sec:layercalculations}
The \textit{Layer Calculations
} \gls{lpp} software module described here is specific to the elastic channels at 355~nm and 532~nm. 
These calculations are used to identify and characterize atmospheric layers, such as the ground layer, cloud layers, and~possibly the stratospheric Junge layer. Raman channels are then analyzed taking into account the height ranges of these layers. The~analysis is based on the algorithms presented in~\citet{Fruck:2022igg}, with~several modifications explained below, mainly to speed up the~computation.

The procedure starts with preparatory steps that include the retrieval of molecular profiles, the~computation of the system constant, and~the identification of free troposphere~ranges.

\medskip\noindent
\textbf{System constant.
} We precalculate a system constant $C_0$, which can be obtained either analytically, if~the system is sufficiently well characterized, or~through an absolute LIDAR calibration~\citep{Fruck:2022igg}:
\begin{align}
    C_0(\lambda) = \ln\left(\ddfrac{E(\lambda) \lambda A}{h f_s}\right) + \ln\left( \avg{\beta(h_\mathrm{LIDAR})}\right) 
     + \ln\left(\xi(\lambda)\right)   \label{eq:c0_lambda}
\end{align}
where $\beta(h_\mathrm{LIDAR})$ is the average molecular backscatter coefficient at the altitude of the site, and~the other parameters have been introduced in Equations~(\ref{eq:return_power_elastic}) and~(\ref{eq:RsigRbg}). 

\subsubsection{Molecular~Profile} 
\label{sec:molecular}

We retrieve the molecular density profiles from the publicly available European Centre for Medium-range Weather Forecasts (ECMWF) ERA5 reanalysis~\citep{web.cds}, provided with a horizontal resolution of $0.25^\circ\times 0.25^\circ$, on~37 pressure levels, a~temporal resolution of one~hour, with~a latency of five~days. 
Geopotential heights $\Phi$ are converted to altitudes above ground $Z$ using the prescription of~\citet{List:1951}:
\begin{equation}
Z = \ddfrac{R_\phi \Phi}{\left(g_\phi R_\phi/g_0 \right)- \Phi}~,
\end{equation}
where $R_\phi$ and $g_\phi$ are the Earth's radius and local gravity at latitude $\phi$, respectively. An~updated version of both can be obtained, for~example, from~the \gls{wgs84} reference ellipsoid~\citep{WGS:1984}.

The density profiles  $n(h)$, normalized to the US~standard density of air at Sea level $n_s$~\citep{uss}, are fitted using Equation~(12) of~\citet{Fruck:2022igg} and transformed into the equivalent molecular \gls{rcs} expectation (see Equation~(16) of \citet{Fruck:2022igg}):
\begin{equation}
F(h)  =  {}~  \ln\left(\frac{n(h)}{n_s}\right)  - \ddfrac{16\pi}{3\cos\theta}\,
\beta_0 \! \int_{h_\textrm{LIDAR}}^{h}\!\!\!\! \frac{n(h')}{n_s} \,\ud h' - \ln\left(\ddfrac{\avg{P(h_\mathrm{LIDAR})}}{\avg{T(h_\mathrm{LIDAR})}}\cdot \ddfrac{T_s}{P_s}\right) \quad, \label{eq:lnnr}  
\end{equation}
where $\theta$ is the LIDAR's zenith pointing angle ($90^\circ-\textit{elevation}$), $\beta_0$ the Rayleigh backscatter coefficient at standard Rayleigh backscatter coefficient at $n_0$~\citep{tomasi,Fruck:2022igg}, $h_\mathrm{LIDAR}$ the altitude of the LIDAR, $T_s$ and $P_s$ the \gls{us} standard atmosphere Sea level temperature and pressure~\citep{uss}, and~$\avg{P(h_\mathrm{LIDAR})}$ and $\avg{T(h_\mathrm{LIDAR})}$ their measured averages at the LIDAR~site.  

The so-called molecular fits are then used to fit a given part of the LIDAR \gls{rcs} to $F(h)$.

\subsubsection{Molecular Atmosphere~Ranges}

To identify those atmospheric ranges, in~which the LIDAR \gls{rcs} can be well described by a purely molecular \gls{rcs} expectation, we slide a window of user-defined size $l$ (by default, 500~m is chosen) through the \gls{rcs} and fit each interval $i$ to a constant $C_i$ plus $F(h_i)$. Because~fitting in \texttt{python} is, however, computationally inefficient, we 
pre-calculate the constants that minimize:
\begin{align}
\chi^2_i &= \sum_{j=i}^{i+l} (S_j - F(r_j\cdot \cos\theta) - C_i)^2 \cdot w_j ~; \quad 
 \mathrm{with}:  {}~ w_j = \ddfrac{1}{(\Delta S_j)^2}  ~, \label{eq:chi2j}
\end{align}
with the mean range $r_j$ evaluated at the centre of each bin $j$. 
Minimizing Equation~(\ref{eq:chi2j})  with respect to $C_i$ leads to the following solutions:
\begin{align}
C_i^\mathrm{min} &= \dfrac{\sum_{j=i}^{i+l} w_j (S_j-F(r_j\cdot \cos\theta))}{\sum_{j=i}^{i+l} w_j}\quad, \label{eq:c0s}\\
\left(\dfrac{\chi^2_i}{N_\mathrm{dof}}\right)^\mathrm{min} &= \dfrac{\sum_{j=i}^{i+l} (S_j - F(r_j\cdot \cos\theta) - C_i^\mathrm{min})^2 \cdot w_j}{l-1}
\label{eq:chismin} \\
s^2_{C_i} &= 2\cdot \left\vert\left.\dfrac{\partial \chi^2_i}{\partial C_i} \right\vert_{C_i = C_i^\mathrm{min}} \right\vert^{-1} 
= \dfrac{1}{\left\vert\sum_{j=i}^{i+l} w_j (S_j-F(r_j\cdot \cos\theta))\right\vert}
\end{align}

\subsubsection{Ground~Layer} 

Experience at the \gls{orm} has shown~\citep{Fruck:2022igg} that under conditions that allow science data taking of \glspl{iact}, the~ground layer reaches (1.5--2)~km into the troposphere, until~it reaches a region of free troposphere without aerosol contamination. Even under the worst circumstances, such a region can always be found~\citep{Fruck:2022igg}. 
Using precomputed molecular fit constants (Equation~(\ref{eq:c0s})) and their corresponding fit chi-squares (Equation~(\ref{eq:chismin})), the~{algorithm} determines the lower limit of the free troposphere. 
To do so, a~search range is predetermined, which starts at the LIDAR's range of full overlap and ends 10~km a.s.l., the~highest altitude found for a free troposphere above the \gls{orm}~\citep{Fruck:2022igg}.
The {algorithm} then iterates through that range and stops when: (a) a  $\left(\chi^2_i / N_\mathrm{dof}\right)^\mathrm{min}$ is found that falls below a predefined limit of 1.0 and (b) the corresponding fitted constant $C_i^\mathrm{min}$ minus its uncertainty $s_{C_i}$ is smaller than the assumed system constant $C_0$. The~latter condition serves to avoid the misinterpretation of well-mixed Saharan dust layers with its molecular prediction, with a slope similar to the \gls{rcs} but~a higher overall level. 

The {algorithm} performs a second iteration so that the lower limit of the free troposphere is not influenced by signal tails or low-quality fits from the ground layer. The~refinement process continues to increase the indices $i$ until $C_i - 1/4\, s_{C_i} > C_{i-1}$, that is, the~signal exhibits a consistently decreasing trend beyond small statistical fluctuations. With~this, the~reference constant $C_\textit{ft}$ for the free troposphere is~found. 

Using the identified ground layer top height $h_\textit{ft}$ (or the start of the free troposphere),  the~{algorithm} calculates aerosol extinction coefficients ($\alpha^\mathrm{aer}_i$) derived through the Klett--Fernald~\citep{klett1981,fernald1972,klett1985} inversion, in~non-logarithmic form, as~pointed out by~\citet{Young:1995}, see also~\citep{Speidel:2023}.
For the inversion, the~reference point is exchanged by its corresponding molecular fit constant $C_i$.

The ground layer \gls{vaod} is then computed as follows:
\begin{equation}
    \textit{VAOD} = \ddfrac{\left(C_0 - C_\textit{ft}\right) \cos\theta}{2}~. \label{eq:vaod_from_c0}
    \end{equation}

\begin{figure}[H]
\centering
\includegraphics[width=0.485\textwidth,trim={0 0 0 2cm},clip]{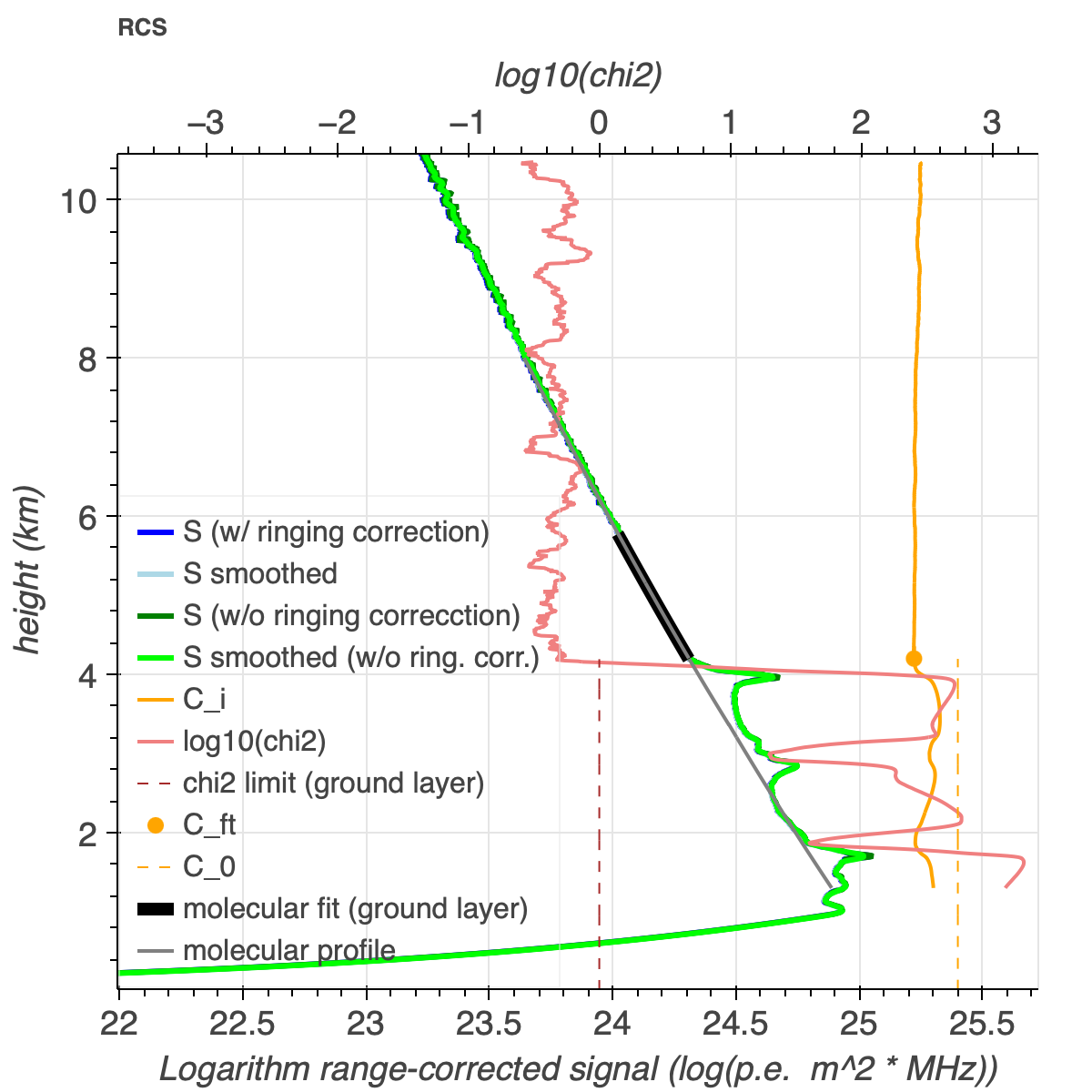}
\includegraphics[width=0.485\textwidth,trim={0 0 0 2cm},clip]{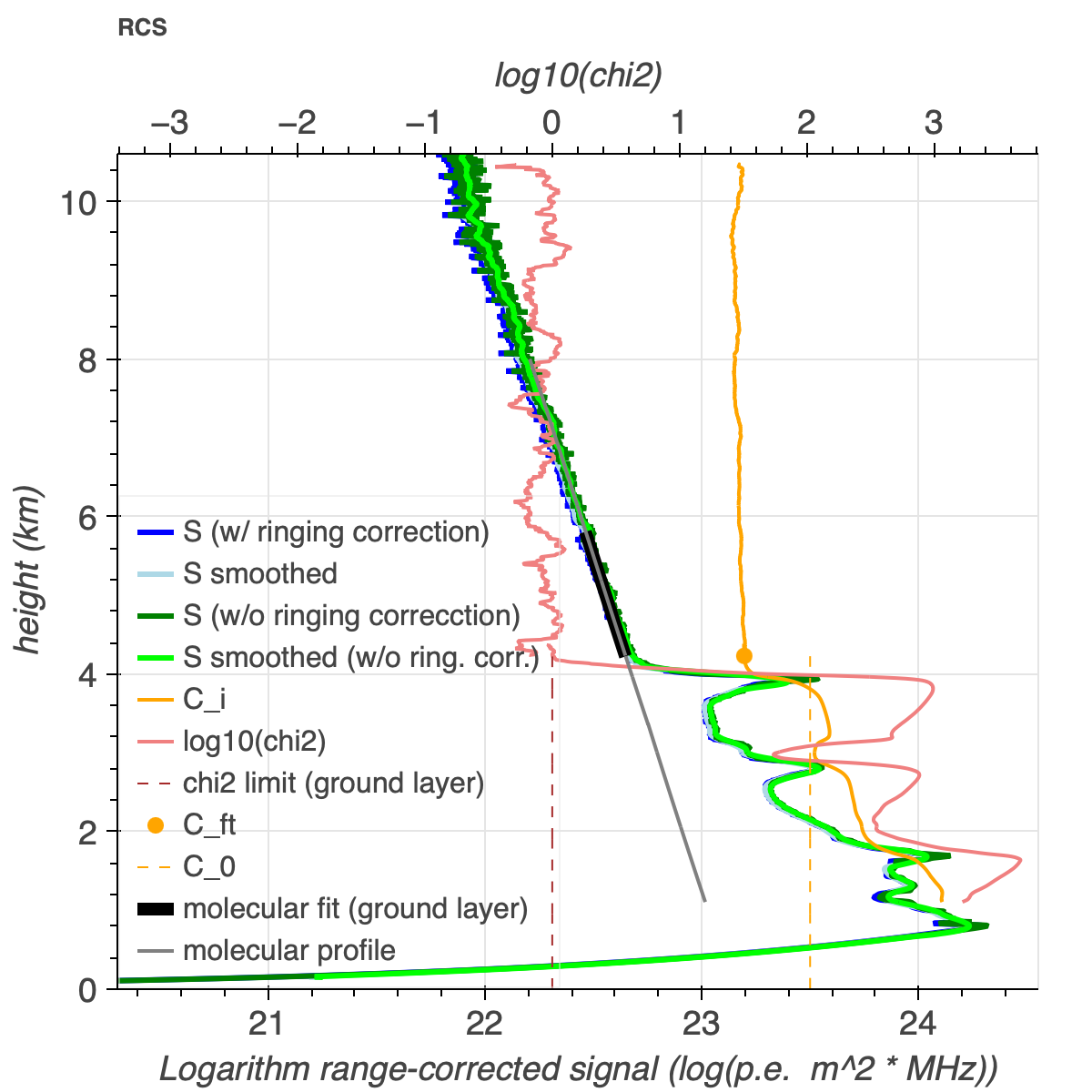}
\caption{Ground 
 Layer calculations for an elastic channel simultaneously observed at 355~nm (Left) and 532~nm (Right). The~blue and green lines show the {logarithms of the range-corrected signals $\mathcal{S}$}, with~and without ringing correction, respectively, and~with and without optional signal smoothing applied. The~purely molecular signal expectation {$F(h)$} is drawn as a black line, the~scaling factor of the signal, $C_i$ {(Equation~(\ref{eq:c0s}))}, to~fit the molecular expectation as a full orange line. The~reference system constant {$C_0$ (Equation~(\ref{eq:c0_lambda}))} is displayed as a dashed orange line. The~results of the molecular fit $\left(\chi_i^2/N_\mathrm{dof}\right)^\mathrm{min}$ (Equation~(\protect\ref{eq:chismin})) are shown in red, as~well as the predefined limit of 1.0 to highlight the free troposphere condition, with~its own scale on an additional horizontal axis on top of the image. The~dark black lines mark the beginning of the free troposphere $h_\textit{ft}$.
\label{f:groundLayer}}
\end{figure}

Figure~\ref{f:groundLayer} highlights several aspects of the performance of the ground-layer detection {algorithm} presented. {Here a ground layer extends from ground to $\sim$4~km height; the signal $\mathcal{S}$ above the ground layer fit perfectly the molecular signal prediction $F(h)$ for that night, location and the corresponding two wavelengths, visible in the fit $\left(\chi_i^2/N_\mathrm{dof}\right)^\mathrm{min}$ (red lines), which  fluctuates around values slightly lower than one (in the case of the 355~nm line) and around one (in the other case of the 532~nm line). All values of $\left(\chi_i^2/N_\mathrm{dof}\right)^\mathrm{min}$ obtained from the ground layer show considerably higher values (note the logarithmic axis). The~corresponding fit constants, $C_i$, appear flat {within 3\% peak-to-peak} in the free troposphere and larger throughout the entire ground layer. Extending the molecular to the ground leads to the reference constant $C_\textit{ft}$ of the free troposphere, which can be used to derive the full \gls{vaod} of the ground layer assuming Equation~(\ref{eq:vaod_from_c0}) and a system constant $C_0$, which needs to be constantly updated and calibrated according to the prescriptions of~\citet{Fruck:2022igg}. 
 } 
 
\subsubsection{Cloud~Layer} 

The \textit{Cloud Layer} {algorithm} identifies and characterizes clouds within the \gls{rcs} data. Clouds are identified by their strong backscatter signals and significant deviations from the molecular~profile. 

An initial search range is chosen from  $h_\textit{ft}$ to a maximum altitude of  $h_\textit{max} = 23$~km above ground, found as the maximum overall cloud top height from a long-term analysis of LIDAR data taken at the \gls{orm}~\citep{Fruck:2022igg}. Note 
 that low clouds lying within or just above the boundary layer without a
  range of free troposphere between, are analysed as part of the boundary layer
  in this scheme. This is, however, rather a semantic than physical difference,
  since the extinction profile will be derived correctly anyhow. Moreover, CTAO
  science observing conditions are normally not fulfilled when low cumulus
  clouds are present. For this reason, such conditions rarely need to be
  characerized as part of the CTAO science data analysis chain.
  
The~cloud analysis follows the following steps: 

\medskip\noindent
\textbf{Search for cloud's lower bound.} The {algorithm} scans the precomputed values $C_i^\mathrm{min}$  and $\chi_i^2/N_\mathrm{dof}$ (see Equations~(\ref{eq:c0s}) and~(\ref{eq:chismin})). 
 If the reduced molecular fit $\chi^2$ exceeds a predefined threshold of 3.5 and at the same time $C_i^\mathrm{min} > C_\textit{thres}$, the~{algorithm} considers this a potential cloud base. Both conditions require a significant positive deviation from the molecular backscatter profile, as~expected for clouds. In~the case of the (search for the) lowest cloud layer,  $C_\textit{thres}$ is chosen to be $C_\textit{ft}$; otherwise, a new reference constant needs to be taken for the free troposphere from above the last cloud layer below.  
    The lower bound of the cloud is then refined by moving back downwards again until the reduced molecular fit $\chi^2$ falls (again) below 1.5 and $C_i^\mathrm{min} < C_\textit{thres} + 1.5 \cdot s_{C_i}$. 
    This part ensures that the signal transitions smoothly to the free troposphere. The~previous step marks the cloud base height $h_\textit{base}$, and~with it a reference $C_\textit{base}$.

\medskip\noindent
\textbf{Identification of cloud's top.} To find the height of the cloud top, the~{algorithm} continues to scan the reduced molecular fit $\chi^2$ upwards and stops when it falls again below 2.2 and $C_i^\mathrm{min} < C_\textit{thres} + 1.5\cdot s_{C_i}$, ensuring that the cloud candidate absorbs light w.r.t. the molecular atmosphere part. From~that point on, the~{algorithm} moves further upwards as long as the $C_i^\mathrm{min}$'s decrease with respect to their immediate predecessor  $C_{i-1}^\mathrm{min}$. This part ensures that possible exponential drops of cloud density on their upper edges are correctly attributed to the cloud and not the free troposphere. The~end of that step results in a 
cloud top height $h_\textit{top}$, and~with it a reference $C_\textit{top}$.
For a detected cloud, the~\gls{vod} is calculated~\citep{benzvi2006}:
\begin{equation}
    \textit{VOD}_{\text{cloud}} = \ddfrac{\left(C_\textit{top} - C_\textit{base}\right)\cos\theta}{2} \label{eq:vod_cloud}
\end{equation}

\medskip\noindent
\textbf{Handling misidentified clouds.} At this point, the~{algorithm} carries out several checks to detect and discard false positives: (i
)
        Clouds with $\textit{\gls{vod}}_\mathrm{cloud} < 10^{-4}$ discarded. 
        (ii) Cloud with $\textit{\gls{vod}}_\mathrm{cloud} < 10^{-2}$ and geometric thickness $(h_\textit{top}-h_\textit{base})<100$~m are considered statistical fluctuations and discarded. 
        (iii) Frequent temperature fluctuations and inversion in the tropopause~\citep{Rodriguez:2013} are excluded by requiring that any cloud with $h_\textit{top}>12$~km above ground (14.2~km a.s.l.) shows a geometric thickness larger than 4~km and $\textit{\gls{vod}}_\mathrm{cloud} > 0.015$.
   
After these tests have been passed, the~cloud is added to the list of detected clouds, and its extinction profile and LIDAR ratio are calculated. 
 
\medskip\noindent
\textbf{Extinction and LIDAR ratio calculation.}
 As in the case of the ground layer inversion, the~reference point $S_i$ at $h_\textit{top}$ is exchanged by $C_\textit{top}$, and~the Klett--Fernald {algorithm}~\citep{klett1981,fernald1972,klett1985} used for the inversion.  The~inversion is performed iteratively refining a global cloud LIDAR Ratio, until~the integrated extinction profile matches $\textit{\gls{vod}}_\mathrm{cloud}$. 
    The LIDAR Ratio is, nevertheless, constrained to between limits of 5~Sr and 120~Sr. If~the ratio fails to converge within a maximum number of iterations, the~{algorithm} uses the nearest limit and rescales the extinction~coefficients.

\begin{figure}[H]
\centering
\includegraphics[width=0.49\textwidth,trim={0 0 0 2cm},clip]{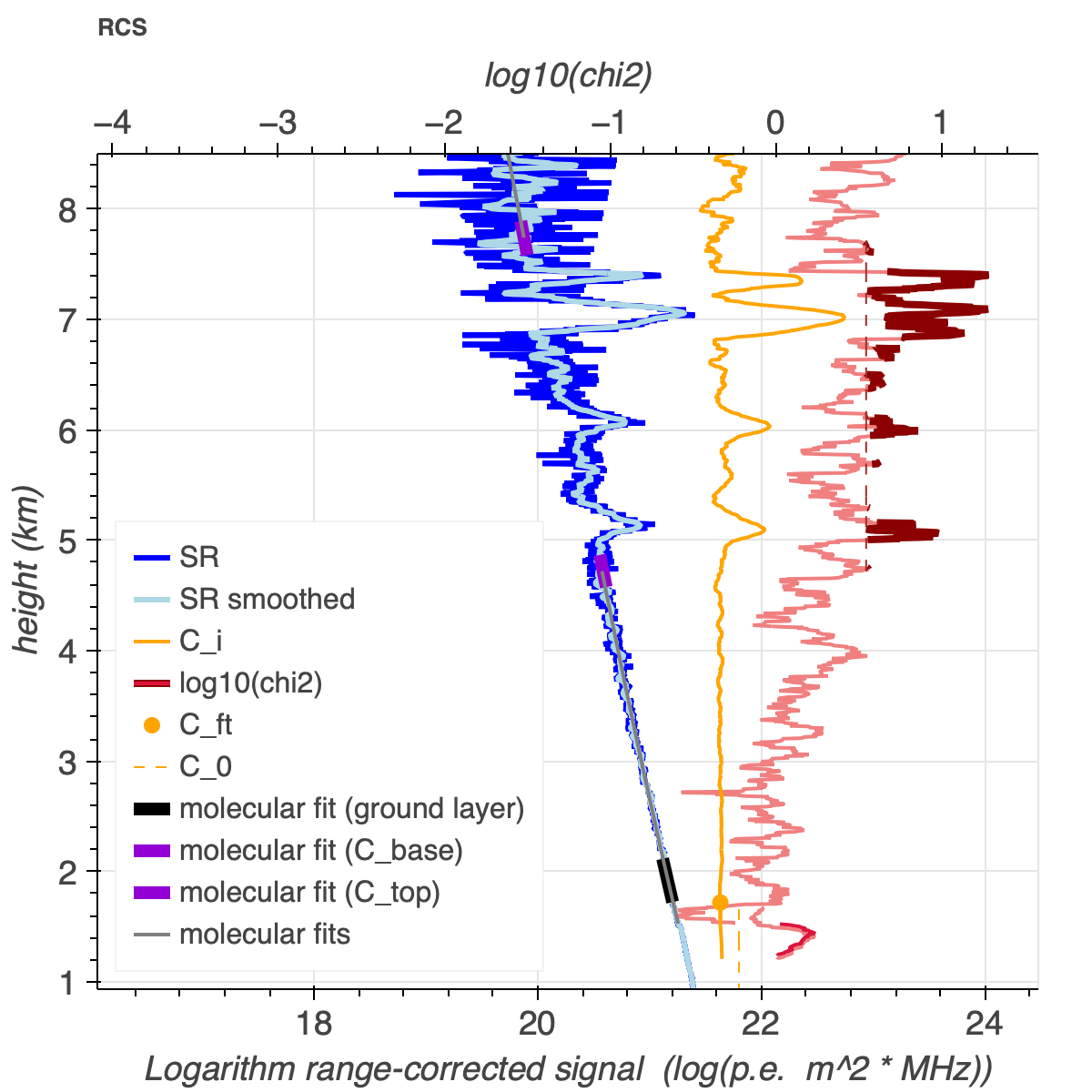}
\includegraphics[width=0.49\textwidth,trim={0 0 0 2cm},clip]{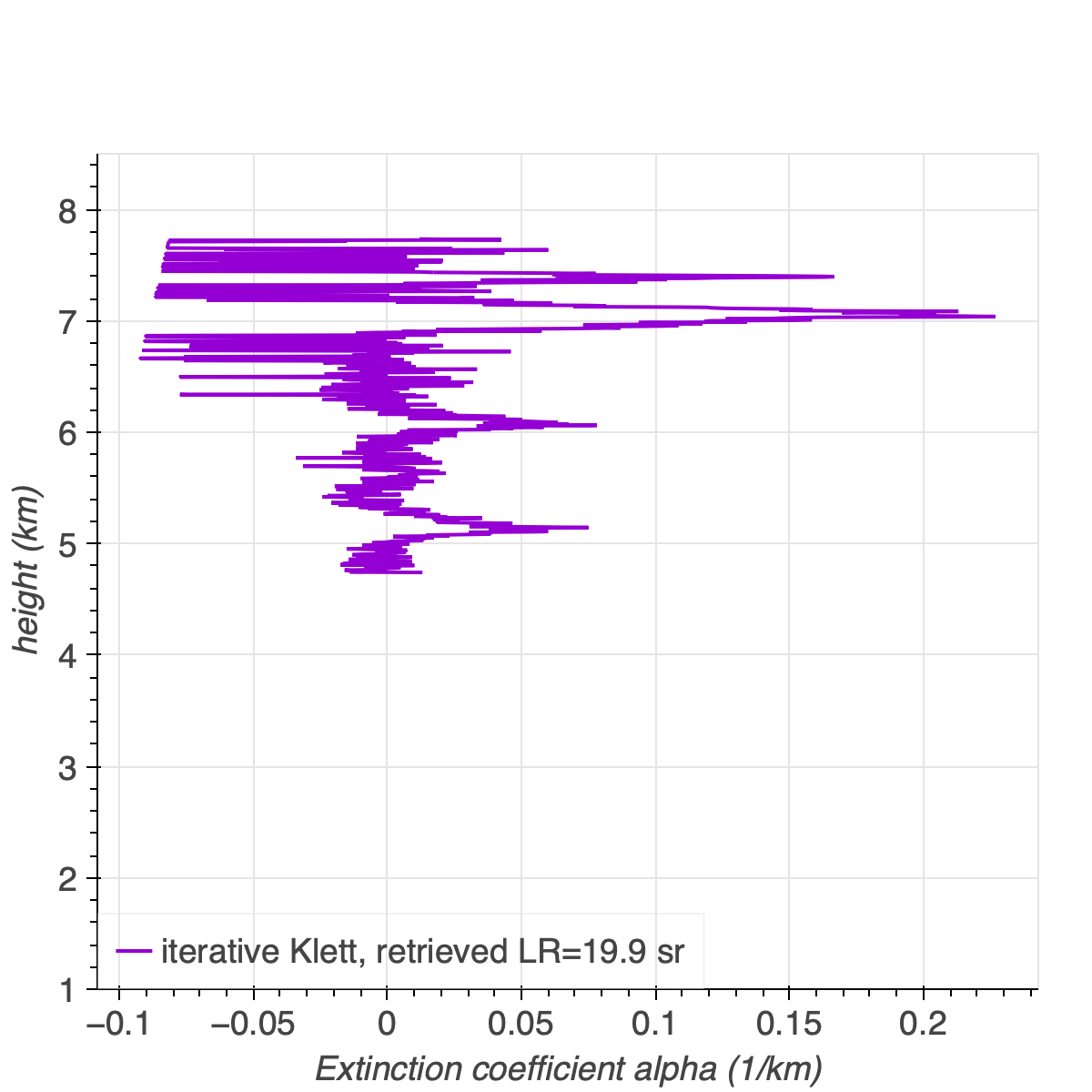}
\caption{Layer 
 Calculations for an elastic channel (355~nm), taken with reduced \gls{pmt} gain during a clear night at La Palma. From a height of about 5~km to 7.5~km, a~cloud is visible with a complicated sub-structure. The~cloud layer analysis has detected it based on the analysis of deviations from the fit to the molecular signal expectation (Equation~(\protect\ref{eq:chismin})), shown in red. The~dark red parts mark those ranges, where $\left(\chi_i^2/N_\mathrm{dof}\right)^\mathrm{min}$ exceeds the threshold of 3.5. The~yellow line marks the fitted values of $C_i$ (Equation~(\protect\ref{eq:c0s})). On the right side, the~extinction coefficients retrieved from an iterative Klett inversion (violet line) are shown, where a typical LIDAR Ratio of 20~sr has been obtained. The~orange line shows the Klett inversion of the ground~layer. }
\label{f:lc-1}
\end{figure}

In Figure~\ref{f:lc-1}, we show the result of a cloud layer identification and inversion on the elastic 355~nm line, obtained during a clear night at the \gls{orm}. {Typical for a clear astronomical night, the~ground layer is hardly detectable here, particularly in the \gls{uv} part of the light spectrum in which the Cherenkov light is detected by the \gls{ctao} telescopes. From~about 5~km above ground extending to $\sim$7.5~km, a~typical structured cirrus cloud has entered the LIDAR's field-of-view. The~cloud layer {algorithm} has correctly identified the existence of a cloud layer identifying those fit $\left(\chi_i^2/N_\mathrm{dof}\right)^\mathrm{min}$ (Equation~(\protect\ref{eq:chismin})), which exceed the threshold of 3.5 (marked as the dark red parts of the $\log10(\left(\chi_i^2/N_\mathrm{dof}\right)^\mathrm{min})$ line) and found the part of the free troposphere just below the cloud (lower dark violet line marked as $C_\mathrm{base}$). Together with the  equivalent part of the free troposphere above the cloud, marked as $C_\mathrm{base}$, the~\gls{vod} could be determined using Equation~(\ref{eq:vod_cloud}). The~iterative Klett inversion converged on a Lidar ratio of 19.9~sr, a~typical value for the cirrus clouds observed above the observatory (see~\citep{Fruck:2022igg}) and the extinction coefficient profile shown on the right side.} 

\subsubsection{Raman~Lines} 

The Raman lines are analyzed in the ranges where the layer calculation {algorithm} has found the ground layer and possible~clouds. 

Above the region of full overlap, the~aerosol extinction profile of a Raman line can be directly inverted~\citep{Ansmann:1992}:
\begin{equation}\label{eq:ext}
    \alpha_\mathrm{aer}(\lambda_0, h_i) = \ddfrac{\ddfrac{\mathrm{d}}{\mathrm{d}R_i}\Big[\ln\left(n(h_i)\right)-\mathcal{S}_i\Big]-\left(\alpha_\mathrm{mol}(\lambda_0, h_i) + \alpha_\mathrm{mol}(\lambda_R, h_i)\right)}{1+ {\left(\ddfrac{\lambda_0}{\lambda_R}\right)}^k} ~,
\end{equation}
where $n(h_i)$ is the molecular density profile (see Section~\ref{sec:molecular}) evaluated at height $h_i = \left(R_i\cos\theta + h_\mathrm{LIDAR}\right)$ a.s.l, $k$ is the \angstrom exponent of the extinction coefficients from the elastic wavelength $\lambda_0$ and its Raman counterpart $\lambda_R$. However, it is well known that the derivative of the logarithm of a strongly falling function amplifies statistical fluctuations, and~some smoothing or filtering of the signal is needed~\citep{Iarlori:2015}. 
We used the Savitzy-Golay filter, (implemented in \texttt{scipy}'s library \texttt{savol\_filter}), both for filtering of the logarithm of \gls{rcs}, as~its derivative. 

\begin{figure}[H]
\centering
\includegraphics[width=0.485\textwidth]{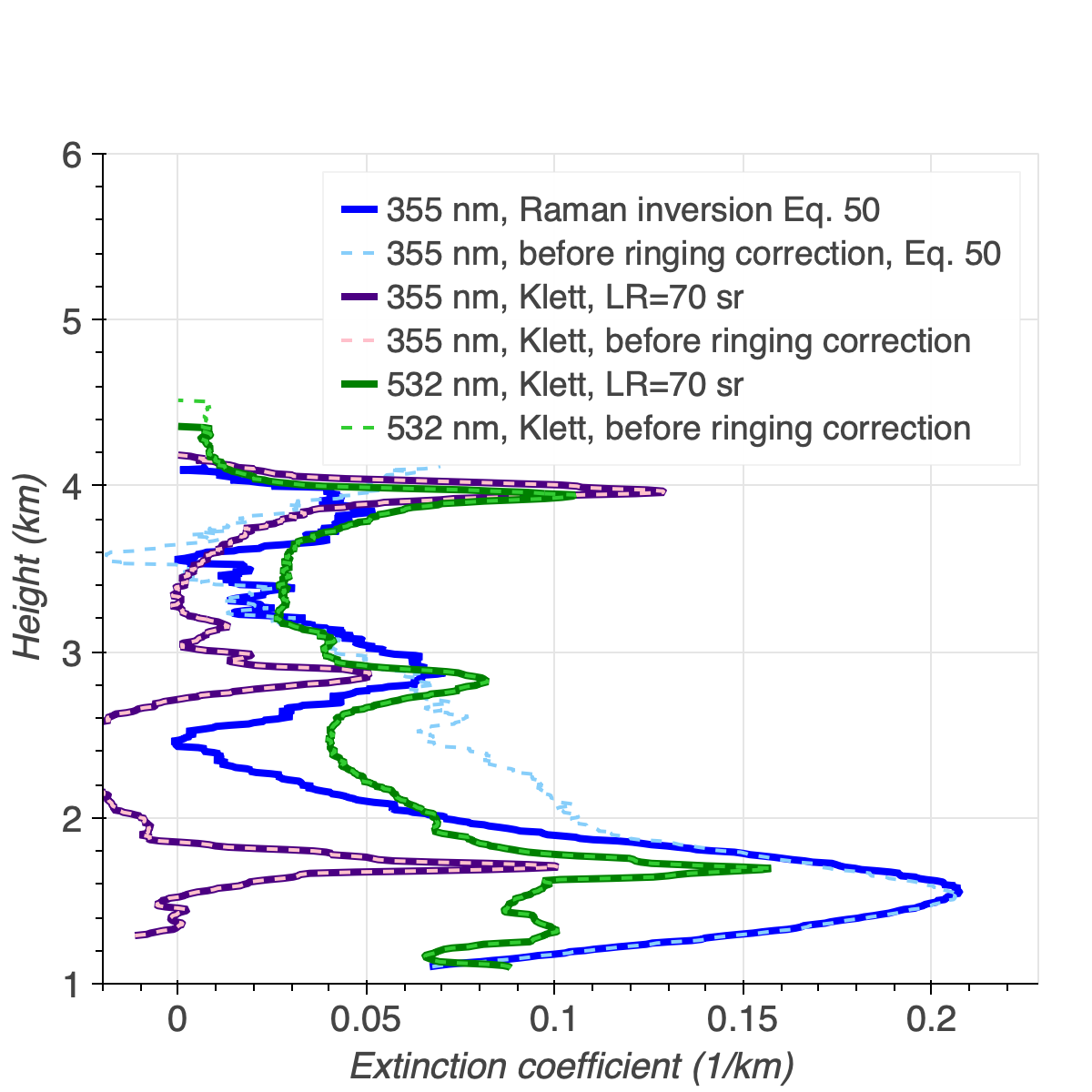}
\includegraphics[width=0.485\textwidth]{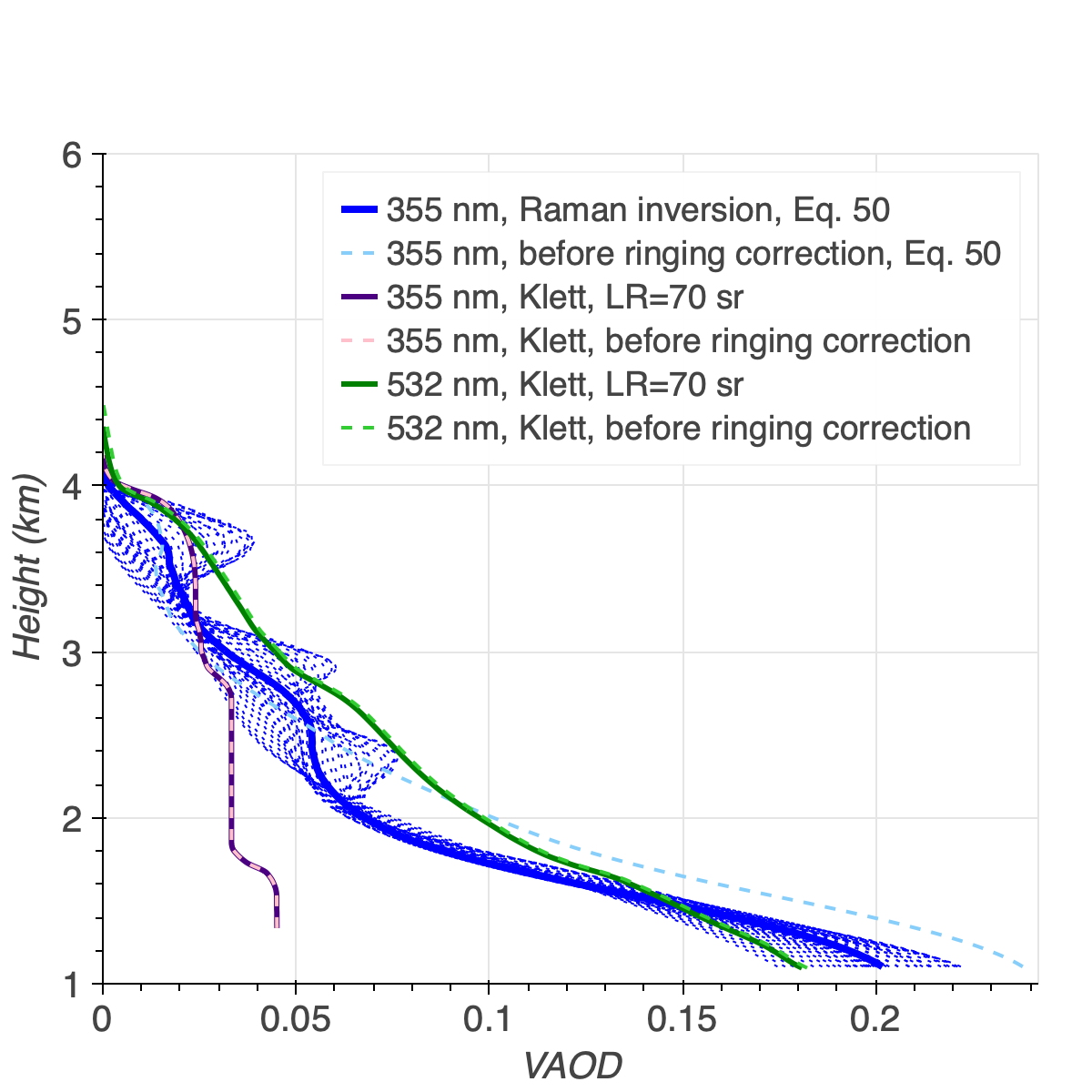}
\caption{Left: A comparison of ground-layer extinction profiles retrieved from the Raman line at 387~nm (using Equation~(\protect\ref{eq:ext}) with $k=1$) (blue) and a Klett--Fernald inversion of the elastic lines (355~nm, violet and 532~nm, green). All three lines are affected by \gls{lotr} ringing, with~the 355~nm line most strongly affected. A~filter window of 1~km has been applied to the Raman inversion, resulting in worse height resolution than the (unfiltered) inverted elastic lines. Right: the same profiles have been integrated from $h_\textit{ft}$ downwards. The~dotted blue lines show each an inversion from the Raman line filtered with a different window in the range from half to twice the default window size and two different filter polynomials.
\label{fig:alphas}}
\end{figure}

Figure~\ref{fig:alphas} shows the different algorithms used for the retrieval of the extinction profile {and the corresponding \glspl{vaod}}, highlighting the effect of \gls{lotr} ringing. One can observe that when the signals are small (here in the {Raman} line), the~relative effects of ringing and their corrections become important: whereas the relatively large signals in the elastic lines have been retrieved correctly with or without ringing correction, only the Raman line with ringing correction yields a correct inversion product. In~the case of the elastic UV line, both approaches yield insufficient (with correction) or even negative extinction coefficients. The~accuracy requirement of $\lesssim$0.03~\gls{rmsd} in the \gls{pbl} \gls{vaod} retrieval is not yet met in this particular case. 
{That inaccuracy can be cured with the use of an absolute LIDAR calibration~\citep{Fruck:2022igg}, which needs quasi-continuous data and was not applied to this data set for that reason.} 

To construct the backscatter profile $\beta_\mathrm{aer}(h)$ from a Raman and an elastic line, we follow the approach of~\citet{Ansmann:1992}, but~allow for the use of arbitrary elastic lines of wavelength $\lambda_e$ in combination with the Raman signal at $\lambda_R$ and its corresponding laser wavelength, $\lambda_0$:
\begin{align}
\label{eq:back}
\beta_\mathrm{aer}(\lambda_e, h) &= \beta^\mathrm{mol}(\lambda_e,h) \cdot \left( 
  \ddfrac{\mathcal{R}_{\lambda_e}(h) \cdot \mathcal{R}_{\lambda_R}(h_0)}{\mathcal{R}_{\lambda_e}(h_0) \cdot \mathcal{R}_{\lambda_R}(h)} \cdot \exp(T/\cos\theta) 
  -1 \right) \nonumber\\
 \mathrm{with:~} T &= \int_h^{h_0}
 \left(
   \alpha^\mathrm{mol}(\lambda_R,h^\prime)
  +\alpha^\mathrm{mol}(\lambda_0,h^\prime) 
  -2\alpha^\mathrm{mol}(\lambda_e,h^\prime)
  \right) ~
  \mathrm{d}h^\prime + \nonumber\\
  {} & {} \qquad + 
  \left(1 + 
   \left(\ddfrac{\lambda_R}{\lambda_0}\right)^k
  -2 \left(\ddfrac{\lambda_R}{\lambda_e}\right)^k
  \right) \cdot \int_h^{h_0}
   \alpha^\mathrm{aer}(\lambda_R,h^\prime) 
  ~
  \mathrm{d}h^\prime \quad. 
\end{align}
Here, $h_0$ must correspond to an altitude in the free troposphere, where no particulates are~present.

\begin{figure}[H]
\centering
\includegraphics[width=0.485\textwidth]{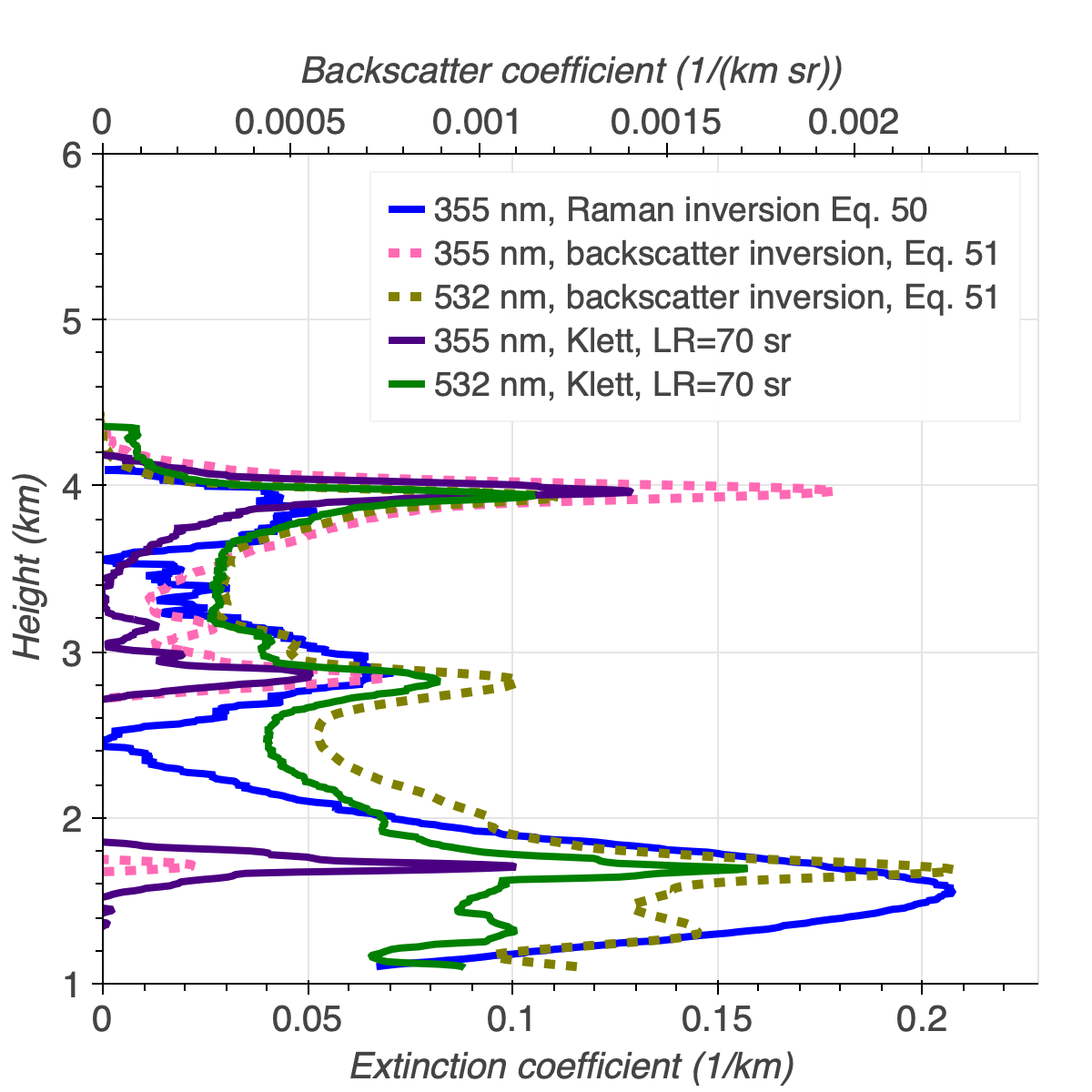}
\includegraphics[width=0.485\textwidth]{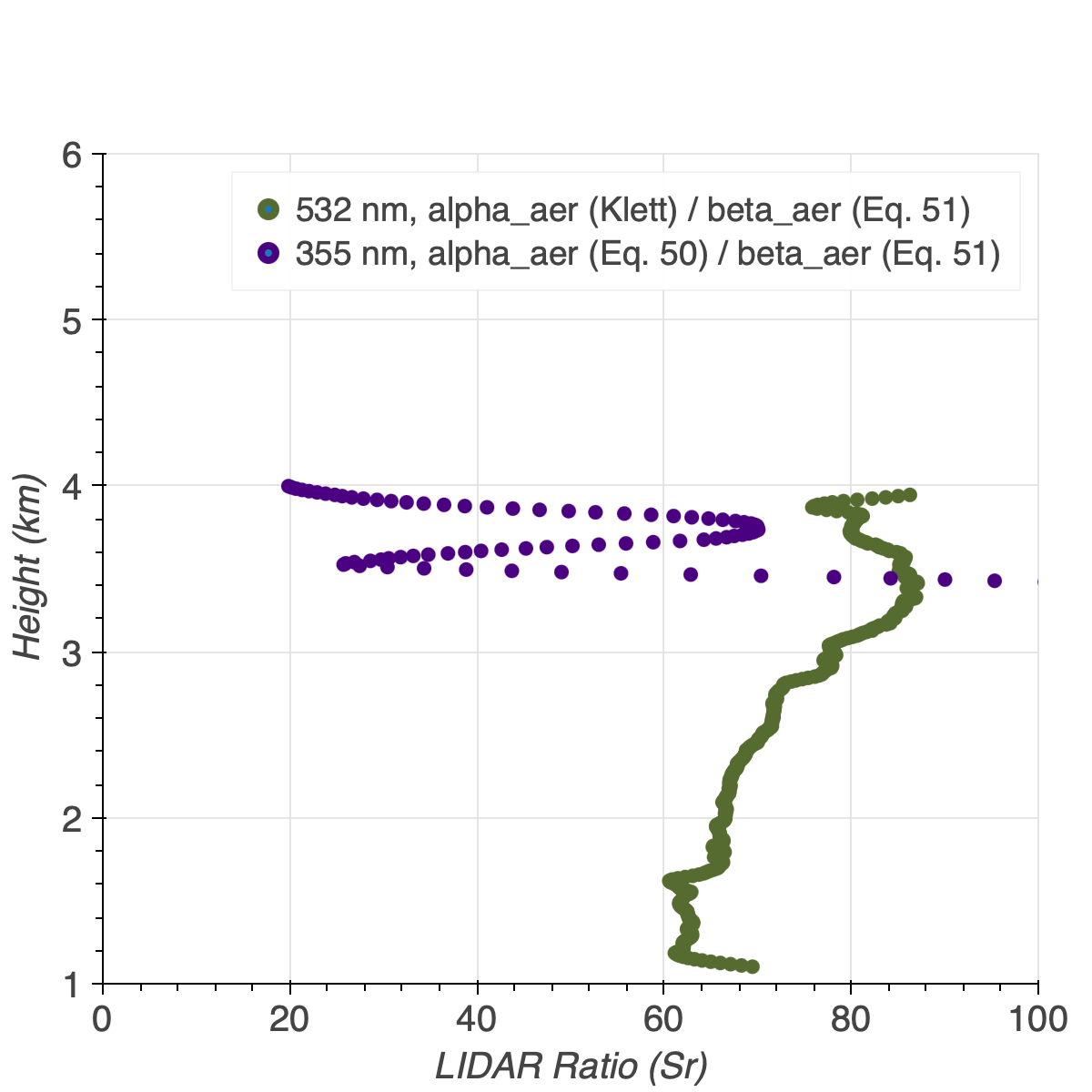}
\caption{Left: A comparison of ground-layer extinction profiles retrieved from the Raman line at 387~nm (using Equation~(\protect\ref{eq:ext}) with $k=1$) (blue) and the Klett--Fernald inversion of the elastic lines (355~nm, violet and 532~nm, green). Backscatter profiles retrieved with Equation~(\protect\ref{eq:back}) are shown for $\lambda_e$ = 355~nm (dashed, pink) and $\lambda_e$ = 532~nm (dashed, olive green). 
All three lines are affected by \gls{lotr} ringing, with~the 355~nm line most strongly affected. A~filter window of 1~km has been applied to the Raman inversion, resulting in worse height resolution than the (unfiltered) inverted elastic lines. The~dim dashed lines show a similar filter applied to the backscatter profiles. Right: the corresponding LIDAR Ratios.
\label{fig:betas}}
\end{figure}

{Figure~\ref{fig:betas} compares the the different algorithms used for retrieval of the extinction and the backscatter profiles for a slightly polluted urban ground layer, highlighting the possibilities and limits of the current \gls{pbrl}: whereas the combination of Raman and backscatter inversions (Equations~(\ref{eq:ext}) and~(\ref{eq:back})) provide consistent results for the (relatively stronger) aerosol signal at 532~nm, the~inversion of the coefficients at 355~nm from a strongly overdriven \gls{lotr} could only be achieved for the highest of the three layers, and~only there with degraded range resolution of the Lidar ratio. These shortcomings can be cured with a more stable gated PMT, particularly for the ringing-sensititive elastic UV line. }

\section{Results}
\label{sec:results}

\begin{figure}[H]
\centering
\includegraphics[width=0.485\textwidth]{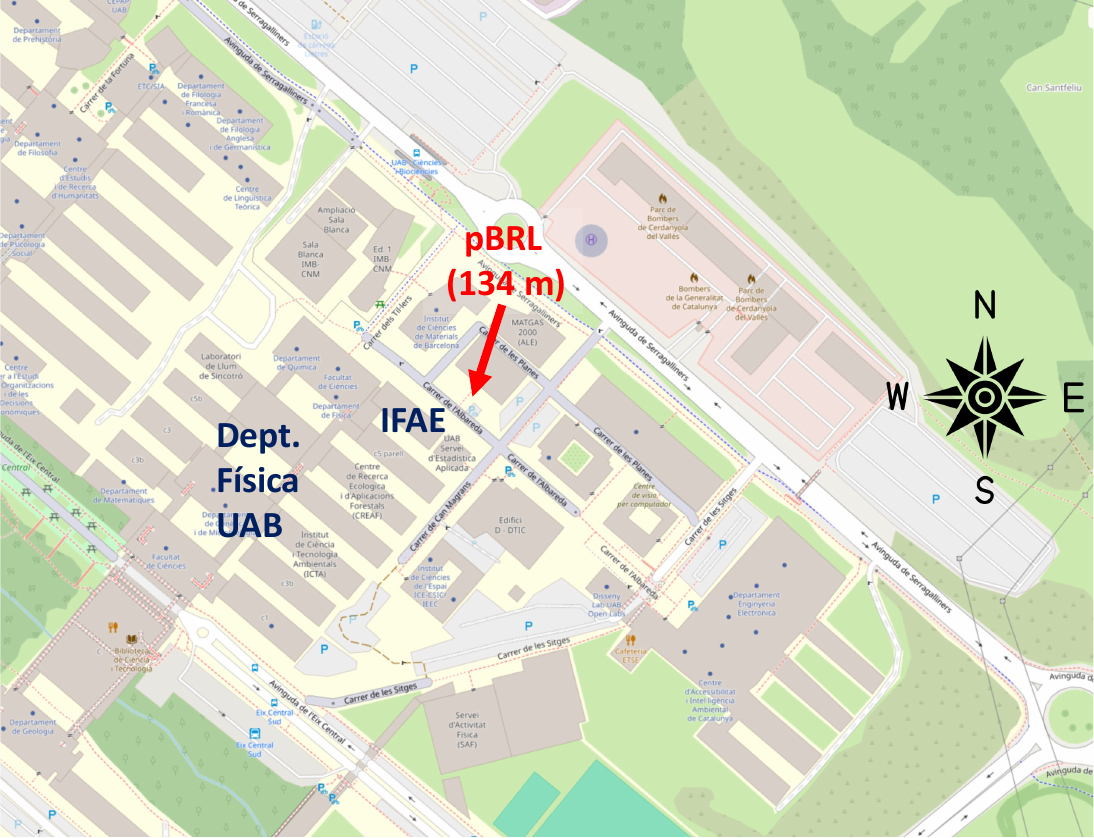}
\includegraphics[width=0.485\textwidth]{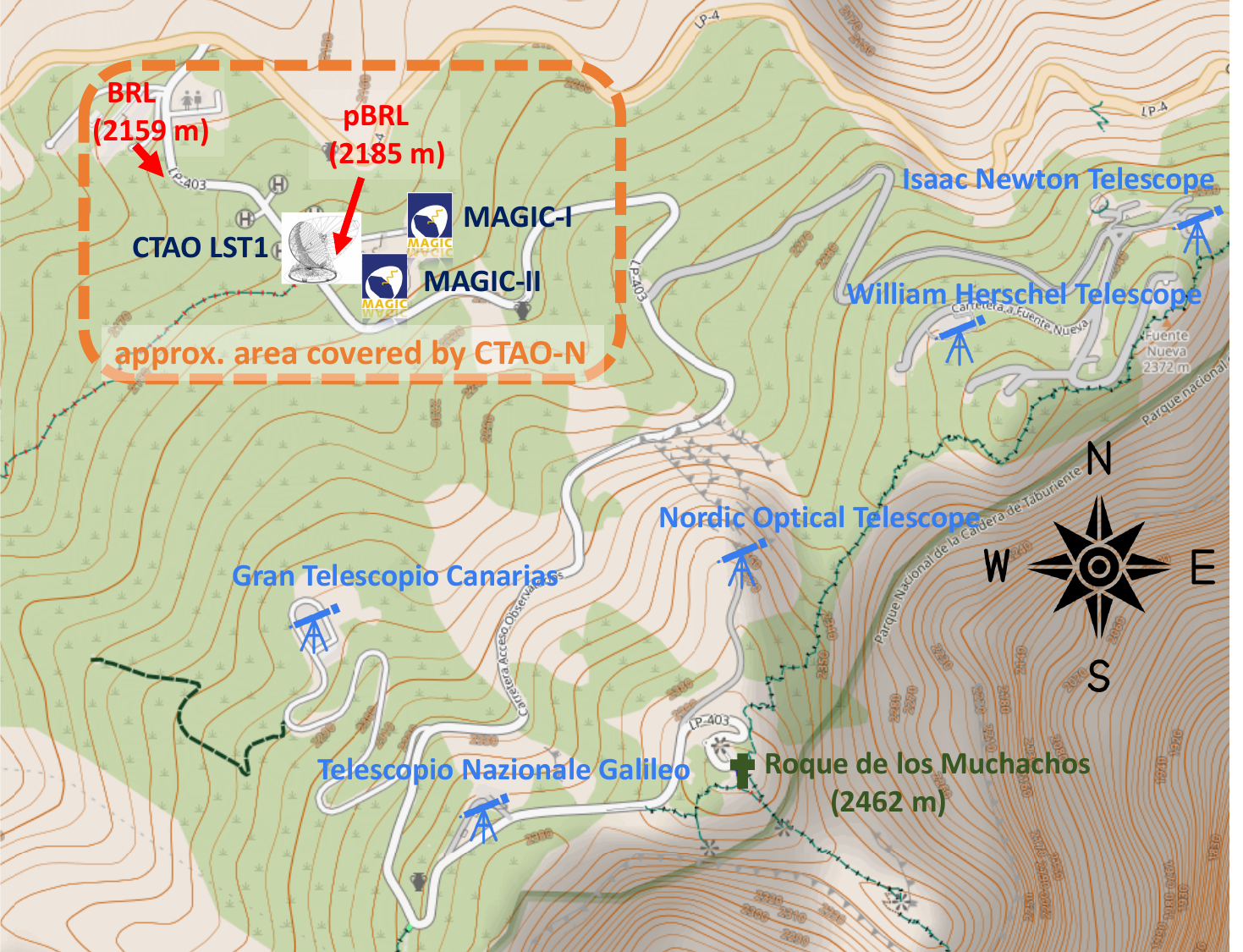}
\caption{\label{fig:locations} {Location 
 of the \gls{pbrl} test sites for the validation campaigns. Left: The \gls{uab} campus (41.5~N, 2.1~E, 134~m a.s.l.). Right: \gls{orm} (28.8~N, 17.9~W,  2185~m a.s.l.). The~final location of the \gls{brl}, as~well as the location of several nearby optical telescopes and the approximate area covered by the future \gls{ctao-n} are also shown in the figure.} }
\end{figure}

In this section, we present results on \gls{pbrl} performance, obtained with field data from commissioning measurements at the \gls{uab} Campus and the pathfinder campaign at the \gls{orm}, from~2021 to 2023, {see Figure~\ref{fig:locations} for both locations}. We discuss different sky conditions, from~clean to hazy, including a night with the presence of volcanic ashes, which furthers scientific interest in our data. Based on the signal preparation presented in Sections~\ref{sec:evaluation_signal} and~\ref{sec:profiles}, we have developed high-level analysis routines to calculate and visualize relevant atmospheric~parameters.

Unfortunately, during~the \gls{orm} campaign, the~\gls{pbrl} operations were limited to astronomical twilight periods on full moon nights. 
This significantly reduced the amount of useful data to a few minutes per night, for~a total of $\sim$10~h of observation in $\sim$30~nights. Furthermore, because~the \gls{pbrl} could not be included in \gls{orm}'s Laser Traffic Control System (\gls{ltcs})~\citep{Summers:ltcs,GaugDoro:2018}, 
we operated the \gls{pbrl} at about 10\% of its laser power. Finally, we also initially operated the elastic line \glspl{pmt} intentionally with reduced high voltages (and therefore gain) to protect the \glspl{lotr} from too high input voltages given the large mirror area and a recent re-aluminization~\citep{technicalpaper}. During~this period, \gls{pc} was therefore not possible for the elastic channels. 
Globally, these limitations resulted in reduced performance, particularly in the far range, during~the initial commissioning period of the \gls{pbrl} at the \gls{orm}.

\subsection{Data~Sets}
\begin{figure}[H]
\includegraphics[width=0.95\textwidth]{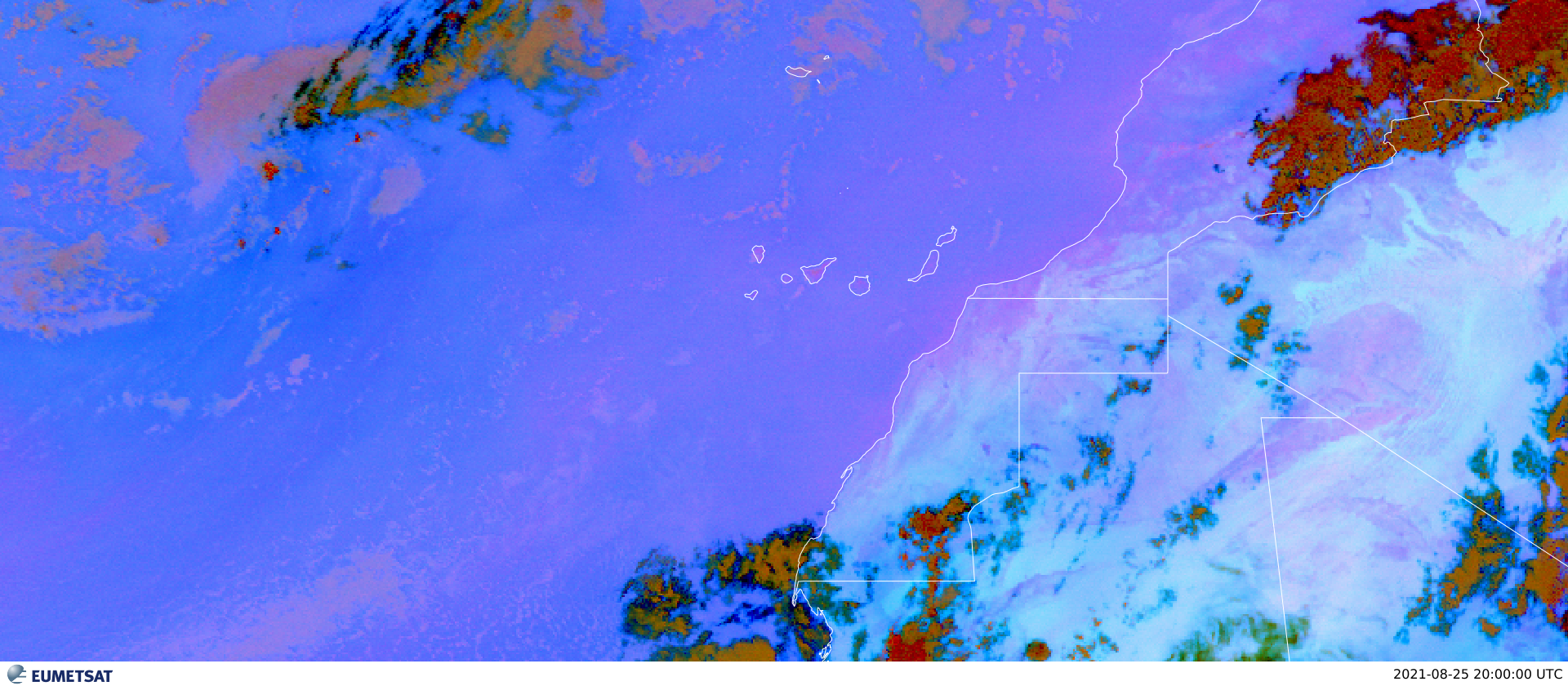}
\caption{Satellite image (\textit{Dust \gls{rgb} - \gls{msg} - 0 degree}) of mineral dust over Western Africa and the Atlantic Ocean, including the Canary Islands. The~island of La Palma is located in most north-western direction of a series of seven islands found in the centre of the image. Dust is colored in pink, clear sky in blue, and~clouds in dark red. The~image was taken on 25 August 2021 at 20:00 UTC, simultaneous to our LIDAR measurements. \textcopyright EUMETSAT~2021.}
\label{fig:eumetsat-calima}
\end{figure}

\begin{figure}[H]
  \includegraphics[width=0.45\linewidth]{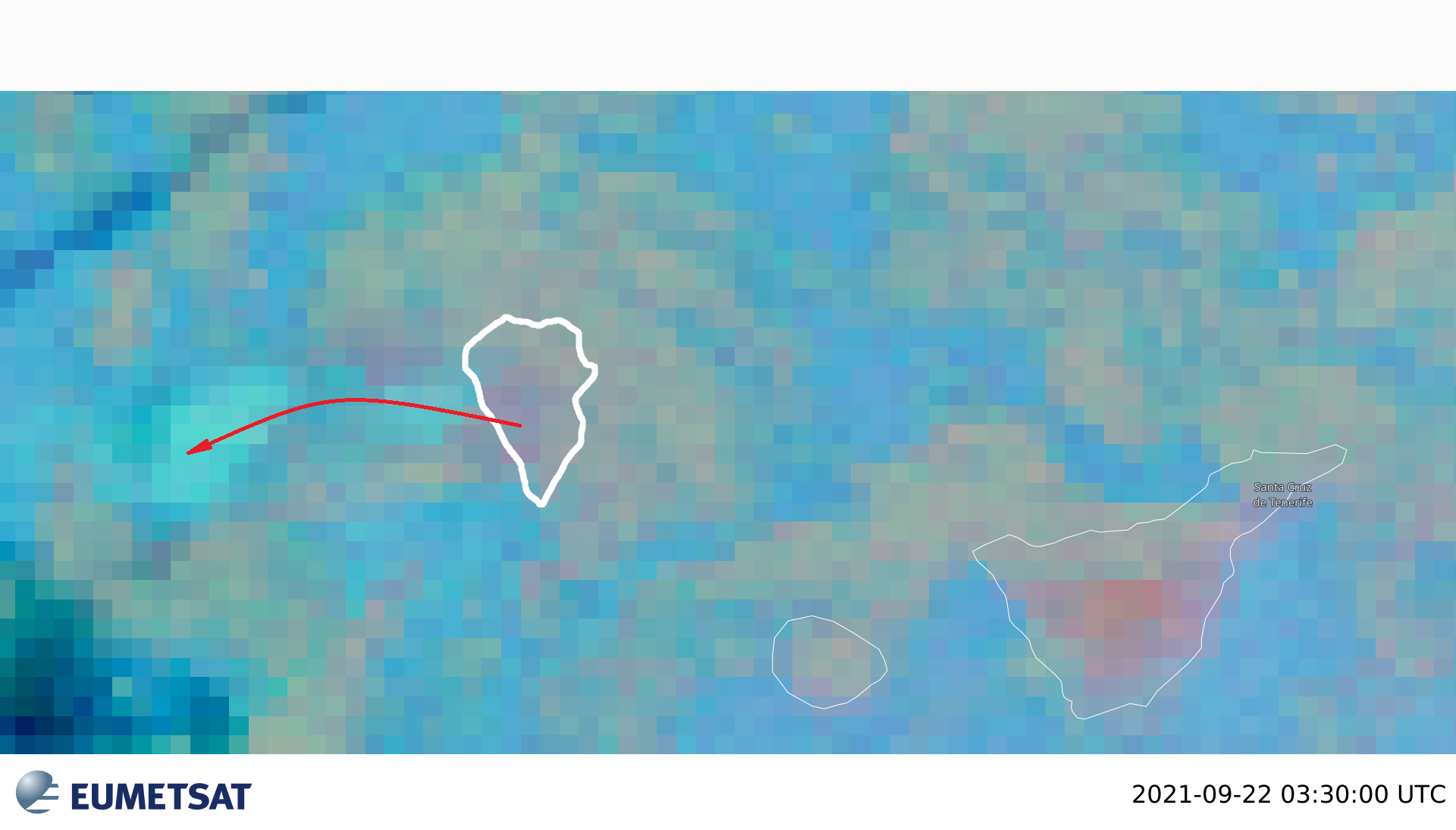}
  \includegraphics[width=0.45\linewidth]{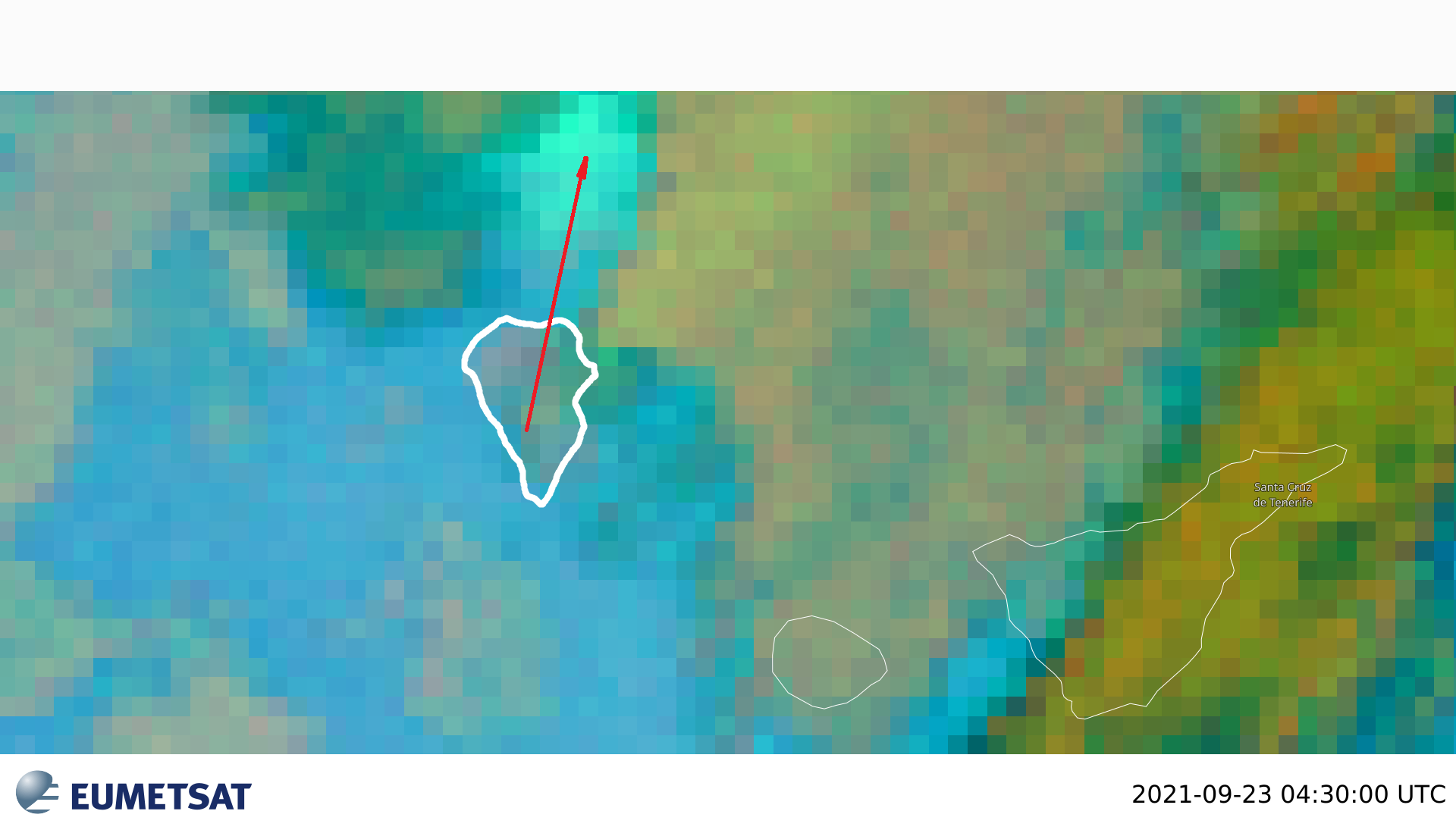}
\caption{Satellite image (\textit{Volcanic Ash \gls{rgb}--\gls{msg}--0 degree}) of the volcanic plume (cyan, corresponding to enhanced SO$_2$ concentration) above La Palma. The~left image was taken on 22 September 2021, at~03:00 UTC; our LIDAR measurement was performed at 19:30 UTC, and~the right image was taken after the measurement, on~23 September 2021, at~04:30 UTC. The~wind direction (on the images indicated with a red arrow) changed in a clockwise manner, which blew the plume above the observatory. \textcopyright EUMETSAT~2021.}
\label{fig:volcano_satellite}
\end{figure}

We show results for four data sets that encompass a typical sky at the \gls{uab} campus, and~different peculiar atmospheric conditions at the \gls{orm}. We characterize the \gls{pbrl} performance based on these nights, which somewhat bracket extreme cases of good and bad atmospheric environments for the science operations of \gls{ctao}. The~data sets~are as follows:
\begin{itemize}
\item \texttt{D-I} The night of 5 July 2018 at the \gls{uab} campus.
 \item \texttt{D-II} The night of 18 March 2022 at \gls{orm}, with~clean atmospheric conditions. This should resemble the standard for \gls{ctao} operations.
 \item \texttt{D-III} During the last two weeks of August 2021, an~approximately ten-day-long Saharan dust intrusion event occurred, the~so-called calima~\citep{Barreto:2022}. Calima breached the usually stable inversion layer and significantly degraded air quality above the observatory. In~the first days, the~mineral dust concentration was so high that multiple scatterings made an accurate analysis of the constituent aerosols almost impossible. In~the last days of the event, when Saharan dust spread over a large part of the Atlantic Ocean (see Figure~\ref{fig:eumetsat-calima}), the~concentration of scatterers decreased. The~analysis presented here refers to data collected in the evenings of 25 and 26 of August 2021. 
\item \texttt{D-IV} On 19 September 2021, the~Tajogaite volcano on the Cumbre Vieja mountain ridge erupted in the southern part
of the La Palma island. The~volcano was located at a distance of about \unit[14]{km} 
toward the south-south-east of the \gls{orm}. In~the following days, a~dust plume spread
over the whole island. During~the measurements in the evening of 22 September 2021, a~vertical scan of the sky was performed. 
Figure~\ref{fig:volcano_satellite} shows satellite data taken during that event.
\end{itemize}

\unskip


\unskip

\subsection{Range-Height-Indication (RHI) Diagram}
\begin{figure}[H]
    \includegraphics[width=0.5\linewidth]{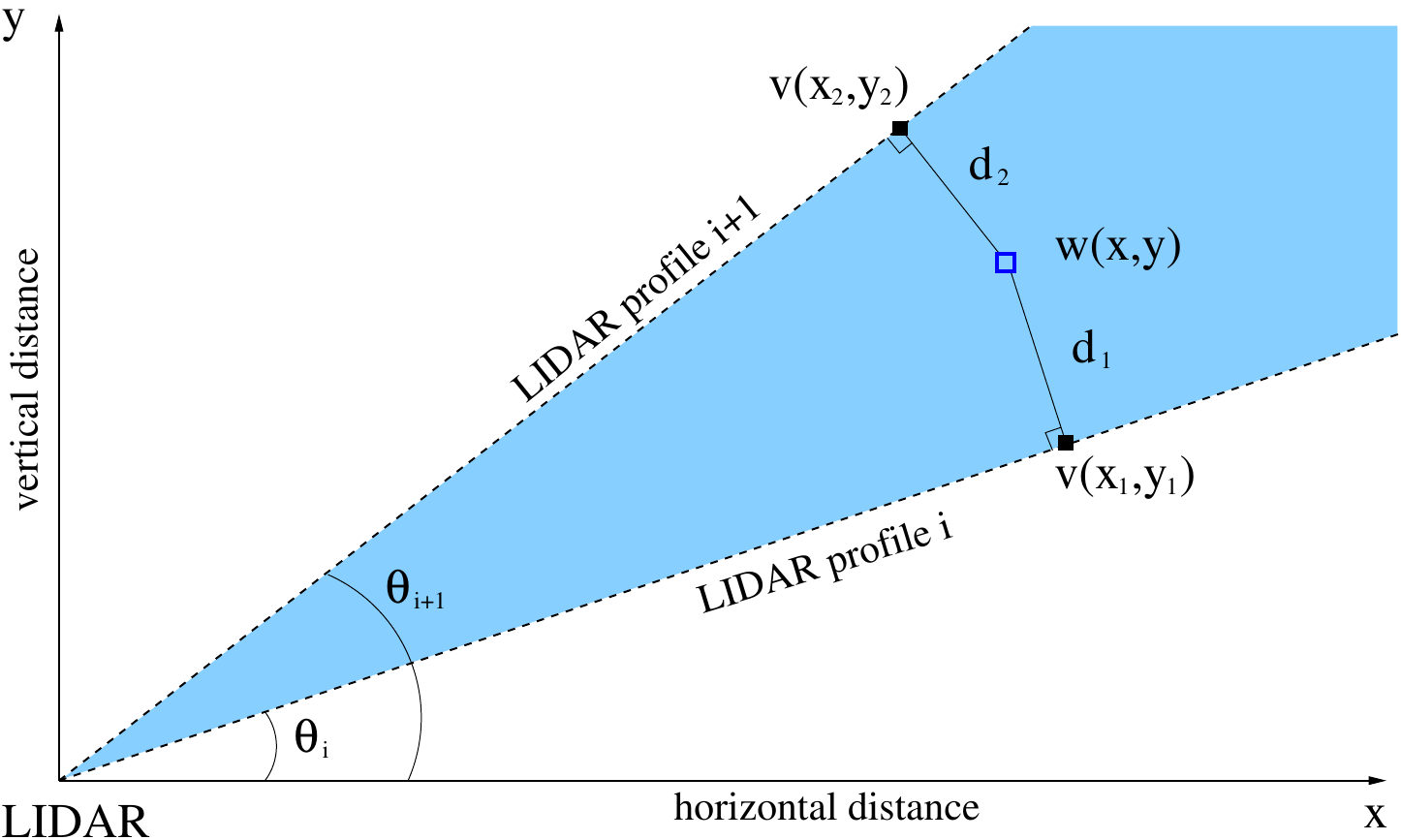}
    \caption{Two-dimensional \gls{rhi} diagrams are obtained using barycentric interpolation for the calculation of a weighted logarithm of range-square corrected LIDAR return values for pixels $w(x,y)$ between LIDAR profiles at two successive zenith angles, highlighted in blue. Values $v(x_1,y_1)$ and $v(x_2,y_2)$ from the two profiles that are closest to $w(x,y)$ are used and weighted by their respective distances from the interpolated pixel. Decreasing the zenith angle step increases the spatial resolution of the~image.}
    \label{fig:RHI_sketch}
\end{figure}

A dedicated routine creates \textit{Range-Height-Indication}  (\gls{rhi}) diagrams in Cartesian coordinates, based on \gls{pbrl} return power data at different discrete zenith angles. In~an \gls{rhi} diagram, the~horizontal axis represents the horizontal distance from the LIDAR and the vertical axis represents
the height above it. LIDAR data obtained at discrete angles fill only a fraction of pixels in an \gls{rhi} plot (only those that exactly correspond to the lines of specific LIDAR profiles) while blank pixels between successive LIDAR profiles are filled using a barycentric interpolation scheme~\cite{min:2004}, as~described in Figure~\ref{fig:RHI_sketch}. The~value at each interpolated pixel is obtained with:
\begin{equation}
    w(x,y) = \frac{v(x_1,y_1)d_2+v(x_2,y_2)d_1}{d_1+d_2},
\end{equation}
where $v(x_1,y_1)$, $v(x_2,y_2)$ are the closest measured logarithms of \gls{rcs} values in adjacent profiles and $d_1$, $d_2$ the shortest distances between the measured points and the coordinates of the interpolated~pixel.

As the \gls{pbrl} has pointing capabilities along zenith and azimuth angles, this allows us to make cross-section scans of the atmosphere, yielding 2D maps of atmospheric features. We present \gls{rhi} diagrams for \texttt{D-III, D-IV}, which are more~interesting.

\begin{figure}[H]
\centering
\includegraphics[width=0.68\textwidth,trim={2.5cm 2.9cm 5.5cm 4.7cm},clip]
{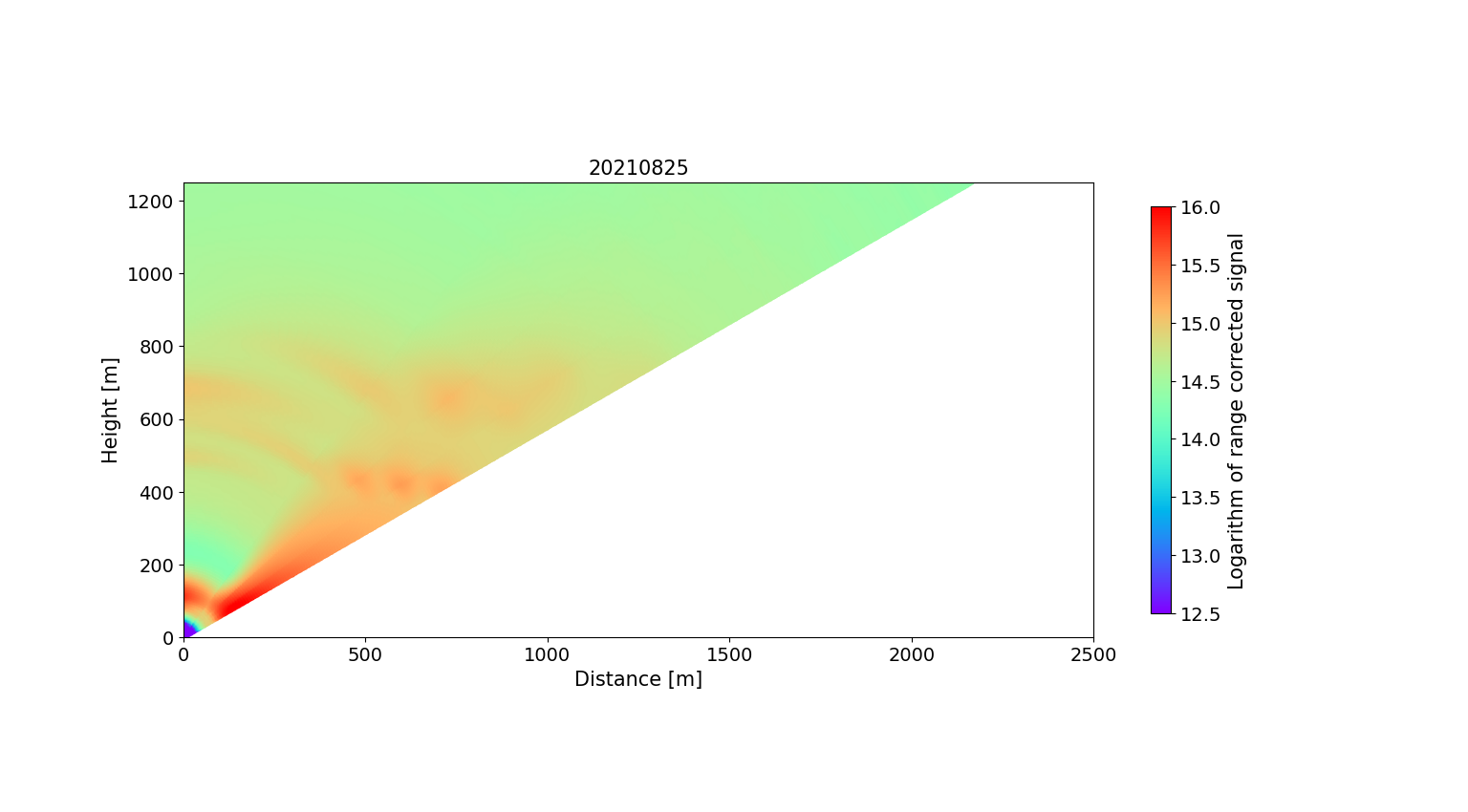}
\includegraphics[width=0.31\textwidth,trim={0.5cm 0cm 0cm 1.3cm},clip]{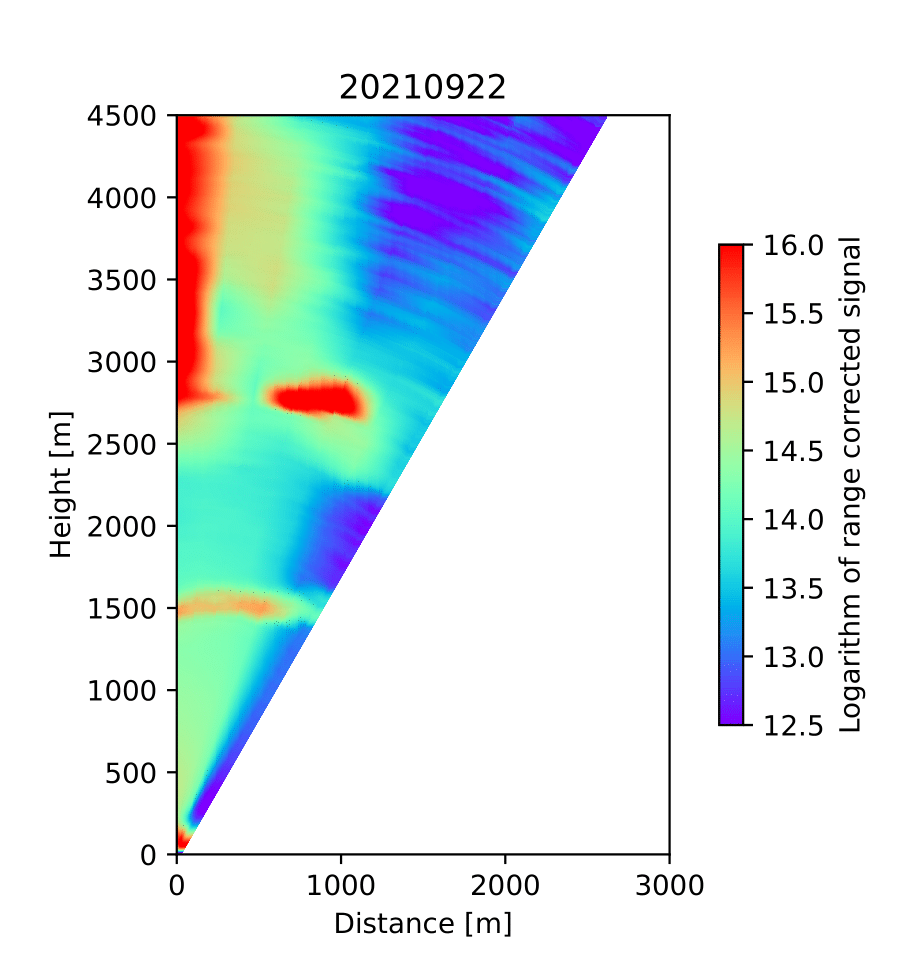}
\caption{Spatial distribution of clouds and aerosol loading above the \gls{orm} (\gls{rhi} diagrams). Left: The night of 25 August 2021 (\texttt{D-III}) with  presence of calima. The~plot is based on five \unit[90]{s} long measurements in the \unit[355]{nm} return channel at different zenith angles ranging from $0^\circ$ to $60^\circ$ in steps that correspond to regular increases from air mass of 1 to air mass of $\sim$2. Right:  The night of  22 September 2021 (\texttt{D-IV}) during the volcano event. The~figure is based on seven \unit[200]{s} long measurements in the \unit[355]{nm} return channel at different zenith angles ranging from $0^\circ$ to $30^\circ$, taken in steps of $5^\circ$. A~barycentric interpolation was used to fill the gaps. The~volcanic ash layer is clearly visible \unit[1500]{m} above the observatory (approximately \unit[3700]{m a.s.l.}).
}
\label{f:rhi-calima}
\end{figure}

Figure~\ref{f:rhi-calima} (left) shows the \gls{rhi} for the calima event, which clearly shows the extension of the atmospheric event. We can see an irregular aerosol density profile that slowly disperses to a clear atmosphere. In~Figure~\ref{f:rhi-calima} (right), we show the same scan for the volcano event. In~the data obtained, two distinct features are visible: an optically thick layer of clouds at altitudes above \unit[2300]{m a.g.l.} directly above the LIDAR, pointing at zenith; 
which dissipates out toward the north and a thinner layer located at an altitude of \unit[1500]{m a.g.l.}, which homogeneously covers the scanned sky. Together with the retrieved \angstrom exponents and LIDAR ratios, the~lower stratified layer is interpreted as the volcanic dust plume moving over the~site. 

\subsection{RCS, Extinction, Backscatter Coefficients, and~\r{A}ngstr\"{o}m Exponent~Profiles}

In the following, we report the logarithm of the \gls{rcs}, the~derived extinction and backscatter coefficients, and the LIDAR ratio and \angstrom exponent profiles for the more interesting data sets \texttt{D-III, D-IV}. 

\begin{figure}[H]
\begin{adjustwidth}{-\extralength}{0cm}
\centering
\includegraphics[width=0.32\linewidth]{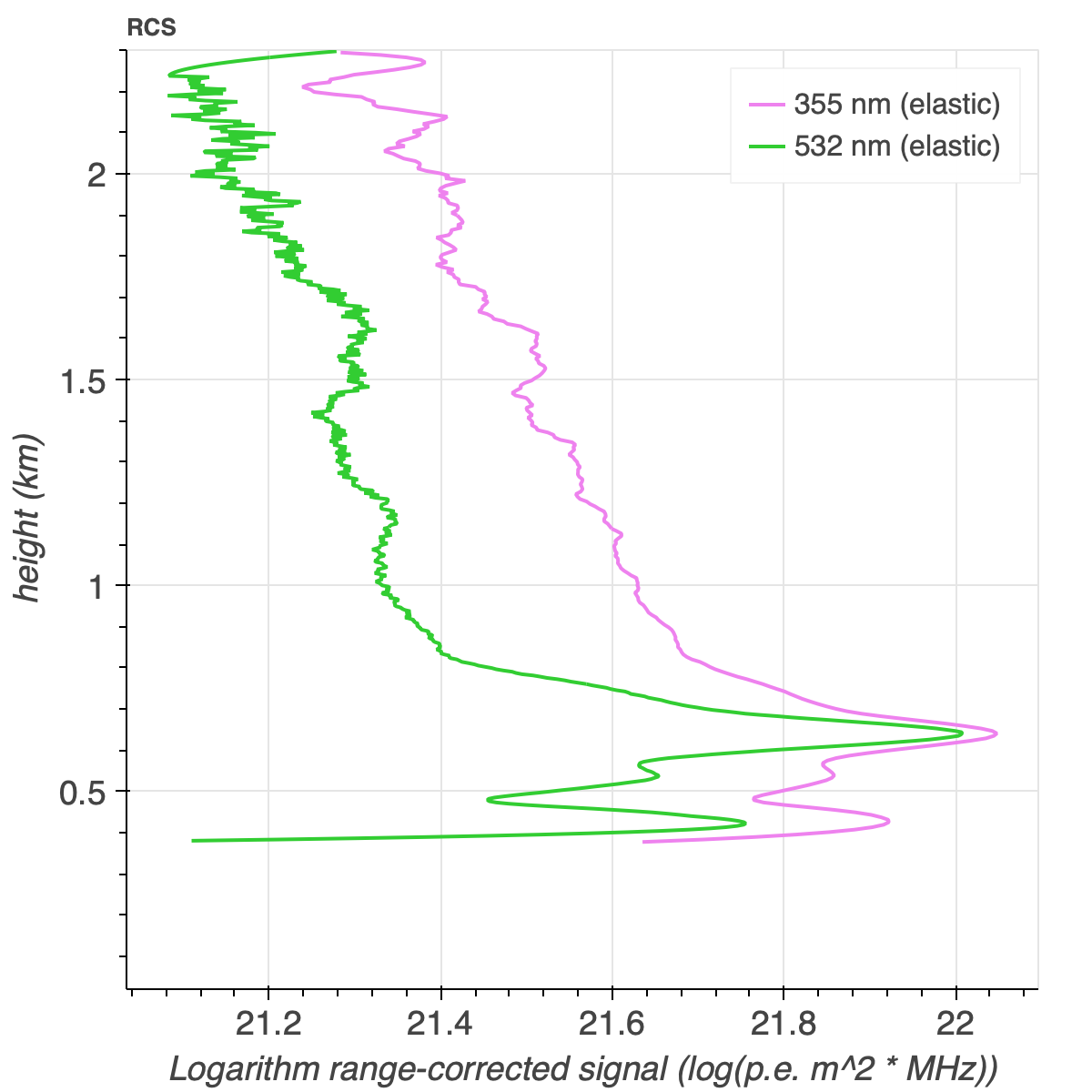}
\includegraphics[width=0.32\linewidth]{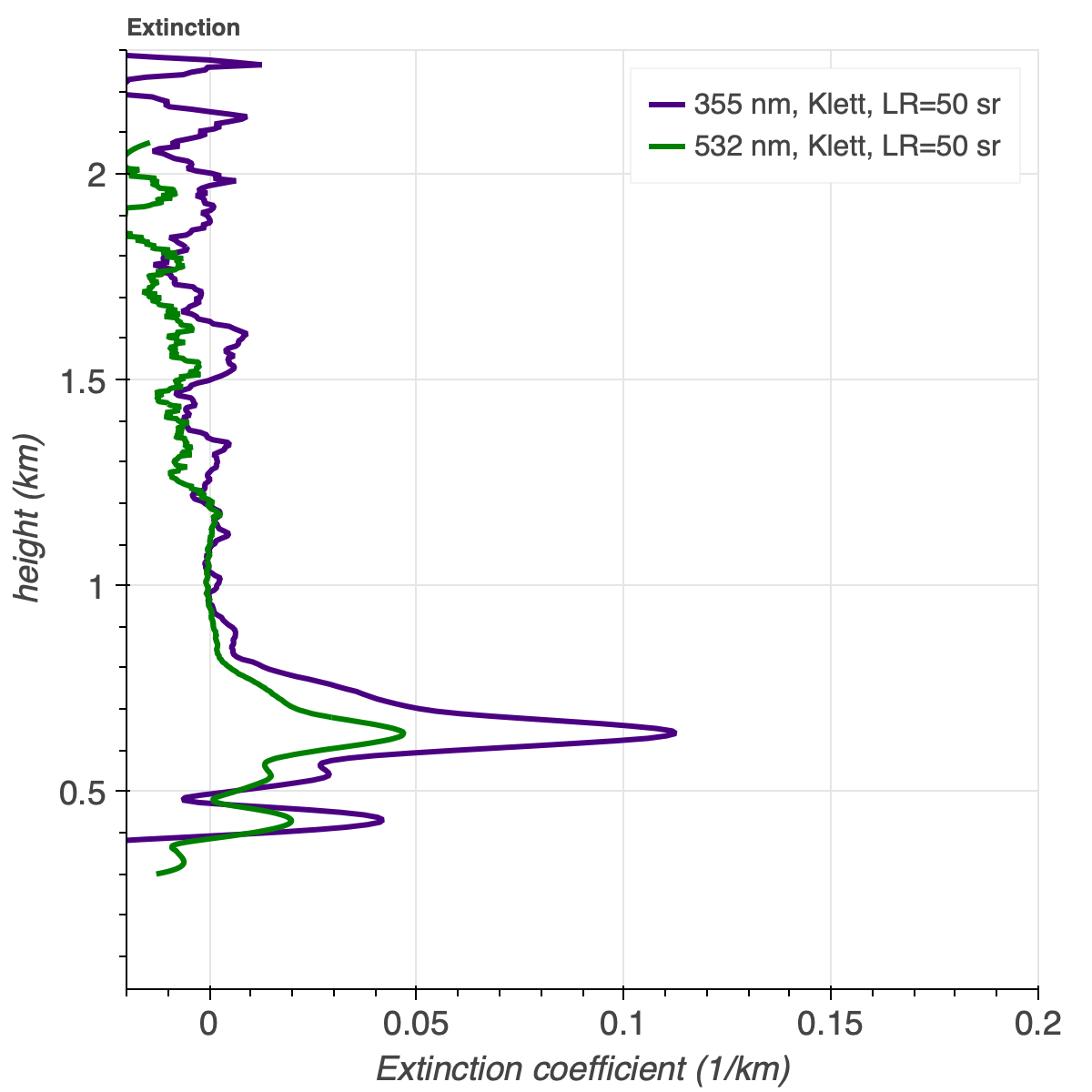}
\includegraphics[width=0.32\linewidth]{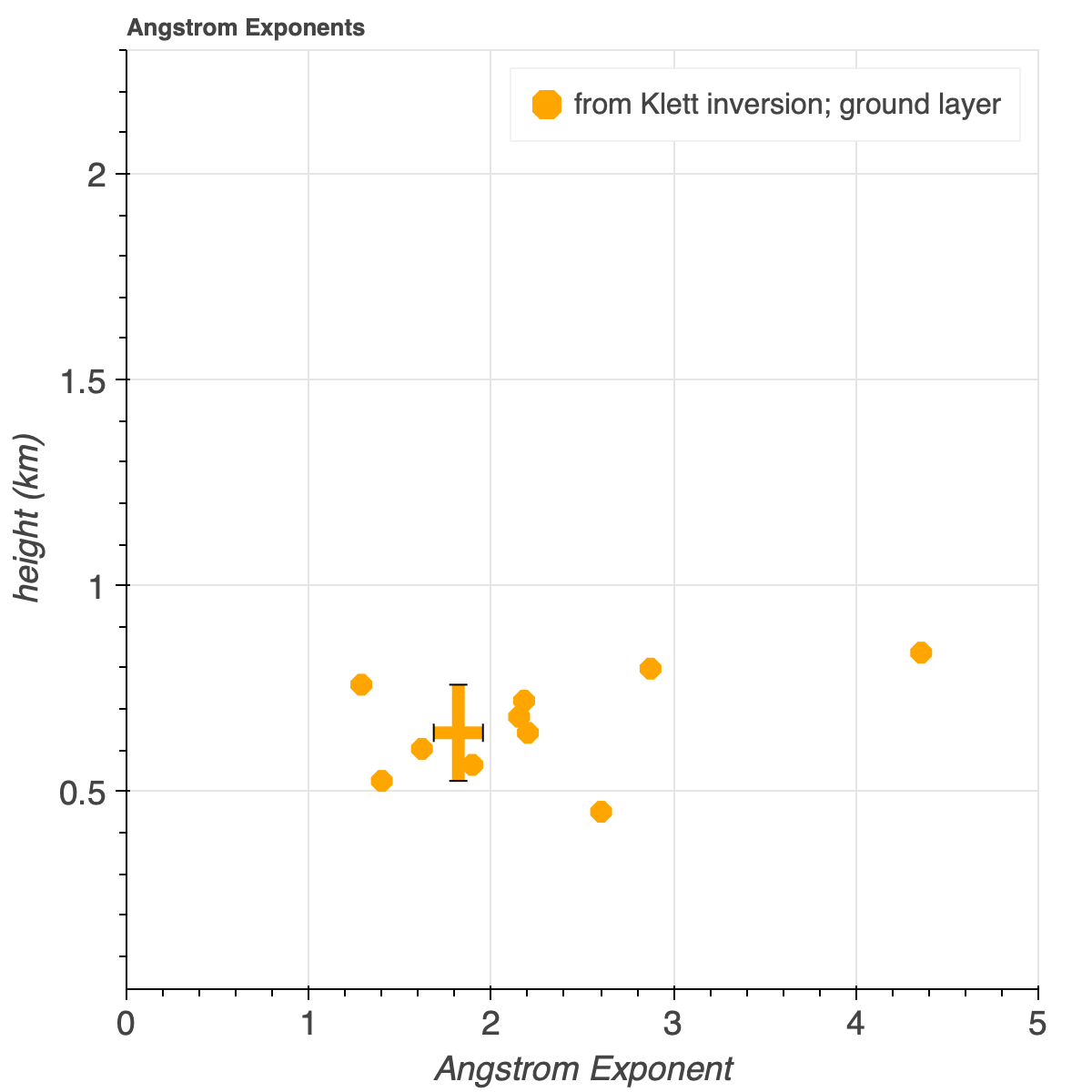}
\end{adjustwidth}
\caption{Atmospheric 
 properties on 25 August 2021 (\texttt{D-III}), during~a moderate intrusion of calima. Left: Logarithm of \gls{rcs} in the two \gls{an} channels:  \unit[355]{nm} ({pink}), \unit[532]{nm} (green), intentionally operated at a very low gain. 
The measurement was performed at a zenith angle $\theta =37^\circ$ with $N_\textit{sh}=900$. 
Center: The extinction coefficient profile of aerosols retrieved from elastic channels using Klett inversion with a supposed LIDAR ratio of 50~sr. A~Savitzky--Golay filter with a moderate filter width of 150~m has been applied.
Right: The \angstrom exponent profile retrieved from the extinction coefficients. The~points show the profile obtained with a minimum height resolution of 40~m, whereas the cross shows the average from the upper layer only, with~a resolution of 200~m. \label{fig:calima_plots}}
\end{figure}

During the calima event (\texttt{D-III}), the~Raman channel readout was unfortunately not yet sufficiently calibrated or reliable, and it has been omitted in this study. Figure~\ref{fig:calima_plots} therefore shows only the two elastic lines
to highlight the \gls{pbrl}'s capacity to retrieve \AA ngstr\"om exponents. We have assumed typical LIDAR ratios of around \unit[50]{sr}~\cite{gross:2013} for this event and found {\angstrom} exponents of 1.82 $\pm$ 0.13, which implies scattering on large, irregularly shaped particles, such as mineral dust~\cite{gross:2013, gutleben:2022, haarig:2022}. 
Furthermore, we found that the layer is structured and contains a lower part showing a slightly higher {\angstrom} exponent of $\sim$2.6.  Toward the upper end of the higher layer, the~{\angstrom} exponent rises to values consistent with accumulation-mode particle sizes.  {Note also the slightly higher {\angstrom} exponent retrieved for the lower layer centred at 450~m, and the significantly increasing {\angstrom} exponent at the upper edge of the second layer, around 800~m. }

Especially interesting is the case on 25 August 2021, where a zenith scan (Figure~\ref{f:rhi-calima}) revealed that the dust was not stratified on the top of the island, consistent with the findings of~\citep{Fruck:2022igg}.

\begin{figure}[H]
\includegraphics[width=0.485\linewidth]{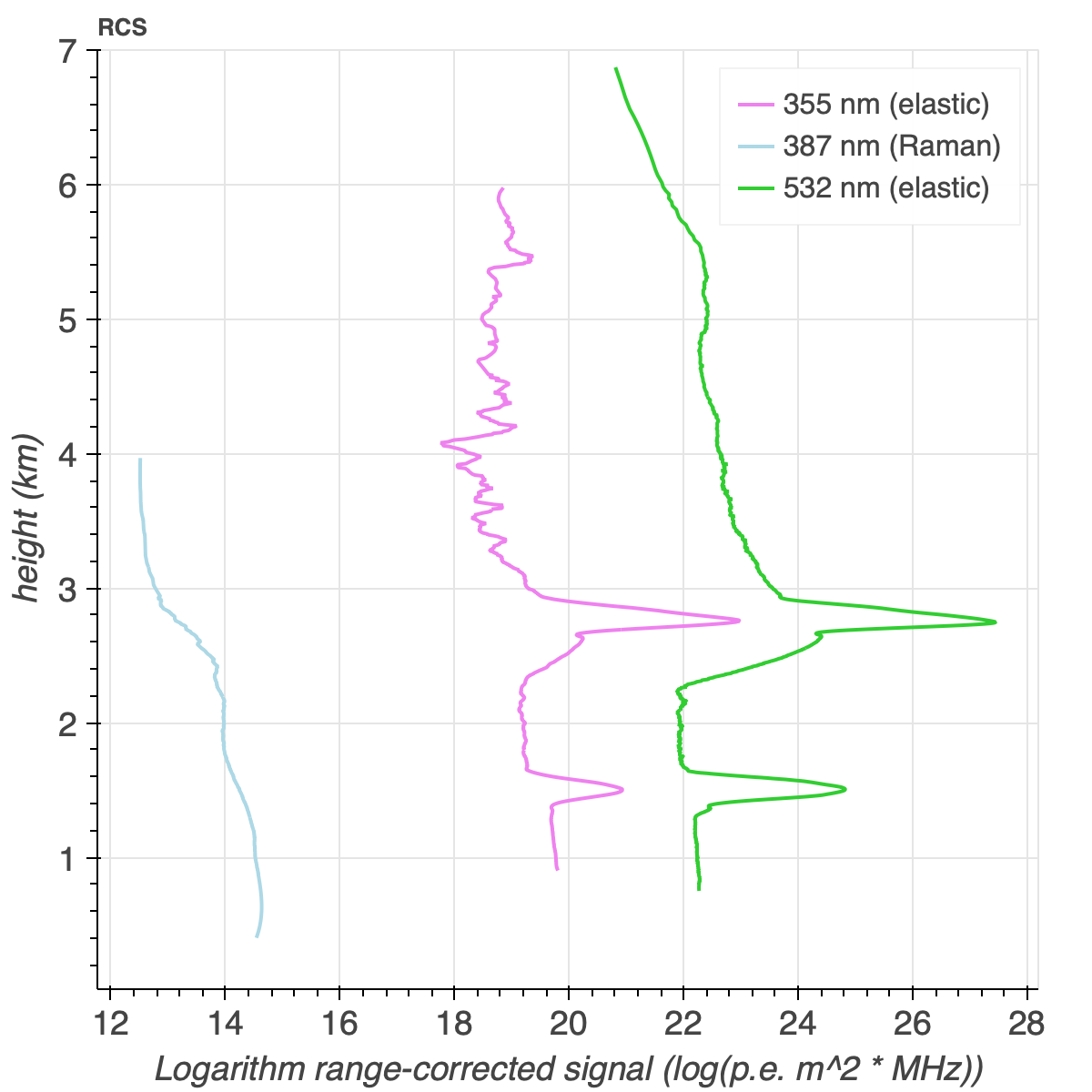}
\includegraphics[width=0.485\linewidth]
{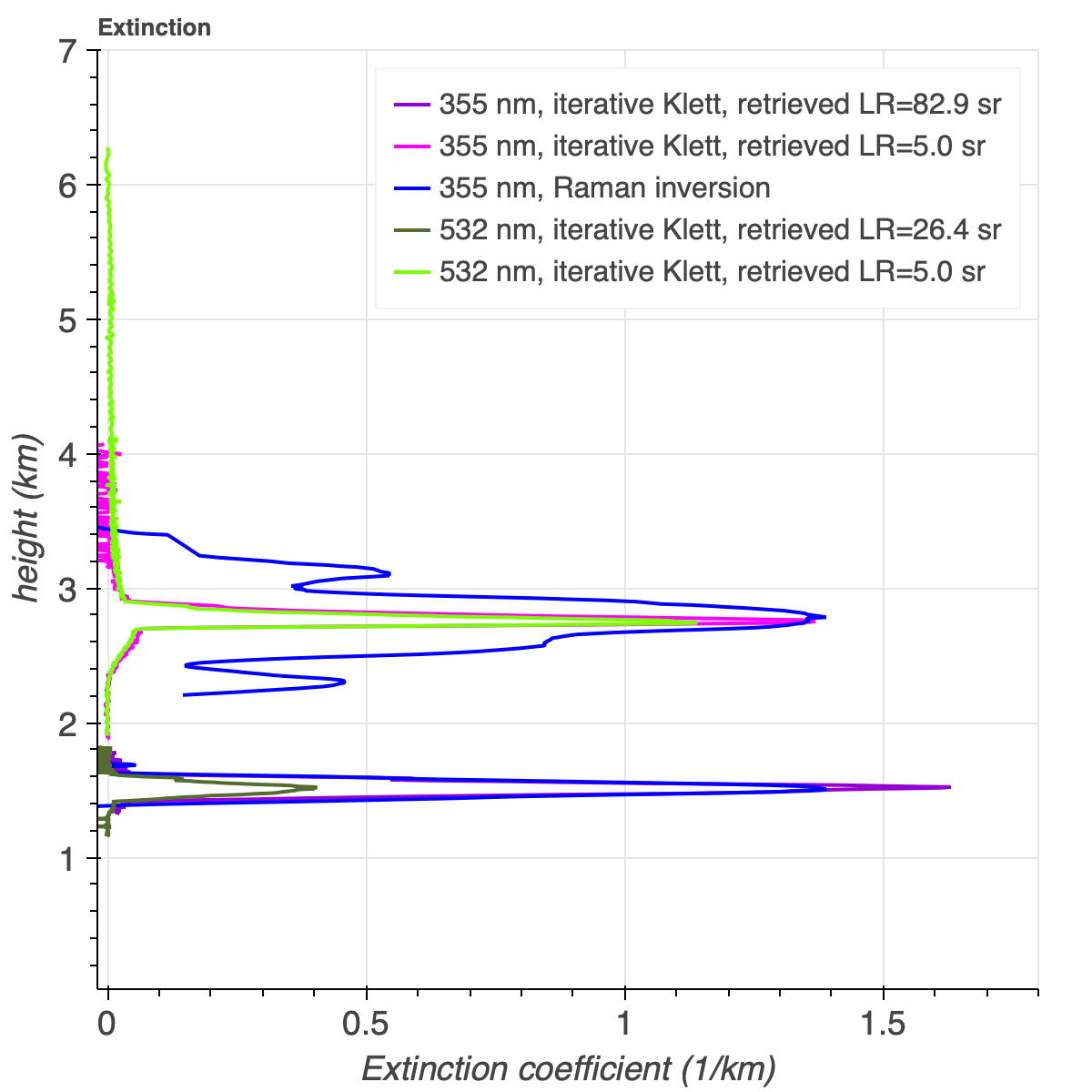}   
\\
\includegraphics[width=0.485\linewidth]{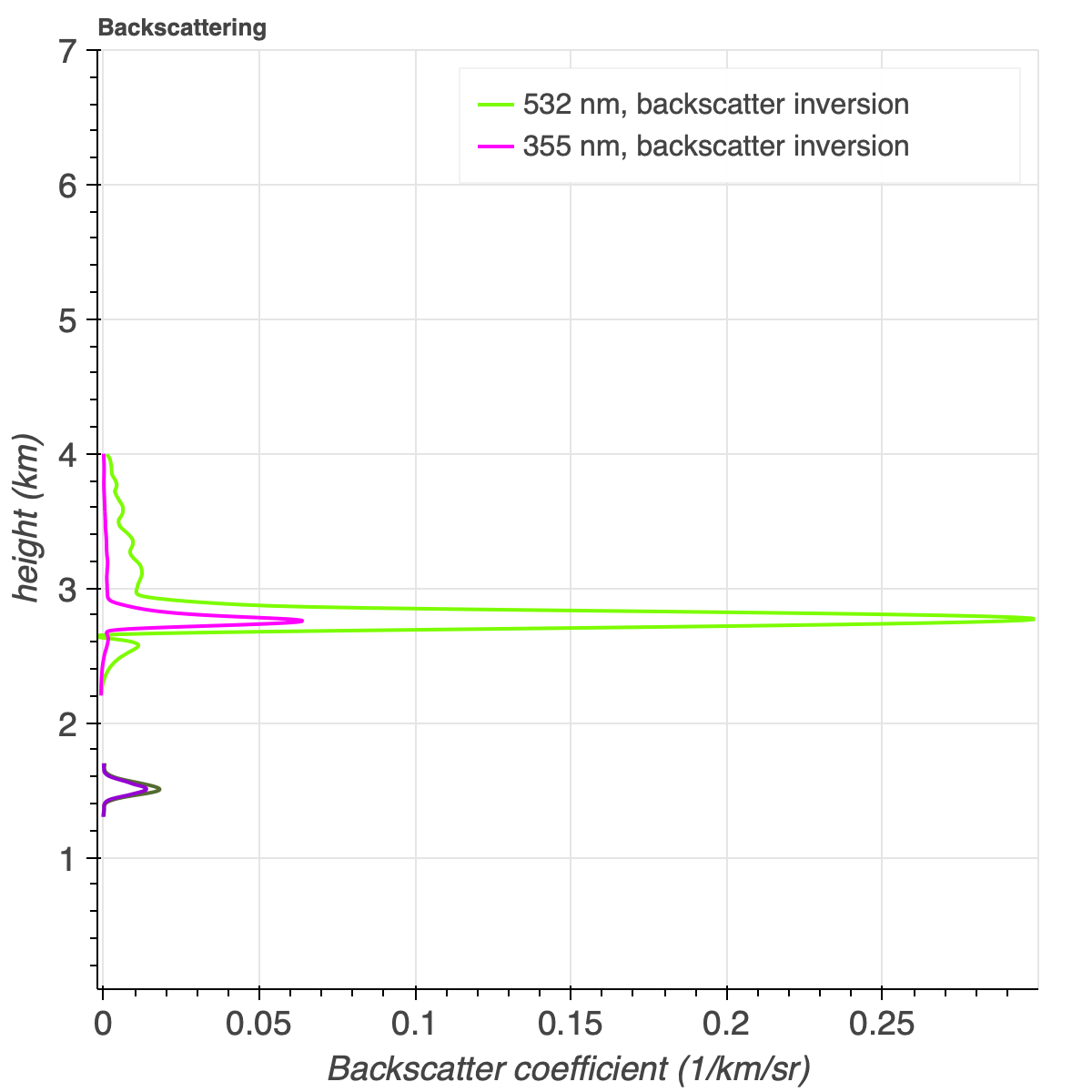}
\includegraphics[width=0.485\linewidth]{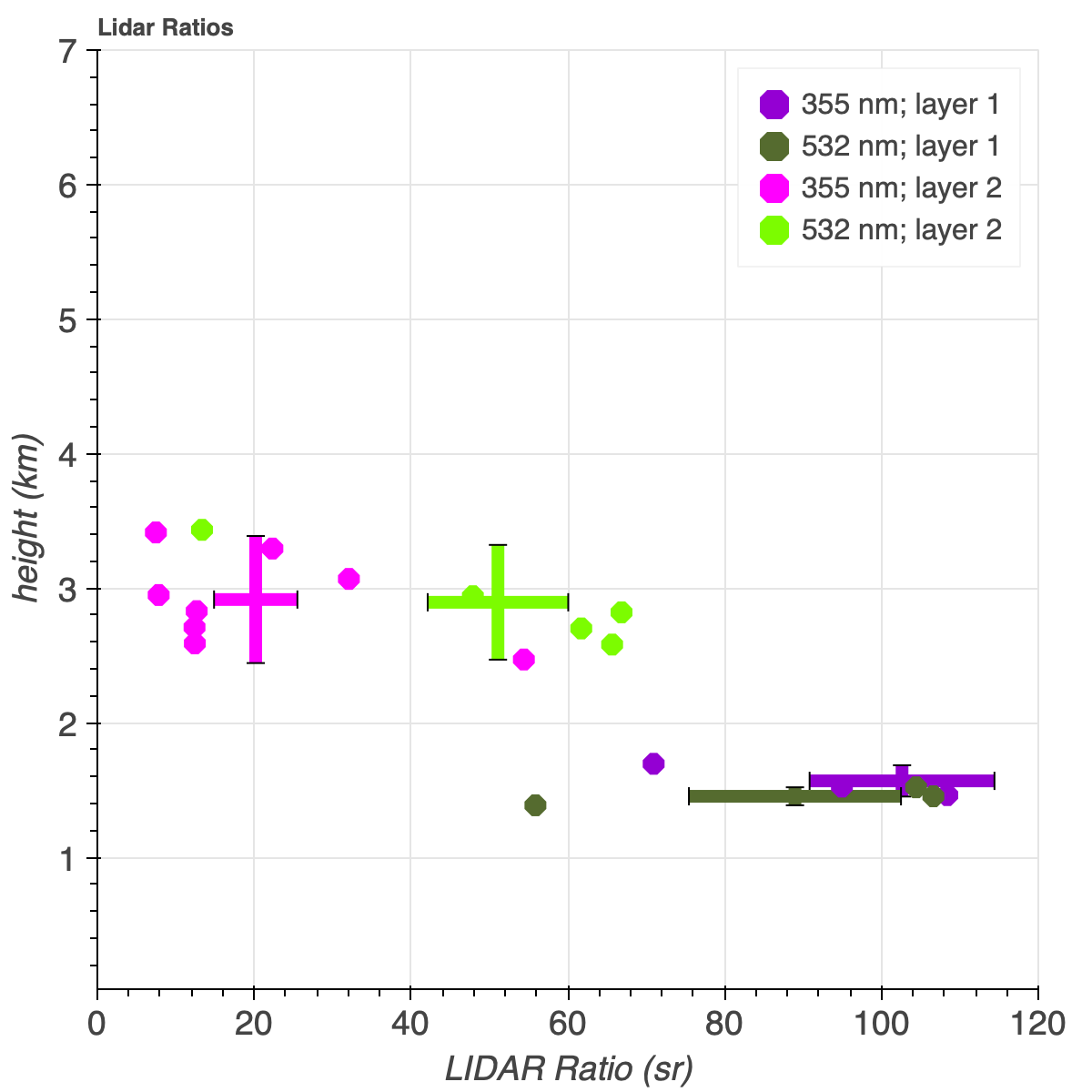}
\caption{Atmospheric 
 properties on 22 September 2021 (\texttt{D-IV}), during~a cloudy night with the Tajogaite volcano dust plume traveling over the observatory. Upper left: Logarithm of \gls{rcs} in the two \gls{an} channels:  \unit[355]{nm} ({pink}), \unit[532]{nm} (green), intentionally operated at a very low gain, and~the Raman channel at \unit[387]{nm} ({blue}). 
The measurement was performed at a zenith angle of $37^\circ$ with $N_\textit{sh}=900$. 
Upper right: The extinction coefficient profile of aerosols retrieved from elastic channels using iterative Klett inversion with LIDAR ratio retrieval, and~the Raman channel retrieval (Equation~(\protect\ref{eq:ext})).  A~Savitzky--Golay filter with a moderate filter width of 50~m has been applied.
Lower left: The backscatter coefficient profile retrieved from the combination of elastic and Raman channels using Equation~(\protect\ref{eq:back}). 
Lower right: The LIDAR ratio profile retrieved from the extinction and backscatter coefficients. The~points show the profile obtained with a minimum height resolution of 100~m, whereas the crosses show the average from the full layer. The~measurement was performed at a zenith angle of 20$^\circ$ with 2001 laser shots at \unit[10]{Hz} rate.\label{fig:volcano_plots}}
\end{figure}

The case of the volcano event 
 (\texttt{D-IV}) has been selected to highlight the \gls{pbrl}'s capacity to derive extinction and backscatter coefficients as well as LIDAR ratio profiles. Figure~\ref{fig:volcano_plots} highlights the
capability of the two elastic lines and a repaired Raman line. The~retrieved extinction profiles show two peaks: a lower one around 1.5~km altitude, where the volcano dust debris is blown over the observatory, and~another structured layer above \unit[2.2]{km}. Both layers are detected in both elastic channels and the Raman channel, though the latter has a poorer range resolution in the upper layer. Analysis of three channels separately reveals excellent agreement of retrieved extinction profiles (if the different range resolutions for the upper layer are taken into account). From~the extinction and backscatter profile, 
we obtain a LIDAR ratio of \unit[(103 $\pm$ 12)]{sr} at 355~nm and \unit[(88 $\pm$ 14)]{sr} at 532~nm for the volcano debris layer. The~LIDAR ratio profile has a range resolution of about 50~m in this case. Note that the less accurate iterative Klett inversion has provided 83~sr for the 355~nm line, but~only 26~sr in the green line, strengthening the shortcomings of an analysis based on elastic channels only. 
For the upper layer, identified as a cloud, the~LIDAR ratio is considerably smaller and, moreover, different for the two wavelengths, comparable with line-mode particulate scatterers. Note that a transition is visible at about \unit[3]{km} within the upper layer from larger scatterers (higher LIDAR ratio at 532~nm) to smaller ones (LIDAR ratio approaching the one of 355~nm).

From the combined results of both ratios and knowledge of the local environment, we can conclude that the lower aerosol loading is composed of volcanic ash that was dispersed from the Cumbre Vieja volcano plume, while the upper part must be due to a cloud with typical characteristics for the La~Palma atmosphere at those altitudes.  Even though, during~the LIDAR measurement, the~plume is found below the thick cloud layer, 
completely covering it from the satellite view, the~clockwise motion of the plume, visible in multiple consecutive satellite pictures, suggests that it was located above the \gls{orm} during the time of~measurement.

\subsection{Temporal Evolution of Atmospheric~Properties}
\begin{figure}[H]
    \includegraphics[width=0.9\linewidth]{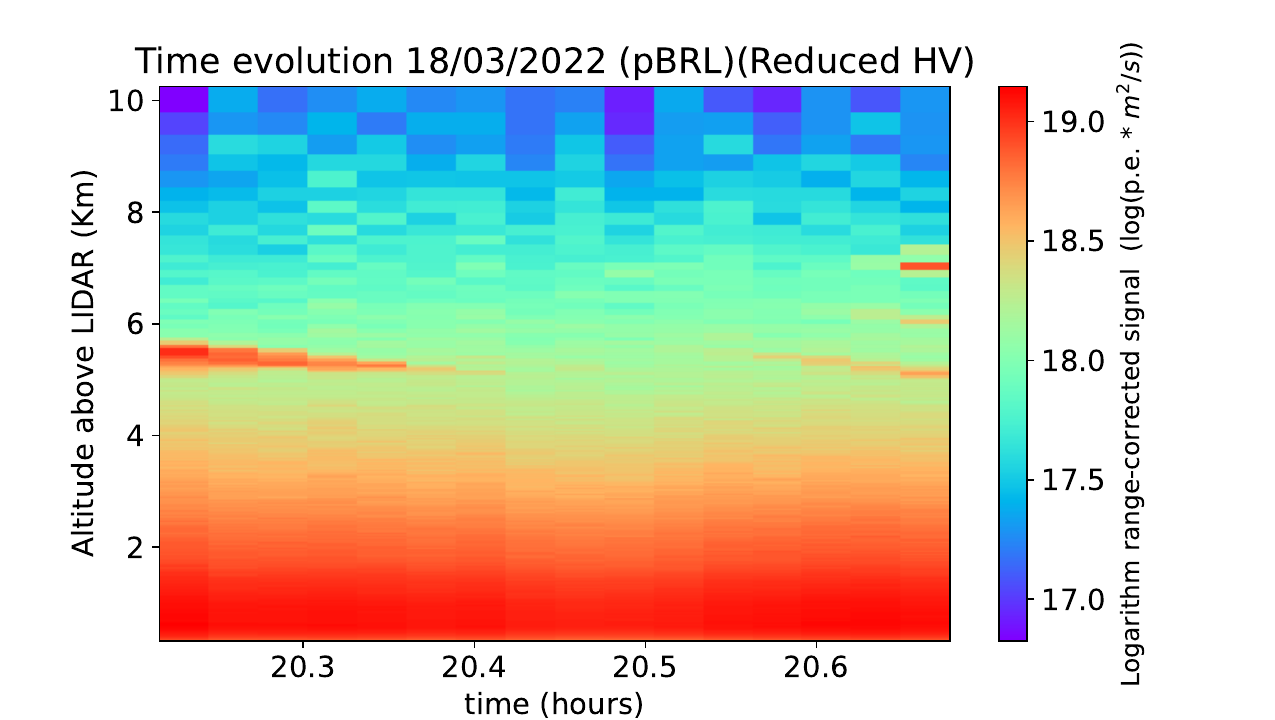}
    \caption{Time 
 evolution plot of aerosols and clouds above \gls{orm} on 18 March 2022 at a zenith angle of $45^{\circ}$ with 1000 shots per time~bin.}
    \label{fig:time_evo_clean}
\end{figure}

Our analysis also allows for a display of the temporal evolution of atmospheric properties in a specific direction with the instrument kept at a constant pointing throughout several consecutive data-taking runs. The~routine joins multiple consecutive data sets and plots a \textit{Time-Height Indication} (\gls{thi}) diagram of the logarithm of the range-corrected signal. The~colour code represents the signal values. An~example is shown in Figure~\ref{fig:time_evo_clean} for the clean night at \gls{orm} \texttt{D-II}. In~Figure~\ref{fig:time_evo_clean} we see a cloud located at $\sim$5.5~km above ground (7.7~km a.s.l.) evaporating before 8.30~pm and reappearing less intense and more dispersed in height some tens of minutes~later.

\unskip

\subsection{Maximum~Range}
\label{sec:range}
\begin{figure}[H]
    \includegraphics[width = 0.55\linewidth]{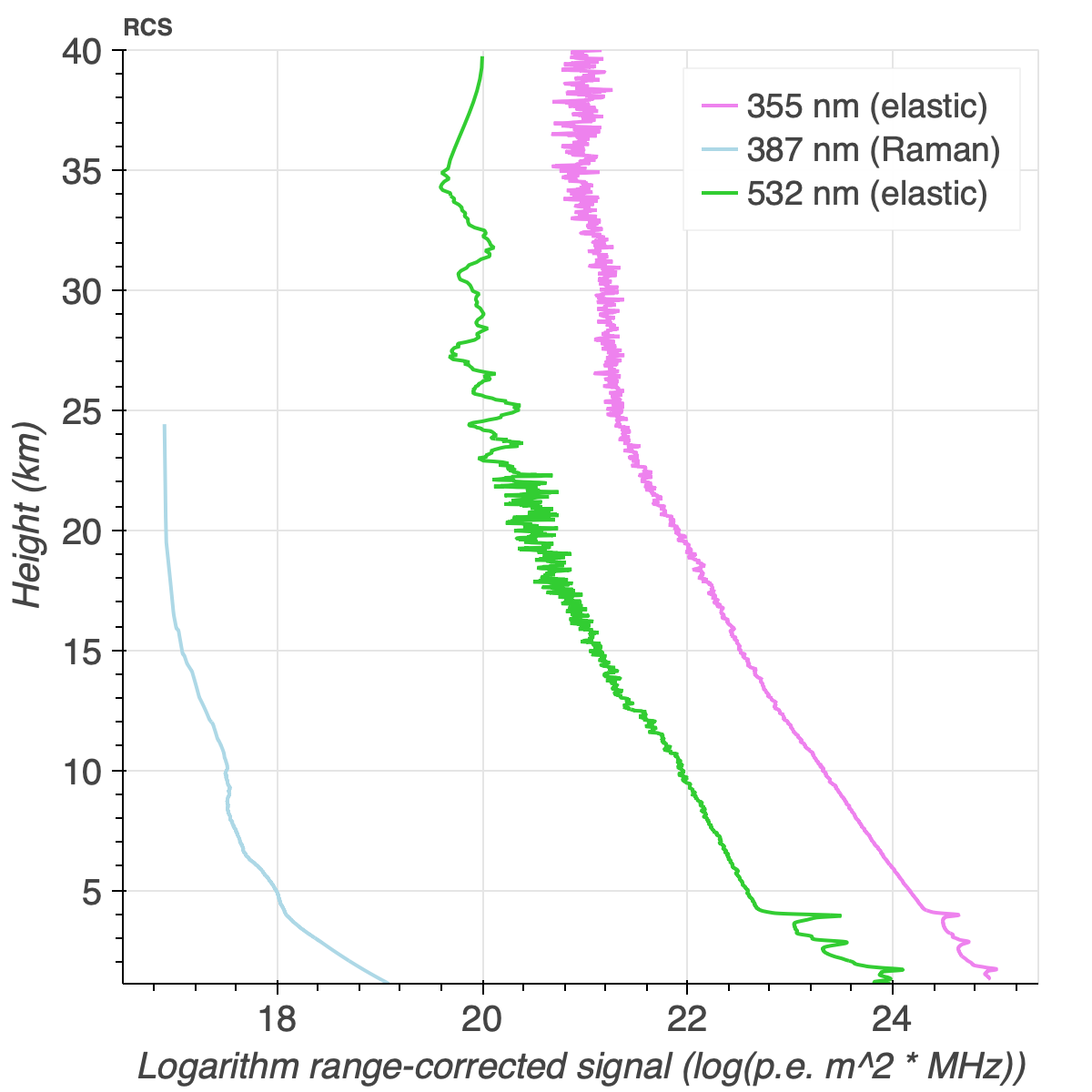}
    \caption{Logarithms 
 of range-corrected signals $\mathcal{S}$ in the two elastic channels at 355 nm (pink) and 532 nm (green) and one Raman channel at 387~nm (blue) for the vertically pointing LIDAR. The~profile has been taken in 50~s ($N_\textit{sh}=$~500 shots, at~10~Hz repetition rate). 
    \label{fig:max_range}
    }
\end{figure}

Because of the {limitations imposed on laser power and \gls{pmt} gain of the elastic channels} of the \gls{orm} data set, the~\gls{pbrl} maximum range is computed using measurements performed at the \gls{uab} campus (\texttt{D-I}, see Figure~\ref{fig:max_range}). These data were obtained before the re-aluminization of the main mirror, with~significantly reduced reflectivity. Nonetheless, the~data taken with the design value \gls{pmt} \gls{hv} settings of 1500~V and fully powered laser produced a clean signal up to the most distant ranges detectable above the ambient photon background from street~lighting.

The data collected are based on an average of 500~shots, taken within \unit[50]{s}. The~maximum range (signal-to-noise ratio greater than 1) in the elastic channels was about \unit[35]{km}, and~for the Raman signal at \unit[387]{nm} it was about \unit[20]{km}. 

\section{Discussion}
\label{sec:discussion}

{Current systems~\citep{Bregeon:2016,Fruck:2022igg} used for atmospheric monitoring of \glspl{iact} use a single-line elastic LIDAR with 25~$\upmu$J power (250~Hz \gls{prf}) or a system of two elastic lines with $O(100)$~mJ laser power and 10~Hz \gls{prf}. Both systems rely on a 60~cm primary mirror with no Raman line detection capabilities, and both versions achieve maximum ranges beyond 20~km; however, only the LIDAR of \citet{Fruck:2022igg} is steerable. While the version based on two elastic lines of \citet{Bregeon:2016} achieves measurements of \angstrom exponents with a resolution of $\sim$0.3 for periods affected by biomass burning, neither of the two systems has been capable of retrieving LIDAR ratios on their own. These limitations have led to excess fluctuations in the reconstruction of gamma-ray fluxes from the Crab Nebula (used as reference standard candle) in excess of 10\% for LIDAR-corrected data (see Figure~5 of \citet{Devin:2019}), 
incompatible with the \gls{ctao} requirements. The~system developed by \citet{Fruck:2022igg}, evaluated on Crab Nebula data~\citep{Schmuckermaier:2023huo}, 
achieves 8\% systematic uncertainty for the \gls{vaod} of the ground layer, still beyond the requirements of \gls{ctao} and limited by the missing wavelength dependence of the aerosol extinction and the lack of precise measurements of the LIDAR ratio. The~considerably worse performance of the MAGIC LIDAR on clouds is due to the LIDAR's pointing offset with respect to the telescopes and will be solved for \gls{ctao} by the introduction of an accompanying stellar photometer~\citep{Ebr:2021,Prester:2024}.}

{In addition to LIDARs fully incorporated in \gls{iact} data analysis,  the~INFN Raman LIDAR (\gls{irl}) has been taking data at the \gls{orm} since 2018 for site characterization purposes. This system operates with a single laser line of 355~nm only (6~mJ pulse energy, 100~Hz \gls{prf}), without~steering capabilities~\citep{Iarlori:2019}. It has proven to retrieve LIDAR ratios above $\gtrsim$1~km height~\citep{Iarlori:2025} but no \angstrom exponents.} 

{Perhaps the solution of a LIDAR system that comes closest to the \gls{brl} needs and is used for aerosol and cloud monitoring for a major astroparticle physics observatory is the LIDAR system of the Pierre Auger Observatory (\gls{pao})~\citep{benzvi2006}. Interestingly, the~\gls{pao} LIDAR was built upon a recycled steerable mirror system~\citep{Aglietta:1993aa}, such as the \gls{pbrl}, consisting of three 80~cm parabolic mirrors with parallel optical axes and a large \gls{fov} ($f/\# = 0.5$). Together with a 100~$\upmu$J pulse energy, 333~Hz \gls{prf} laser operating at 351~nm and only one elastic line, the~\gls{pao} LIDAR covers ranges from 750~m to 25~km. Additional Raman capabilities had been added through a separate Raman LIDAR (operating at 355~nm, 10~mJ pulse energy, 20~Hz \gls{prf} and using a single 50~cm mirror), which reaches $\sim$7~km height. The~performance of the \gls{pbrl} comes close to that of the two \gls{pao} LIDARs combined, with~the benefit of comprising both elastic and Raman lines in one single system. }

{We briefly mention that the \gls{orm} had also been characterized by another system capable of Raman line detection at 607~nm~\citep{Sicard:2010}. Nonetheless, the~signal-to-noise ratios were too low in the Raman channel for a useful application of the Raman inversion to the data taken for the site characterization of the \gls{orm} (see Section~2.1 of~\citep{Sicard:2010}). Uncertainties on the elastic line \glspl{vaod} were provided, but~are published with three decimals, an~unrealistically good accuracy. }

{A detailed and systematic comparison of quasi-simultaneous data taken by the photometer of~\citet{Ebr:2021}, 
the AERONET sun photometer data~\citep{Holben:1998} taken at the \gls{orm}, the~MAGIC LIDAR~\citep{Fruck:2022igg}, and~the \gls{irl}~\citep{Iarlori:2019} is beyond the scope of this paper, but~is foreseen for a future publication. }

{The \gls{pbrl} described in this work and in~\citet{technicalpaper}, and~the \gls{brl} in construction, are therefore significantly improving upon the current generation of \glspl{rl} for \glspl{iact}, designed with special attention on meeting the \gls{ctao} requirements, presented in Section~\ref{sec:introduction}. As~we have shown, this is obtained with a specific instrument design (using a 1.8~m diameter mirror and Raman lines, mostly discussed in our accompanying paper Ref.~\citep{technicalpaper}), but~also improving on data reconstruction algorithms as presented in Section~\ref{sec:methods}, especially with new and innovative methods using robust statistics, a~greatly improved likelihood-based gluing technique including methods for the correction of baseline ringing, a~new rebinning {algorithm}, and~novel ways of layer detection and treatment. {The analysis can be further improved through the production of ``pull''-distributions of LIDAR signals compared with molecular signal predictions for a large set of free troposphere data (see, for instance, Figures~12 and~13 of~\citet{Fruck:2022igg}) and thus improve on the assessed accuracy of the molecular profile predictions. Finally, ground-layer \glspl{vaod} require an absolute LIDAR calibration to achieve the requirement of $<3$\%~\gls{rmsd} accuracy over the full range of \glspl{vaod} allowed for \gls{ctao} science data observations. }}

{In Table~\ref{tab:discussion_performance}, we present a synoptic view of the \gls{ctao} requirements and the current compliance of our \gls{pbrl}, together with expectations for the final \gls{brl}. As~a general comment and a reminder to the reader, such a comparison is hindered by two shortcomings: on the one hand, we could test the \gls{pbrl} at its final site only for a very limited time due to interoperability constraints with other instruments and have collected only a limited number of test cases. On~the other hand, most of the time, the~\gls{pbrl} operated with reduced performance to guarantee the safety of the operating personnel and equipment. Therefore, the~performance achievements of the \gls{pbrl} shown in this paper should be considered conservative. We consider that the \gls{pbrl} already meets almost all \gls{ctao} requirements, but~an improved and upgraded \gls{brl} is designed to reach full compliance, based on the \gls{pbrl} experience and lessons learnt. For~this, a~higher \gls{prf} laser is foreseen for the final \gls{brl} delivered to the \gls{ctao}, as~well as a \gls{pmt} with 14\% \gls{pde} at 607~nm. Similarly, gated \glspl{pmt} are foreseen for the elastic lines to eliminate the undesired baseline ringing and stay within the recommended operation ranges of the \glspl{lotr}.}

\begin{table}[H]
\small
    \caption{\gls{ctao} requirements 
 validation for the \gls{pbrl} (this work) and the planned upgraded final version (\gls{brl}). Legend: * Expected. \checkmark:~Validated. \checkmark!:~Partially validated. \texttimes:~Not~met.}
    \label{tab:discussion_performance}
\begin{adjustwidth}{-\extralength}{0cm}
    \begin{tabular}{m{3.5cm}<{\raggedright}m{2.5cm}<{\raggedright}ccp{8.7cm}}
          \toprule     
         \textbf{\gls{ctao} Requirement} & & \textbf{\gls{pbrl}} & \textbf{\gls{brl}} * &  \textbf{Comments} \\
         \midrule
         \multicolumn{2}{l}{\textbf{Cloud}} & & & \\
         ~Altitude Range & 2--20~km & \checkmark!  & \checkmark  &  {\small \gls{pbrl} reaches $\sim$35~km with the elastic lines at 355~nm and 532~nm, see Figure~\ref{fig:max_range} (limited by street lighting and obtained before mirror re-aluminization). This corresponds to the requirement met for elevations higher than 35$^\circ$ for the \gls{pbrl}.} \\
         ~\gls{vod} & 0.01--0.7 &  \checkmark & \checkmark & {\small The \gls{pbrl} is able to detect and resolve clouds down to \glspl{vod} of 0.01, see Figure~\ref{f:lc-1} and \glspl{vod} of at least 0.2 see~\ref{fig:volcano_plots} with greatly reduced voltage settings. Higher \glspl{vod} have not been observed, but~should be easily detectable with canonical \gls{pmt} gains of a factor of $O(100)$ higher.} \\
         ~\gls{vod} \gls{rmsd} & $<$0.03 & \checkmark! & \checkmark &{\small  Iterative Klett analysis converges to correct LIDAR ratios (also in Ref.~\citep{Fruck:2022igg}, where method was first implemented) above a sensitivity of $\gls{vod}>0.01$. Formally, the~analysis code still needs to be validated with dedicated simulations and real data sets on clouds through dedicated cross-calibrations (in preparation).} \\
         ~Base/Height \gls{rmsd} & $<$300~m & \checkmark & \checkmark & {\small Estimated $O(100)$~m, depending on cloud height and optical thickness (see Figures~\ref{f:lc-1},\ref{fig:volcano_plots}).}\\
         \midrule
         \multicolumn{2}{l}{\textbf{\gls{pbl}}} & & & \\
         ~Altitude Range & 0.5--9~km   & \texttimes & \checkmark & {\small  Current data sets do not cover full limiting altitude ranges,  no showstoppers detected to be reached for \gls{brl}, with~full operation of the near-range channels.}\\
         ~\gls{vaod} Range/\gls{rmsd} & 0.03--0.7/$<$0.03  & \checkmark! &\checkmark & {\small Currently only achieved for the elastic lines over the full altitude range, and~for elevations higher than 40$^\circ$. The~Raman line analysis retrieves the correct LIDAR ratio, but~is limited \gls{rmsd}$ \gtrsim 0.05$  by residual ringing (see Figures~\ref{fig:alphas} and ~\ref{fig:volcano_plots}). For~these limiting cases, a~continuous absolute calibration of the LIDAR (following Ref.~\citep{Fruck:2022igg}) is needed.  To~improve \gls{rmsd} for the \gls{brl}, gated \glspl{pmt} and fully operative near-range optics will be used~\cite{technicalpaper}.}\\
        ~Height \gls{rmsd} & $<$300~m & \checkmark & \checkmark & {\small  Less stringent than requirement on \gls{vod} \gls{rmsd}, see Figures~\ref{fig:alphas} and~\ref{fig:volcano_plots}.}\\
         ~\angstrom \gls{rmsd} & $<$0.3 & \checkmark & \checkmark & {\small \gls{pbrl}: requirement met with spatial resolution better than 100~m, see Figure~\ref{fig:calima_plots}. 
         Accuracies needs to be validated by dedicated cross-calibrations with other instruments (in preparation).}\\
         \midrule
         \multicolumn{2}{l}{\textbf{Pointing}} & & & \\
         ~~Elevation & $>$25$^\circ$ & \checkmark!  & \checkmark  &  {\small Accompanying technical paper Ref.~\citep{technicalpaper}.}\\
         ~~Azimuth & $0^\circ$--$360^\circ$ & \checkmark  & \checkmark  & {\small Accompanying technical paper Ref.~\citep{technicalpaper}.}\\
        \midrule    
        \multicolumn{2}{l}{\textbf{Obs. Time}} & & & \\
         ~~Extinction Profiles & $<$1~min & \checkmark   & \checkmark  &  {\small \gls{pbrl}: Obtained in 50~s with 500 shots at 10~Hz repetition rate, see Figure~\ref{fig:max_range}. \gls{brl}: Further improvements expected due to higher laser \gls{prf}. }\\
         \bottomrule
     \end{tabular}
    \end{adjustwidth}
\end{table}

{Finally, we refer the reader interested in the future interplay of the \gls{brl} with the stellar photometer~\citep{Ebr:2021} and the incorporation of the related data products (profiles of extinction coefficients and \angstrom exponents in the case of the \gls{brl}) into the calibration pipeline of the \gls{ctao} to \citet{Prester:2024} and references therein. }

\section{Conclusions}
\label{sec:conclusion}

The Cherenkov Telescope Array Observatory (\gls{ctao}) requires accurate measurements of atmospheric conditions on its two sites and includes Raman LIDARs, among~other common array elements~\citep{Gaug:2017Atmo,Ebr:2021,Prester:2024} for continuous atmospheric characterization.  The~LIDARs need to operate at wavelengths covered by the observed spectrum of Cherenkov light in the range from 300~nm to 700~nm and
need to be able to point at, or~close to, the~observed field of view of the \gls{ctao} telescopes in order to characterize its optical properties: mainly the vertical aerosol optical depths of the ground layer, clouds, possible stratospheric debris, and~their wavelength~dependence. 

We presented the prototype of a Raman LIDAR solution, the~so-called pathfinder Barcelona Raman LIDAR (\gls{pbrl}) in a second article within this special issue~\citep{technicalpaper}.  In~this document, we have tested and analyzed the performance of the \gls{pbrl} at two different test sites: the campus of the Universitat Autònoma de Barcelona (\gls{uab}) and the \gls{lst}-1 site, located within the northern site of the \gls{ctao} at the Observatorio del Roque de los Muchachos (\gls{orm}), La Palma, Spain. The~tests were carried out under various atmospheric conditions, including clear nights, episodes of dust intrusions (calima), and even during the passage of a dust plume from a nearby volcanic eruption. The~measurements allowed us to assess the robustness and stability of the instrument's performance, even though all of them were strongly hampered by strong background light, either from street lighting at UAB or the full moon at the \gls{orm}. During~these observations, the~\gls{pbrl} operated with two elastic lines at 355~nm and 532~nm and one Raman line at 387~nm. A~second Raman line at 607~nm requires a dedicated \glspl{pmt} with acceptable photon-detection efficiency in that range, which was not available for the tests. Similarly, the~near-range optics described in~\citet{technicalpaper} could not yet be used. Nevertheless, the~performance of this yet incomplete set of detection lines could be positively assessed: the Raman line could be used for independent and reliable extinction profiles of the ground layer above 400~m distance, with~a range resolution of only 50~m for the strongly absorbing layer of volcanic dust debris and about 500~m for the cloud layer located above it. In~addition, the~Raman line successfully helped us determine the backscatter profiles and associated LIDAR ratios with similar range resolutions, which already meet the strict \gls{ctao} requirements. 

Both elastic lines could be used to determine the extinction profile of different types of ground and dust layers and clouds. For~layers limited by the free troposphere from below and above, the~extinction profile, together with an estimate of the LIDAR ratio, could be delivered independently of the Raman signal, using the technique of iterative Klett inversion. The~accuracy of that LIDAR ratio retrieval was, however, still hampered by the low gain at which the \glspl{pmt} were operated at the \gls{orm} and by the strong moonlight. Nevertheless, \angstrom exponents could be successfully obtained with resolutions significantly better than the \gls{ctao} requirement of 0.3~\gls{rmsd} (although we have not yet been able to determine its accuracy with sufficient precision). The~data analysis was able to carry out such atmospheric characterization for ranges above about 400~m, independent of the LIDAR's zenith pointing angle. For~even lower distances, the~two elastic near-range lines will need to be used in the future and be inverted using the LIDAR ratios obtained above, 
where the Raman line is already~available. 

The instrument in its current configuration reaches a maximum range of about 35~km in the elastic channels and a maximum range of 20~km in the Raman channel. 
When operated at maximum gain, strong overdriving of the Licel transient receiver readout (\gls{lotr}) has resulted in some low-amplitude ringing of the photon-counting baseline, which can be corrected by our analysis, but~should be remedied in the future by upgraded hardware, such as a laser of lower power but~higher repetition rate and~additional protection to avoid the direct diffuse reflection of laser light from the guiding mirrors onto the primary mirror. Gated \glspl{pmt} for the elastic lines are an additional~improvement. 

To obtain these results, a~complete data processing and analysis software suite has been developed: the~LIDAR PreProcessing (\gls{lpp}) software. The~\gls{lpp} includes advanced statistical techniques for offset determination, signal gluing, ringing detection, absolute calibration, and~elastic and Raman signal inversion, several of which are original and presented for the first time. 
{In particular, the use of a correct and working likelihood for signal gluing, with~or without the presence of photon counting baseline ringing and the incorporation of photon counting efficiency, background fitting, the~correct introduction of the PMT excess noise, and the correct retrieval of fit parameter uncertainties, using the correct Poissonian variance from \citet{Garwood:1936}, goes much beyond the method developed by \citet{Veberic:2012}. Also novel is the use of robust statistics to obtain a reliable criterion for the number of time slices after the signal used for background estimation, as~well as the use of precalculated molecular fit $\chi^2$’s for layer detection, in~addition to the novel dynamic re-binning {algorithm}. 
}

The findings obtained from these tests will be valuable for the upgraded final version of the \gls{brl}, which is designed to become a reliable tool for accurate atmospheric monitoring of the \gls{ctao}, and thus ensure the quality of gamma-ray observations and the overall scientific output of the~observatory.

\vspace{6pt}

\authorcontributions{
Conceptualization, 
R.G., M.D., M.G.; 
methodology, 
O.B., A.C.-O., M.D., M.G., R.G.;
software, 
P.J.B.-R., A.C.-O., R.G., M.G.;  
formal analysis, 
P.J.B.-R., A.C.-O., M.D., M.G., R.G.; 
data acquisition, 
O.B., P.G.C., A.C.-O., S.M.Ç., M.D., L.F., M.G., R.G., C.M., M.M., S.S., S.U., M.Z., M.Ž.;
hardware contribution, 
O.B., M.D., L.F., M.G., M.M., S.S., M.Z.; 
investigation, 
O.B., P.G.C., A.C.-O., M.D., L.F., M.G., R.G., D.K., M.M., S.S., S.U., M.Ž.;
resources, 
O.B., M.D., L.F., M.G., M.M., S.S., M.Z.;
data curation,
P.J.B.-R., M.G., R.G., M.Ž.;
writing---original draft preparation, 
P.J.B.-R., A.C.-O., M.D., M.G., R.G., M.Ž.;
writing---review and editing, all authors; 
supervision, 
O.B., M.D., M.G., M.M., S.S.;
project administration, 
O.B., M.D., L.F., M.G., M.M.;
funding acquisition, 
O.B., M.D., L.F., M.G., M.M., S.S.
All authors have read and agreed to the published version of the~manuscript.}

\funding{This project has received funding from the European Union's Horizon Europe Research and Innovation Programme under Grant Agreement No 101131928; by the Spanish grants PID2022-139117NB-C41 and PID2022-139117NB-C43, funded by MCIN/AEI/10.13039/501100011033/FEDER, UE, the~Departament de Recerca i Universitats de la Generalitat de Catalunya (grant SGR2021 00607), and~by ``ERDF A way of making Europe'',  the~CERCA program of the Generalitat de Catalunya, and the European Union NextGenerationEU/PRTR. In~Slovenia, it was funded by the Slovenian Research and Innovation Agency, grants P1-0031, J1-3011, and I0-E018. M.D. acknowledges funds from the 2012 ``Bando Giovani Studiosi'' of the University of Padova.   R.Gr.~acknowledges funding from the FSE under the program Ayudas predoctorales of the Ministerio de Ciencia e Innovación  PRE2020-093561.
}

\dataavailability{The data supporting the conclusions of this article will be made available by the authors on request. A public image of the \gls{lpp} v1.0.0 can be found at \url{https://gitlab.cta-observatory.org/cta-array-elements/ccf/lpp_deployment}. To obtain access to the source code, please contact bauza@ieec.cat or markus.gaug@uab.cat.} 

\acknowledgments{This work would have been impossible without the support of our colleagues from the \gls{magic} and \gls{lst} collaboration and the \gls{ctao} Consortium, which we gratefully acknowledge.
We thank the Instituto de Astrof\'isica de Canarias for the excellent working conditions at the Observatorio del Roque de los Muchachos on La Palma. 
We also thank the funding agencies and institutions mentioned in the above section (Funding) for the financial support.}

\conflictsofinterest{Author Camilla Maggio was employed by the company CAEN Tools for Discovery  S.p.A solely during the editing of this manuscript. Author Paolo G. Calisse is employed by Cherenkov Telescope Array Observatory gGmbH, (CTAO gGmbH). The remaining authors declare that the research was conducted in the absence of any commercial or financial relationships that could be construed as a potential conflict of interest.} 

\begin{adjustwidth}{-\extralength}{0cm}

\reftitle{\highlighting{References} 
}

\PublishersNote{}

\end{adjustwidth}
\end{document}